\pdfoutput=1
\documentclass[11pt]{article}
\usepackage{fullpage}
\usepackage{nsd}

\usepackage{mdframed}

\title{A~Little Derandomization Goes a~Long Way:\\Unconditional Circuit Lower Bounds using Low-Degree PIT}
\author{
   Nikolai Chukhin
   \thanks{JetBrains Research. Email: \url{buyolitsez1951@gmail.com}}
   \and
   Karthik Gajulapalli
   \thanks{University of Central Florida. Email: \url{karthik.complexity@gmail.com}}
   \and
   Alexander Golovnev
   \thanks{Cornell University. Email: \url{alexgolovnev@gmail.com}}
   \and
   Ivan Mihajlin
   \thanks{JetBrains Research. Email: \url{ivmihajlin@gmail.com}}
}
\date{}
\begin{document}
\pagenumbering{roman}
\maketitle

\begin{abstract}
We~prove unconditionally that either $\cc{NE}$ contains a~language requiring Boolean circuits of~size at~least $2^n/(2n)$, or~$\cc{VP}^0_{\Z}\ne\cc{VNP}^0_{\Z}$, a~constant-free version of~Valiant's conjecture over the integers.
The Boolean alternative implies that $\NP\not\subseteq\SIZE[n^k]$ for every constant $k\ge1$.
Our proof gives a~fine-grained variant of~the connection between Polynomial Identity Testing ($\cc{PIT}$) and circuit lower bounds established by~Kabanets and Impagliazzo (2004).
We~show that the specific identity tests required by~this argument can be~performed over the integers without a~derandomization assumption.

Over a~fixed finite field~$\F$, we~first give a~deterministic $O(st^{d-1})$-time algorithm for $\cc{PIT}$ on~size-$s$ circuits computing degree-$d$ polynomials in~$t$ variables.
We~further show that, for any fixed $d\ge2$ and $\varepsilon>0$, improving this running time by~a~factor of~$t^{\varepsilon}$, even nondeterministically, implies either the same Boolean lower bound or~an~arithmetic circuit lower bound of~$t^{1+\varepsilon/2}$ for a~polynomial family from $\cc{VNP}_{\F}$.
If~nondeterministic $\cc{PIT}$ instead runs in~time $s^{o(d)}t^{(1-\varepsilon)d}$ for infinitely many degrees~$d$, then the arithmetic lower bound strengthens to~$\cc{VP}_{\F}\ne\cc{VNP}_{\F}$.
Over the integers, we~give a~deterministic $\cc{PIT}$ algorithm for syntactically set-multilinear circuits running in~time $s^{O(1)}t^{5d/6}$, which yields the unconditional result above.

Combining the Kabanets--Impagliazzo argument with range avoidance, we~show that an~$\NP$ algorithm for $\cc{PIT}$ over~$\Z$ gives a~purely arithmetic win--win: either $\cc{VP}_{\Z}\ne\cc{VNP}_{\Z}$, or~constant-degree polynomials of~maximum circuit complexity can be~constructed in~$\cc{FNP}$.

Along the way, we~show that arithmetic circuits of~size $s(N)$ for the permanent, or~for an~explicit family in~$\cc{VNP}$, give perfectly complete input-oblivious Merlin--Arthur protocols running in~time $s(O(n))\,n^{O(1)}$ for $3$-$\cc{UNSAT}$ with $O(n)$ clauses; thus ruling out faster protocols yields arithmetic lower bounds that scale with the protocol's running time.
\end{abstract}

\thispagestyle{empty}
\newpage

\pagenumbering{arabic}

\tableofcontents
\clearpage
\section{Introduction}\label{sec:introduction}

Proving circuit lower bounds is one of the central challenges of complexity
theory.
Such bounds establish computational hardness directly: they show that every
circuit solving a given problem must be large.
A boolean circuit is a computation built from simple boolean operations, such as AND,
OR, and NOT gates and its size is the
number of gates.
Every polynomial-time algorithm can be simulated by polynomial-size
circuits with very small overhead, so a superpolynomial circuit lower bound for $\cc{SAT}$ would
separate $\cc{P}$ from $\NP$.

Despite the simplicity of this model, our understanding of unrestricted
circuits remains very limited.
For a function in $\cc{P}$, the strongest known lower bound is
$3.5n-o(n)$, due to Mishra and Santhanam~\cite{MS26}, improving the
earlier bounds $(3+1/86)n-o(n)$~\cite{FGHK16} and
$3.1n-o(n)$~\cite{LY22}.
Moving to the much larger class
$\cc{E}^{\NP}=\cc{DTIME}[2^{O(n)}]^{\NP}$, which allows exponential-time
computation with an $\NP$ oracle, did not help: until very recently, no
superlinear lower bound was known even there~\cite{GKST18}.
This changed with a recent breakthrough of OpenAI~\cite{OpenAI26}: a
deterministic $2^{O(n/\sqrt{\log n})}$-time algorithm for $3$-$\cc{SAT}$ on
$n$ variables with $O(n)$ clauses, refuting the Exponential Time
Hypothesis.
By known connections between satisfiability algorithms and circuit lower
bounds~\cite{IP01,W13,JMV18}, this algorithm yields the first superlinear
lower bound for $\cc{E}^{\NP}$: a language that requires circuits of size
$\Omega(n\sqrt{\log n})$ at infinitely many input lengths.
Yet even an algorithmic advance of this magnitude lifts the lower bound
only by a $\sqrt{\log n}$ factor above linear, and the following modest improvement remains open.

\begin{mainopenproblem}\label{op:boolean_lb}
Is there a language in $\cc{E}^{\NP}$ that requires Boolean circuits
of size $\Omega(n\log n)$?
\end{mainopenproblem}

The algebraic analogue of a Boolean circuit replaces Boolean gates by
arithmetic operations.
An \emph{arithmetic circuit} over a ring $R$ computes from variables and
constants of $R$ using addition and multiplication; each gate computes a
polynomial, and the circuit computes the polynomial of its output gate.
Valiant's classes $\cc{VP}$ and $\cc{VNP}$ formalize the distinction
between efficient algebraic computation and algebraic
counting~\cite{V79a,V82}.
The class $\cc{VP}_R$ consists of families of polynomials with polynomially
many variables and polynomially bounded degree that have polynomial-size
arithmetic circuits over $R$.
The class $\cc{VNP}_R$ consists of the families $\{P_t\}$ of the form
\[
    P_t(x)=\sum_{y\in\{0,1\}^{m(t)}} Q_t(x,y),
\]
where $m(t)$ is polynomially bounded and $\{Q_t\}\in\cc{VP}_R$.
Thus $Q_t$ plays the role of an efficient verifier, and $P_t$ sums it over
exponentially many Boolean witnesses $y$, much as a $\#\cc{P}$ function
counts the accepted witnesses of a polynomial-time verifier.

The central example is the permanent,
\[
    \PERM_m(X)=\sum_{\sigma\in S_m}\prod_{i=1}^{m}X_{i,\sigma(i)}.
\]
On a $0$--$1$ matrix, it counts the perfect matchings of the corresponding
bipartite graph, and computing it is
$\#\cc{P}$-complete~\cite{V79}; this connects the permanent to counting
satisfying assignments and other counting problems.
Over every field $\F$ of characteristic different from two, the permanent
is also $\cc{VNP}_\F$-complete under projections~\cite{V79a}, so proving
that it requires superpolynomial-size arithmetic circuits over $\F$ is
equivalent to separating $\cc{VP}_\F$ from $\cc{VNP}_\F$.
The algebraic structure of polynomials provides additional tools for
proving lower bounds, yet the strongest known lower bound against general
arithmetic circuits for an explicit polynomial of polynomially bounded
degree is $\Omega(t\log t)$, where $t$ is the number of
variables~\cite{S73,BS83,SY10}, and no better bound is known over any
field.

\begin{mainopenproblem}\label{op:arithmetic_lb}
For a fixed finite field $\F$, is there an explicit family of polynomials
in $t$ variables, of degree polynomially bounded in $t$, whose arithmetic
circuit complexity over $\F$ is $\omega(t\log t)$?
\end{mainopenproblem}

Both targets are far less ambitious than the long-term goals of near-maximal
Boolean hardness, on the scale $2^n/n$, and exponential arithmetic hardness
of the $m\times m$ permanent, on the scale $2^{\Omega(m)}$.
Yet even they are beyond current techniques.

Perhaps surprisingly, both questions are closely connected to a central
problem in derandomization.
\emph{Polynomial identity testing} ($\cc{PIT}$) asks whether a given
arithmetic circuit computes the zero polynomial; testing whether two
circuits compute the same polynomial reduces to this problem by taking
their difference.
Randomness makes this task easy: evaluate the circuit at a random point
from a sufficiently large set, and a nonzero polynomial is unlikely to
vanish.
Over small finite fields one can evaluate in an extension field, and over
the integers one can compute modulo a random prime to keep the numbers
short~\cite{DL78,Z79,S80}.
Thus $\cc{PIT}$ has a randomized polynomial-time algorithm, whereas the
best known deterministic algorithms for general circuits take exponential
time.

Kabanets and Impagliazzo~\cite{KI04} showed that derandomizing this single
problem would already imply major circuit lower bounds, even if the
derandomized algorithm is nondeterministic and runs in subexponential time.
Together with subsequent refinements~\cite{KMS12,AM11}, their argument
gives the following statement.

\begin{theorem}[\cite{KI04,KMS12,AM11}]\label{thm:KI}
Assume that $\cc{PIT}_{\Z}\in\cc{NSUBEXP}$.
Then at least one of the following holds:
\begin{enumerate}
    \item $\cc{NEXP}\cap\cc{coNEXP}\not\subseteq\PPoly$;
    \item $\PERM\notin\cc{VP}_{\Z}$, and hence
          $\cc{VP}_{\Z}\ne\cc{VNP}_{\Z}$.
\end{enumerate}
Under the stronger assumption $\cc{PIT}_{\Z}\in\NP$, the Boolean
alternative can be strengthened to
\[
    \cc{NE}\cap\cc{coNE}
    \not\subseteq\SIZE\!\left[\frac{2^n}{2n}\right].
\]
\end{theorem}

Here $\cc{NSUBEXP}=\bigcap_{\varepsilon>0}\cc{NTIME}[2^{N^\varepsilon}]$,
where $N$ is the input length.
Thus the polynomial-time hypothesis yields either an arithmetic separation
or a nearly maximal Boolean lower bound for a class contained in
$\cc{E}^{\NP}$.\footnote{For the subexponential-time statement, the
comparison is different: neither containment between $\cc{NEXP}$ and
$\cc{E}^{\NP}$ is known. In contrast,
$\cc{NE}\cap\cc{coNE}\subseteq\cc{E}^{\NP}$.}

The use of the integers is essential: the argument relies on the $\#\cc{P}$-completeness of the $0$--$1$ permanent over~$\Z$. Over a fixed finite field of characteristic~$p$, the permanent of a $0$--$1$ matrix gives only this count modulo~$p$. For $p=2$, it equals the determinant and is computable in polynomial time. For odd~$p$, it is not known to be $\#\cc{P}$-hard, or even $\NP$-hard under deterministic reductions~\cite{CIKK15}.

This argument also verifies identities involving the permanent, whose degree grows with the matrix dimension.\footnote{The
$m\times m$ permanent has degree $m=\sqrt t$ in $t=m^2$ variables.
For qualitative polynomial-time or subexponential-time hypotheses,
polynomially bounded degree can be reduced to a degree bound of
$\sqrt t$ by adjoining polynomially many unused variables.
This padding does not preserve the quantitative running times studied
here.}
How much of this identity-testing power is actually needed?
In particular, could faster algorithms for polynomials of constant degree
already imply new lower bounds?
Over a fixed finite field, constant-degree $\cc{PIT}$ is already in
deterministic polynomial time: computing, gate by gate, the coefficients of
all monomials of degree at most $d$ solves it in time $O_d(st^d)$ for a
circuit of size $s$ in $t$ variables.
The question is therefore quantitative.

\begin{quote}
\textbf{Question 1.}
How weak a derandomization of $\cc{PIT}$ suffices to imply new circuit
lower bounds? In particular, can a non-trivial speed-up for constant-degree
polynomials, for which polynomial-time algorithms are already known, have
such consequences?
\end{quote}

One can also ask whether the derandomization assumption can be removed
altogether.
Here the precedents concern classes whose verifiers may use randomness.
If the permanent has polynomial-size arithmetic circuits over $\Z$, then
Merlin can send such circuits and Arthur can check them using the downward
self-reducibility of the permanent and randomized identity
tests~\cite{KI04}; this places $\cc{P}^{\#\cc{P}}$, and hence
$\Sigma_2^{\cc{P}}$, in $\cc{MA}$.
Combined with Kannan's theorem~\cite{K82}, this gives a win--win: either
$\PERM\notin\cc{VP}_\Z$, or $\cc{MA}\not\subseteq\SIZE[n^k]$ for every
constant $k$.
Santhanam~\cite{Santhanam09} removed the arithmetic alternative at the cost
of one bit of advice, proving that $\cc{MA}/1\not\subseteq\SIZE[n^k]$ for
every constant $k$.
At exponential time, Kabanets and Impagliazzo~\cite{KI04} also proved,
unconditionally, that either $\cc{NEXP}^{\cc{RP}}\not\subseteq\PPoly$ or
the permanent requires superpolynomial-size arithmetic circuits
over~$\Z$; again, the Boolean alternative concerns a class with randomness.

More generally, fixed-polynomial lower bounds are known for
$\Sigma_2^{\cc{P}}$~\cite{K82}, for symmetric polynomial time
$\cc{S}_2^{\cc{P}}$~\cite{C07,CHLR26} and its oblivious
counterpart~\cite{GLV24}, and for $\cc{PP}$~\cite{V05,A06}; moreover, symmetric exponential time requires near-maximum circuit
size~\cite{CHLR26}.
Each of these classes uses a second quantifier, randomness, or counting.
For $\NP$ and $\cc{NE}$, whose verifiers are deterministic and use a single
quantifier, no such bounds are known.

This leaves a sharper goal: remove the identity-testing assumption from the
Kabanets--Impagliazzo theorem while keeping the Boolean lower bound for a
class with a deterministic verifier.

\begin{quote}
\textbf{Question 2.}
Can we prove an unconditional Boolean--arithmetic win--win in which the
Boolean lower bound holds for $\cc{NE}$, without randomized verification or
advice?
\end{quote}

We answer both questions.
For Question~1, we show that, over a fixed finite field, shaving a
polynomial factor off the best known deterministic running time for
constant-degree $\cc{PIT}$, even with a nondeterministic algorithm, would
resolve Open Problem~\ref{op:boolean_lb} or Open
Problem~\ref{op:arithmetic_lb}.
For Question~2, we show that over the integers the identity tests needed by
our argument can be performed unconditionally, which gives the main theorem
below.

\subsection{Our Results}\label{section:our_results}

Our main result answers Question~2.
It~is~an~unconditional win--win between nearly maximal Boolean hardness
for~$\cc{NE}$ and a~constant-free version of~Valiant's conjecture.
An~arithmetic circuit over~$\Z$ is~\emph{constant-free} if~the only
constants at~its leaves are $1$ and~$-1$.
Let $\cc{VP}^0_{\Z}$ be~the class of~polynomial families computed by
constant-free circuits of~polynomially bounded size and formal
degree,\footnote{Every leaf, including a~constant, has formal degree one;
a~sum gate has the maximum of~the formal degrees of~its children, and
a~product gate has their sum. Bounding the formal degree also bounds the
bit lengths of~the coefficients computed at~all gates.}
and let $\cc{VNP}^0_{\Z}$ consist of~exponential sums of~$\cc{VP}^0_{\Z}$
families over polynomially many Boolean variables, as~in~the definition
of~$\cc{VNP}$.

\begin{restatable}{maintheorem}{ThmUnconditionalDichotomy}
\label{theorem:unconditional_dichotomy}
	At~least one of~the following holds:
	\begin{enumerate}
		\item $\cc{NE}\not\subseteq\SIZE[2^n/(2n)]$;
		\item $\cc{VP}^0_{\Z}\ne\cc{VNP}^0_{\Z}$.
	\end{enumerate}
\end{restatable}

We remark that replacing $\cc{NE}$ even by $\cc{NTIME}[2^{o(n)}]$ in the Boolean circuit lower-bound alternative of \cref{theorem:unconditional_dichotomy} would already yield the unconditional separation $\cc{VP}^0_{\Z}\ne\cc{VNP}^0_{\Z}$. Indeed, the assumption $\cc{VP}^0_{\Z}=\cc{VNP}^0_{\Z}$ implies $\NP\subseteq\PPoly$~\cite{B09}, and by a simple padding argument, $\cc{NTIME}[2^{o(n)}]\subseteq\SIZE[2^{o(n)}]\subseteq\SIZE[2^n/(2n)]$.

Both alternatives in \cref{theorem:unconditional_dichotomy} are far beyond current techniques.
The first is~a~nearly maximal circuit lower bound for~$\cc{NE}$.
Since $\cc{NE}\subseteq\cc{E}^{\NP}$, it~resolves
Open Problem~\ref{op:boolean_lb}, and by~padding it~implies
$\NP\not\subseteq\SIZE[n^k]$ for every constant~$k$.
The second is~a~constant-free version of~Valiant's conjecture.
It~follows from $\cc{VP}_{\F}\ne\cc{VNP}_{\F}$ for any single
field~$\F$ (\Cref{proposition:vp0_to_vpf}), and from
$\PERM\notin\cc{VP}_{\Z}$, since the permanent belongs
to~$\cc{VNP}^0_{\Z}$.
It~also follows from the Boolean hypothesis
$\cc{PP}\not\subseteq\PPoly$: if~$\cc{VP}^0_{\Z}=\cc{VNP}^0_{\Z}$, then the
permanent has constant-free circuits of~polynomial size and formal degree,
and therefore $\cc{PP}\subseteq\PPoly$~\cite[Lemma~2.12]{B09}.

Compared with \Cref{thm:KI}, \Cref{theorem:unconditional_dichotomy}
requires no derandomization hypothesis.
Relative to~the strongest form of~\Cref{thm:KI}, which assumes
$\cc{PIT}_{\Z}\in\NP$, the price is~that the hard language lies
in~$\cc{NE}$ rather than $\cc{NE}\cap\cc{coNE}$, and that the arithmetic
alternative is~the constant-free separation rather than
$\PERM\notin\cc{VP}_{\Z}$.
In~contrast to~the unconditional results recalled in~the introduction,
the Boolean alternative concerns a~class whose verifier is~deterministic
and uses no~randomness, no~advice, and no~second quantifier.

\subsubsection{The Route to the Main Theorem: Constant-Degree Identity Testing}
\label{section:results_constant_degree}

To~prove \Cref{theorem:unconditional_dichotomy}, we~first answer
Question~1.
We~determine how fast an~algorithm for constant-degree identity testing
must be~before it~implies new lower bounds.
We~then show that, over the integers, the required speed-up can be
achieved unconditionally for the identities that arise in~our proof.
We~consider the following promise problem.

\begin{mdframed}[nobreak=true]
    \textbf{Problem: Promise $f(t)$-$\cc{PIT}_{\mathcal R}$}
    \paragraph{Input:} An~arithmetic circuit $C$ over a~ring $\mathcal R$ of~size $s$ with $t$ input variables, promised to~compute a~polynomial of~degree at~most $f(t)$.
    \paragraph{Output:} Yes, if~$C$ computes the zero polynomial.
\end{mdframed}
Here, the size of~a~circuit is~the number of~its gates together with the
bit-sizes of~all constants used by~the circuit.
The promise concerns only the output polynomial; intermediate gates may
compute polynomials of~larger degree.
We~write $d$-$\cc{PIT}_{\mathcal R}$ when the degree bound is~a~constant~$d$,
and $\cc{PIT}_{\mathcal R}$ for the same problem without a~promise on~the
degree.
In~all statements over a~finite field~$\F$, the field is~fixed, and the
constants hidden in~the $O(\cdot)$ notation may depend on~it.

\paragraph{A~simple improvement.}
Over a~fixed finite field, computing, gate by~gate, the coefficients
of~all monomials of~degree at~most~$d$ solves $d$-$\cc{PIT}_{\F}$ in~time
$O_d(st^d)$, as~noted in~the introduction.
Our first result saves a~factor of~$t$.
The algorithm evaluates the circuit only at~points with at~most $d-1$
nonzero coordinates.
The only nonzero polynomials of~degree at~most~$d$ that vanish at~all
these points are homogeneous multilinear of~degree~$d$, and such
polynomials are detected through their first-order partial derivatives.
The construction of~Baur and Strassen~\cite{BS83} computes all these
derivatives simultaneously, with only a~constant-factor increase in
circuit size (\Cref{section:low_degree_pit}).

\begin{restatable}{maintheorem}{ThmLowDegreePITUpper}
\label{theorem:low_degree_pit_upper}
	Let $\F$ be a fixed finite field, and let $d\ge1$
	be a constant.
	Then $d$-$\cc{PIT}_{\F}$ on circuits of size~$s$
	with $t$ variables can be solved deterministically
	in time $O_d(st^{d-1})$.
\end{restatable}

Over~$\Z$, there is~no~analogous baseline.
Even a~circuit without variables can compute an~integer with
exponentially many bits, and testing whether it~computes zero is
polynomial-time equivalent to~general $\cc{PIT}_{\Z}$~\cite{ABKM09}.
For this reason, our results on~constant-degree identity testing are
stated over fixed finite fields.

\paragraph{Slightly faster algorithms imply strong lower bounds.}
\Cref{theorem:low_degree_pit_upper} is~a~simple improvement, so~it~is
natural to~ask how much further one can go.
We~show that any further polynomial saving, even by~a~nondeterministic
algorithm, would imply circuit lower bounds beyond current techniques.

\begin{restatable}{maintheorem}{ThmPITFixedDegree}
\label{theorem:pit_fixed_degree}
	Let $\F$ be a finite field.
	Suppose that
	$d$-$\cc{PIT}_{\F}\in\NTIME[s^a t^b]$
	for a fixed integer~$d$ and constants $a\ge1$ and $b\ge0$
	satisfying $a+b<d$.
	Then at least one of the following holds:
	\begin{enumerate}
		\item $\cc{NE}\not\subseteq\SIZE[2^n/(2n)]$;

		\item There is an explicit family $\{P_N\}$ of polynomials
		in $O(N)$ variables, depending only on the characteristic
		of~$\F$, such that, for every
		$1<\kappa<(d-b)/a$,
		\[
			\operatorname{asize}_{\F}(P_N)>N^\kappa
		\]
		for infinitely many~$N$.
	\end{enumerate}
\end{restatable}

Here $\operatorname{asize}_{\F}(P)$ denotes the smallest size of an arithmetic circuit over $\F$ computing~$P$.

The running time of~\Cref{theorem:low_degree_pit_upper} corresponds
to~$a=1$ and $b=d-1$, exactly at~the threshold $a+b=d$.
Hence saving any factor~$t^{\varepsilon}$ already suffices.
If~$d$-$\cc{PIT}_{\F}$ can be~solved in~nondeterministic time
$O(st^{d-1-\varepsilon})$ for some $d\ge2$ and $\varepsilon\in(0,1)$,
then either $\cc{NE}\not\subseteq\SIZE[2^n/(2n)]$, or
$\operatorname{asize}_{\F}(P_N)>N^{1+\delta}$ for every
$0<\delta<\varepsilon$ and infinitely many~$N$.
Since the best known lower bound for explicit polynomials
is~$\Omega(t\log t)$~\cite{S73,S75,BS83}, the first alternative resolves
Open Problem~\ref{op:boolean_lb}, and the second resolves
Open Problem~\ref{op:arithmetic_lb}.

The family $\{P_N\}$ is~the \emph{universal polynomial}
$\mathcal U=\{\mathcal U_N\}$ of~\Cref{section:universal_polynomial},
whose reduction modulo the characteristic of~$\F$ lies
in~$\cc{VNP}_{\F}$.
Fixing all but $N$ of~its $10N+1$ variables appropriately turns
$\mathcal U_N$ into the generating polynomial of~the satisfying
assignments of~any prescribed $3$-CNF formula on~$N$ variables.
Consequently, all polynomials in~our proofs are projections
of~$\mathcal U$ (see \Cref{section:proof_overview}).

\paragraph{A~sufficient condition for $\cc{VP}_{\F}\ne\cc{VNP}_{\F}$.}
Since $\mathcal U$ does not depend on~the degree~$d$, savings for
infinitely many degrees combine.
If~the exponent of~$s$ grows sublinearly in~$d$ and the exponent of~$t$
is~at~most $(1-\varepsilon)d$, then every polynomial size bound
for~$\mathcal U$ is~excluded.
The arithmetic alternative then becomes a~separation of~$\cc{VP}_{\F}$
from~$\cc{VNP}_{\F}$.

\begin{restatable}{maincorollary}{CrlyPITNPSize}
\label{corollary:pit_np_size}
	Let $\F$ be a finite field.
	Suppose that there are a constant $\varepsilon\in(0,1)$
	and a function $a(d) \geq 1$ with $a(d)=o(d)$ such that
	\[
		d\text{-}\cc{PIT}_{\F}
		\in
		\NTIME\!\left[s^{a(d)}t^{(1-\varepsilon)d}\right]
	\]
	for infinitely many integers~$d$.
	Then at least one of the following holds:
	\begin{enumerate}
		\item $\cc{NE}\not\subseteq\SIZE[2^n/(2n)]$;

		\item $\cc{VP}_{\F}\ne\cc{VNP}_{\F}$.
	\end{enumerate}
\end{restatable}

The second alternative also implies $\cc{VP}_{\Z}\ne\cc{VNP}_{\Z}$, since
a~separation over any finite field gives one over the
integers~\cite{V79a}.

Our proofs connect faster identity testing to~faster nondeterministic
algorithms for Boolean circuit satisfiability.
Assuming that $\mathcal U$ has small circuits, we~use the hypothesized
$\cc{PIT}$ algorithm to~decide unsatisfiability of~Boolean circuits
of~size $2^{\rho n}$ in~nondeterministic time $O(2^{(1-\rho)n})$, for
a~constant $\rho>0$.
Similarly to~the proofs of~Williams~\cite{W13,W14}, Murray and
Williams~\cite{MW18}, and Chen~\cite{Che25}, this algorithm certifies
truth tables of~high circuit complexity at~infinitely many input lengths.
The same small circuits yield Merlin--Arthur protocols
for~$\Sigma_2\cc{P}$, and these truth tables derandomize them, which gives
$\cc{NE}\not\subseteq\SIZE[2^n/(2n)]$ (\Cref{section:ne_lower_bound}).

\paragraph{Closing the loop over the integers.}
To~prove \Cref{theorem:unconditional_dichotomy}, we~obtain the speed-up
required by~\Cref{corollary:pit_np_size} unconditionally, over the
integers, for the identities that arise in~our proof.
These identities are set-multilinear.
Let the variables be~split into disjoint blocks $X_1,\ldots,X_d$ of~at
most $m$ variables each.
A~polynomial is~\emph{set-multilinear of~type} $B\subseteq[d]$ if~each
of~its monomials is~the product of~exactly one variable from every
block~$X_j$ with $j\in B$.
A~circuit is~\emph{syntactically set-multilinear} if~its gates~$\alpha$
can be~assigned types $B(\alpha)\subseteq[d]$ such that a~leaf labeled
by~a~variable of~$X_j$ has type~$\{j\}$, a~leaf labeled by~a~nonzero
integer has type~$\varnothing$, a~leaf labeled by~zero has any type, both
children of~a~sum gate have its type, and the children of~a~product gate
have disjoint types whose union is~its type.
Then every gate~$\alpha$ computes a~set-multilinear polynomial~$f_\alpha$
of~type~$B(\alpha)$.
For a~polynomial~$f$ over~$\Z$, let $\|f\|_1$ be~the sum of~the absolute
values of~its coefficients, and let $f_\Phi$ be~the polynomial computed
by~a~circuit~$\Phi$.
A~polynomial of~type~$[d]$ has up~to~$m^d$ coefficients; we~test whether
it~is~zero while saving a~factor of~$m^{\lceil d/6\rceil}$ over computing
all of~them.

\begin{restatable}{maintheorem}{ThmZeroTest}\label{theorem:zero_test}
	Let $d\ge2$, and let the variables be~split into $d$~blocks of~at~most $m$ variables each.
	There is~a~deterministic algorithm that, given an~integer $L\ge1$ and a~syntactically set-multilinear circuit~$\Phi$ over~$\Z$ with $s$ gates in~these variables whose output gate has type~$[d]$, behaves as~follows.
	\begin{itemize}
		\item If~$f_\Phi\ne0$, then it~rejects.
		\item If~$f_\Phi=0$ and every gate~$\alpha$ of~$\Phi$ satisfies $\|f_\alpha\|_1\le2^L$, then it~accepts.
		\item If~$f_\Phi=0$ but some gate violates this bound, then it~may accept or~reject.
	\end{itemize}
	On~every input, it~runs in~time $m^{d-\lceil d/6\rceil}\cdot(dsL\log(2m))^{O(1)}$.
\end{restatable}

The algorithm uses the fact that, over the integers, a~polynomial is~zero
if~and only if~the sum of~the squares of~its coefficients is~zero.
It~computes this sum without expanding the polynomial.
Its running time has the form $s^{O(1)}t^{(1-\varepsilon)d}$ with
$\varepsilon=1/6$, which is~exactly the regime
of~\Cref{corollary:pit_np_size}.
To~prove \Cref{theorem:unconditional_dichotomy}, we~run the argument
behind \Cref{corollary:pit_np_size} over~$\Z$ with two changes.
First, we~replace~$\mathcal U$ by~a~partition polynomial
in~$\cc{VNP}^0_{\Z}$ that yields every polynomial we~need after
substituting only the constants $0$ and~$1$ and variables.
The assumption $\cc{VP}^0_{\Z}=\cc{VNP}^0_{\Z}$ then gives constant-free
circuits of~polynomial formal degree for all of~them, so~every gate
computes a~polynomial whose coefficients have polynomial bit length, as
\Cref{theorem:zero_test} requires.
Second, we~replace the hypothesized $\cc{PIT}$ algorithm
by~\Cref{theorem:zero_test}, applied to~the set-multilinear part of~each
identity.
\Cref{theorem:zero_test} may be~of~independent interest.
We note that Raz and Shpilka~\cite{RS05} give a deterministic polynomial-time algorithm for testing \emph{pure} set-multilinear circuits,
in~which all product gates split types along a~single tree, but the circuits arising in~our proof are not pure.

\subsubsection{Further Applications}\label{section:results_applications}

\paragraph{Fast oblivious Merlin--Arthur protocols from small circuits.}
Our second application concerns input-oblivious Merlin--Arthur
protocols~\cite{FSW09,GM15}.
In~such a~protocol, Merlin's message depends only on~the input length.
For every input length, a~single honest message must make Arthur accept
every yes-instance of~this length, while on~every no-instance, every
message is~rejected with probability at~least $2/3$.
The protocol is~\emph{perfectly complete} if~the honest message makes
Arthur accept with probability one.
The argument of~Kabanets and Impagliazzo~\cite{KI04} shows that small
arithmetic circuits for the permanent give such protocols for~$\coNP$:
Merlin sends the circuits, and Arthur checks them with the downward
self-reduction of~the permanent.
Combined with the reduction of~Ben-Dor and Halevi~\cite{BH93}, these
protocols lose only a~constant factor in~the number of~variables.
We~show that the same holds for the universal polynomial~$\mathcal U$
over every finite field, and for a~language in~$\oplus\cc P$ that
computes the bits of~$\mathcal U$.\footnote{Here, $\oplus\cc P$ is~the class of~languages~$L$ for which there is~a~nondeterministic polynomial-time machine that has an~odd number of~accepting computations on~an~input~$x$ if~and only if~$x\in L$.}
Here, $L_\oplus\in\oplus\cc P\cap\cc{DTIME}[2^{O(M/\log M)}]$ is~a~fixed
language whose strings of~length $M_N=\Theta(N\log N)$ encode a~point and
a~coordinate of~a~value of~$\mathcal U_N$ over~$\F_{2^{O(\log N)}}$
(\Cref{section:oma}).

\begin{restatable}{maintheorem}{ThmOMAFromCircuits}\label{theorem:oma_from_circuits}
	Let $c\ge1$ be~an~integer, let $\F$ be~a~finite field, and let $\mathcal U$ be~the universal family constructed for the characteristic of~$\F$.
	Let $s\colon\mathbb N\to\mathbb N$ be~a~nondecreasing function whose value can be~computed from~$N$ in~time polynomial in~$N$, and such that $s(N)\le2^N$ for all sufficiently large~$N$.
	Suppose that one of~the following conditions holds for all sufficiently large~$N$:
	\begin{enumerate}
		\item $\operatorname{asize}_{\Z}(\PERM_N)\le s(N)$;
		\item $\operatorname{asize}_{\F}(\mathcal U_N)\le s(N)$;
		\item $L_\oplus$ has Boolean circuits with at~most $s(N)$ gates on~inputs of~length~$M_N$.
	\end{enumerate}
	Then there is~a~constant $D\ge1$, depending only on~$c$ and~$\F$, such that $3$-\UNSAT{} on~$n$ variables with at~most $cn$ clauses has a~perfectly complete input-oblivious Merlin--Arthur protocol running in~time $s(Dn)n^D$.
\end{restatable}

Read contrapositively, \Cref{theorem:oma_from_circuits} is~a~trade-off
for every time bound~$T$.
If~$3$-\UNSAT{} on~$n$ variables with at~most $cn$ clauses has no~such
protocol running in~time $T(n)n^{O(1)}$, then $\PERM_N$ and~$\mathcal U_N$
require circuits of~size $T(\Omega(N))$ for infinitely many~$N$, and
$L_\oplus$ requires circuits of~size $T(\Omega(M/\log M))$ for infinitely
many input lengths~$M$.
The families $\PERM$, $\mathcal U$, and~$L_\oplus$ do~not depend on~$T$,
and the lower bounds scale linearly with the number of~variables of~the
formula.
For example, for $T(n)=2^{n/\log n}$, either $3$-\UNSAT{} on~$n$ variables
with at~most $cn$ clauses has a~perfectly complete input-oblivious
Merlin--Arthur protocol running in~time $2^{O(n/\log n)}$, or~$\PERM$
and~$\mathcal U$ require circuits of~size $2^{\Omega(N/\log N)}$ for
infinitely many~$N$, and $L_\oplus\notin\SIZE[2^{o(M/\log^2M)}]$.
The algorithm of~OpenAI~\cite{OpenAI26} decides $3$-$\cc{SAT}$ with $O(n)$
clauses deterministically in~time $2^{O(n/\sqrt{\log n})}$.
This is~a~perfectly complete input-oblivious protocol with an~empty
message, so~the first alternative of~this example asks for a~protocol that
is~faster by~a~factor of~$\sqrt{\log n}$ in~the exponent.
At~the other end, suppose that $3$-\UNSAT{} on~$n$ variables with at
most $cn$ clauses has no~perfectly complete input-oblivious
Merlin--Arthur protocol running in~polynomial time.
Then \Cref{theorem:oma_from_circuits} with $s(N)=N^k$ for every
constant~$k$ gives $\cc{VP}_{\mathcal R}\ne\cc{VNP}_{\mathcal R}$ for
every~$\mathcal R$ that is~a~finite field or~$\Z$.
Unlike earlier results of~this kind~\cite{CRTY23}, these trade-offs do
not assume an~exponential-time hypothesis (see \Cref{section:related_work}).
Over a~finite field of~odd characteristic, the permanent is
$\cc{VNP}_{\F}$-complete under projections of~polynomial
dimension~\cite{V79a}.
Hence the second condition also follows from
$\operatorname{asize}_{\F}(\PERM_N)\le s(N)$, with $s(n^{O(1)})$ in~place
of~$s(Dn)$ in~the running time.

\paragraph{A~purely arithmetic win--win.}
Our third application combines the Kabanets--Impagliazzo framework with
range avoidance.
Under the assumption $\cc{PIT}_{\Z}\in\NP$, it~yields arithmetic lower
bounds in~both alternatives.
We~begin with the following consequence of~their argument.

\begin{theorem}[Kabanets--Impagliazzo~\cite{KI04}]
\label{theorem:KI_dichotomy}
Assume that $\cc{PIT}_{\Z}\in\NP$.
Then at least one of the following holds:
\begin{enumerate}
    \item $\cc{VP}_{\Z}\ne\cc{VNP}_{\Z}$; or
    \item $\cc{P}^{\#\cc{P}}\subseteq\NP$, and hence $\cc{PH}=\NP$.
\end{enumerate}
\end{theorem}

The collapse in~the second alternative already yields strong Boolean
lower bounds.\footnote{Refinements of~the Kabanets--Impagliazzo argument
give $\cc{NE}\cap\cc{coNE}\not\subseteq\SIZE[2^n/(2n)]$~\cite[Theorem~4.1]{AM11}.
Under the collapse, the counting argument extends to~circuits with any
fixed oracle $\mathcal A\in\cc{PH}$, where an~oracle gate with $q$ inputs
has cost $\max\{1,q\}$ and the hard language may depend on~$\mathcal A$.}
We~show that it~also yields strong arithmetic lower bounds.
We~reduce the task of~constructing hard constant-degree polynomials to
range avoidance: a~string outside the range of~a~suitable circuit, which
maps short descriptions of~small arithmetic circuits to~their
coefficient vectors, describes a~polynomial that no~small circuit
computes.
When $\cc{PH}=\NP$, nonmembership in~the range of~a~circuit has
polynomial-size, polynomial-time verifiable certificates, which gives
an~$\cc{FNP}$ algorithm for range avoidance.

\begin{restatable}{maincorollary}{CrlyStrongAvoidPITWinWin}
\label{corollary:strong_avoid_pit_winwin}
Assume that $\cc{PIT}_{\Z}\in\NP$.
Then at least one of the following holds:
\begin{enumerate}
    \item $\cc{VP}_{\Z}\ne\cc{VNP}_{\Z}$; or
    \item For every fixed $d\ge 1$, there is a single-valued $\cc{FNP}$ construction
    of polynomials $G_t\in\Z[x_1,\ldots,x_t]$ of degree at most $d$
    satisfying
    \[
        \operatorname{asize}_{\Z}(G_t)=\Theta_d(t^d)
    \]
    for all sufficiently large $t$.
    The coefficients of $G_t$ have $O_d(\log t)$ bits, and the resulting
    polynomial families belong to $\cc{VP}_{\Z}$.
\end{enumerate}
\end{restatable}

Single-valuedness means that the algorithm has an~accepting computation
and that all accepting computations output the same coefficient list.
This is~a~very strong arithmetic win--win: both alternatives are
arithmetic lower bounds far beyond current techniques.
The first is~a~superpolynomial separation.
The second gives, for every fixed degree~$d$, constant-degree polynomials
of~optimal circuit complexity $\Theta_d(t^d)$, far above the
$\Omega(t\log t)$ benchmark.
These polynomials are explicit in~a~nondeterministic sense, and they lie
in~$\cc{VP}_{\Z}$ itself.
The only remaining assumption is~$\cc{PIT}_{\Z}\in\NP$, and the argument
uses it~only to~verify the identities of~the downward self-reduction of
the permanent.

\paragraph{A~Kabanets--Impagliazzo theorem over finite fields.}
As~mentioned in~the introduction, the Kabanets--Impagliazzo argument
relies on~the $\#\cc{P}$-hardness of~the permanent over~$\Z$, and does
not apply over a~fixed finite field.
Carmosino, Impagliazzo, Kabanets, and Kolokolova~\cite{CIKK15} discuss
this obstacle.
Under nondeterministic subexponential-time identity testing over
a~finite field~$\F$, they obtain an~arithmetic lower bound over~$\F$ for
a~multilinear polynomial whose values can be~verified in~nondeterministic
exponential time.
Replacing the permanent by~the universal polynomial, our framework gives
a~direct analogue of~the Kabanets--Impagliazzo theorem over every fixed
finite field.
The arithmetic alternative is~the standard separation
of~$\cc{VP}_{\F}$ from~$\cc{VNP}_{\F}$.
The hypothesis concerns only circuits whose degree is~at~most their
number of~variables (see \Cref{section:KI_finite_fields}).

\begin{restatable}{maintheorem}{ThmKIFiniteFields}\label{theorem:KI_finite_fields}
	Let $\F$ be~a~fixed finite field.
	If~$t$-$\cc{PIT}_{\F}\in\cc{NSUBEXP}$, then
	$\NEXP\not\subseteq\PPoly$ or~$\cc{VP}_{\F}\ne\cc{VNP}_{\F}$.
\end{restatable}

\paragraph{Derandomization as~a~route to~lower bounds.}
The algorithmic method of~Williams~\cite{W13,W14} shows that beating
exhaustive search, or~\emph{perebor}~\cite{Tra84}, even slightly,
implies circuit lower bounds.
In~particular, refuting the Exponential Time Hypothesis (ETH), the Strong
Exponential Time Hypothesis (SETH), or~their nondeterministic variants would yield circuit lower bounds beyond the reach of~current
techniques~\cite{W13,JMV18,MW18,CGIMPS16}.
This implication has often been read in~reverse.
Since we~do~not know how to~prove such lower bounds, the corresponding
algorithms were taken to~be~out of~reach, and the implication served
as~a~barrier: the lower bound, it~seemed, would have to~come before the
algorithm.

A~recent wave of~algorithmic breakthroughs invites us~to~revisit this
belief. Algorithms for sparse $3$-$\cc{SAT}$ found independently by~OpenAI~\cite{OpenAI26} and by~Anthropic~\cite{Ant26} refute the Exponential Time Hypothesis. An~algorithm discovered by~Anthropic and developed by~Alman and Vassilevska Williams~\cite{AVW26} refutes the $3$SUM and APSP hypotheses of~fine-grained complexity. The algorithm of~\cite{OpenAI26,Ant26} already implies a~superlinear lower bound for~$\cc{E}^{\NP}$, beyond the reach of~all previous techniques.
Thus, the lower bound need not come first: it~can come as~a~consequence of~the algorithm.

The same belief has shaped the study of~derandomization.
Polynomial identity testing has seen substantial progress for restricted
models of~computation (see \Cref{section:related_work}).
For general circuits, however, the Kabanets--Impagliazzo
theorem~\cite{KI04} was read as~a~warning.
An~efficient deterministic identity test would prove lower bounds beyond
current techniques, so~it~seemed out of~reach, and focusing on~restricted
models was the natural response.

Our results make the case for targeting general circuits directly.
\Cref{theorem:unconditional_dichotomy} establishes unconditionally a~form
of~the Kabanets--Impagliazzo conclusion.
We~obtain it~the way the algorithmic method obtains lower bounds: by
designing a~polynomial identity test that beats brute force on~exactly
the instances our argument needs.
These instances are structured integer circuits, and the natural next
step is~to~achieve such speed-ups for general constant-degree circuits
over a~finite field.
Shaving any polynomial factor off \Cref{theorem:low_degree_pit_upper},
even nondeterministically, would resolve
Open Problem~\ref{op:boolean_lb} or Open Problem~\ref{op:arithmetic_lb}
(\Cref{theorem:pit_fixed_degree}).
Saving a~constant fraction of~the exponent of~$t$, for infinitely many
degrees, would separate $\cc{VP}_{\F}$ from~$\cc{VNP}_{\F}$ (\Cref{corollary:pit_np_size}), which in~turn
implies the arithmetic alternative
of~\Cref{theorem:unconditional_dichotomy}.

\paragraph{Use of~AI~tools.}
We~used AI~tools substantially in~obtaining these results.
Starting from weaker versions of~our results, the model GPT~6~Astra
suggested \Cref{theorem:low_degree_pit_upper}, found and proved
\Cref{theorem:pit_fixed_degree}, and improved
\Cref{corollary:pit_np_size,corollary:strong_avoid_pit_winwin} to~the
versions stated above.
\Cref{theorem:unconditional_dichotomy}, together with the zero test
of~\Cref{theorem:zero_test}, was found by~Claude Opus~5.5 in~an~automated
research session.
We~describe the role of~these models in~\Cref{section:ai_methodology}.

\subsection{Related Work}\label{section:related_work}

\paragraph{Unconditional circuit lower bounds.}
The class $\cc{S}_2\cc{P}$ of~symmetric polynomial time contains both
$\cc{P}^{\NP}$ and $\cc{MA}$, and is~contained in
$\Sigma_2\cc{P}\cap\Pi_2\cc{P}$.
Its fixed-polynomial lower bounds~\cite{C07,CHLR26}, and those for its
oblivious counterpart $\cc{O}_2\cc{P}$~\cite{GLV24}, therefore strengthen
Kannan's lower bounds for~$\Sigma_2\cc{P}$~\cite{K82}.
Fixed-polynomial lower bounds for~$\NP$ would be~substantial progress
even within this line of~work.
\Cref{theorem:unconditional_dichotomy} shows that either they hold, or
the constant-free version of~Valiant's conjecture holds.
In~the arithmetic setting, Fournier, Perifel, and de~Joannis
de~Verclos~\cite{FPV15} showed, under the Generalized Riemann Hypothesis,
that for every~$k$ there is~a~family of~polynomials with coefficients
in~$\cc{MA}$ that requires arithmetic circuits of~size larger than
$O(n^k)$ over~$\mathbb C$.

\paragraph{Identity testing and circuit lower bounds.}
Jansen and Santhanam~\cite{JS12} studied identity testing for
constant-free circuits whose degree is~bounded by~their size, obtaining
stronger trade-offs through a~model that combines Boolean and arithmetic
computation.
They also proved unconditionally that
$\cc{a}\cdot\cc{NEXP}^{\cc{RP}}$ has no~constant-free circuits
of~polynomial size, and that $\cc{a}\cdot\cc{MA}$ has no~arithmetic
circuits of~size $n^k$ for every fixed~$k$.\footnote{For a~complexity
class~$\mathcal C$, the class $\cc{a}\cdot\mathcal C$ consists of~the
families of~integer polynomials whose evaluation problem lies
in~$\mathcal C$; see~\cite{JS12} for the precise definition.}
Like the unconditional result of~Kabanets and Impagliazzo recalled in~the
introduction, these lower bounds concern classes with randomness.
Carmosino, Impagliazzo, Kabanets, and Kolokolova~\cite{CIKK15}
established connections between average-case identity testing and
average-case lower bounds.
Andrews, Kush, and Tell~\cite{AKT25} obtained polynomial-time identity
testing over characteristic zero from nearly necessary lower-bound
assumptions in~a~model of~uniform arithmetic networks.
B{\"u}rgisser~\cite{Burgisser00} showed that, over a~field of~positive
characteristic, $\cc{VP}=\cc{VNP}$ implies
$\PPoly=\cc{NP/poly}=\cc{PH/poly}$.
In~characteristic zero, the same implication is~known only under the
Generalized Riemann Hypothesis.
Beyond hardness versus randomness, $\cc{PIT}$ is~a~central verification
primitive in~proof complexity.
For example, the Ideal Proof System of~Grochow and Pitassi gives another
route from proof lower bounds to~$\cc{VP}\ne\cc{VNP}$~\cite{GP18}.
Dutta, Saxena, and Thierauf~\cite{DST24} showed that lower bounds for
highly restricted algebraic representations already imply
$\cc{VP}\ne\cc{VNP}$.

\paragraph{Deterministic identity testing for restricted models.}
Raz and Shpilka~\cite{RS05} gave deterministic polynomial-time
identity-testing algorithms for noncommutative algebraic branching
programs.
Their methods also apply to~read-once oblivious algebraic branching
programs and to~pure set-multilinear circuits.
For constant-depth arithmetic circuits, the lower bounds of~Limaye,
Srinivasan, and Tavenas~\cite{LST25}, combined with the
hardness-to-randomness connection of~Chou, Kumar, and
Solomon~\cite{CKS18}, give deterministic subexponential-time identity
testing over characteristic zero.
Forbes~\cite{F24} subsequently extended the constant-depth lower bounds
to~every field.
Closer to~\Cref{theorem:zero_test}, Arvind, Chatterjee, Datta, and
Mukhopadhyay~\cite{ACDM18} beat brute force for depth-$3$ circuits.
They do so by~computing the sum of~the squares of~the coefficients as
a~sum of~inner products.
Similar sums of~squares have been used for noncommutative
circuits~\cite{AJS09,LMP19}.
For broader accounts, see~\cite{SY10,S14,KS19}.

\paragraph{Black-box derandomization.}
In~the black-box setting, the connection to~lower bounds is~more direct:
an~explicit hitting set for a~circuit class yields an~explicit polynomial
that the class cannot compute~\cite{HS80,A05}.
Moreover, bootstrapping results show that slightly nontrivial hitting sets
for circuits with few variables already yield nearly polynomial-size
hitting sets for general circuits~\cite{KST19,GKSS22}.
Our hypotheses instead concern white-box algorithms, possibly
nondeterministic, that are given a~circuit and decide whether it~computes
zero.
This is~a~weaker requirement: a~hitting set gives such an~algorithm, but
the converse is~not known.
White-box algorithms can also be~easier to~obtain.
For example, read-once oblivious algebraic branching programs have
polynomial-time white-box tests~\cite{RS05}, while only
quasipolynomial-size hitting sets are known, even when the variable order
is~known~\cite{FS13}.
\Cref{theorem:zero_test} is~itself a~white-box algorithm of~this kind.
White-box algorithms do~not directly yield hard polynomials.
Nevertheless, our results show that even slight white-box speed-ups
already imply strong circuit lower bounds.

\paragraph{Formulations, algorithms, and protocols.}
Our encoding of~satisfiability by~low-degree polynomials was inspired by
the polynomial formulations of~\cite{BGKMS25}, used in~\cite{BKMRRS24} to
study barriers to~fine-grained lower bounds.
Our satisfiability algorithms connect to~the
algorithms-to-lower-bounds methods of~Williams~\cite{W13,W14}, Murray and
Williams~\cite{MW18}, and Chen~\cite{Che25}.
Input-oblivious proof systems were studied by~Fortnow, Santhanam, and
Williams~\cite{FSW09}, and by~Goldreich and Meir~\cite{GM15}.
Chen, Rothblum, Tell, and Yogev~\cite{CRTY23} studied the oblivious
Merlin--Arthur exponential-time hypothesis, which asserts that
$3$-\UNSAT{} has no~input-oblivious Merlin--Arthur protocol running in
time $2^{\varepsilon n}$ for a~constant $\varepsilon>0$.
They derived from it~that $\cc{E}\not\subseteq\SIZE[2^{\delta n/\log n}]$
for a~constant $\delta>0$.\footnote{Theorem~A.1 of~\cite{CRTY23} is~stated for Merlin--Arthur protocols, but its proof also applies to~input-oblivious protocols: Merlin supplies a~circuit for a~fixed language at~the relevant input length, and this circuit does not depend on~the particular input.}
This hypothesis implies the exponential-time hypothesis, which the
algorithm of~OpenAI~\cite{OpenAI26} refutes.
In~contrast, \Cref{theorem:oma_from_circuits} gives a~trade-off for every
time bound, and its Boolean alternative gives a~hard language
in~$\oplus\cc P$.

\paragraph{Range avoidance and explicit constructions.}
The range avoidance problem $\cc{Avoid}$ asks, given a~circuit
$C\colon\{0,1\}^n\to\{0,1\}^m$ with $m>n$, for a~string outside its
range~\cite{KKMP21}.
It~captures many explicit construction problems.
Efficient algorithms for~$\cc{Avoid}$, even with an~$\NP$ oracle, would
give truth tables of~high circuit complexity, rigid matrices,
pseudorandom generators, Ramsey graphs, and optimal
codes~\cite{K21,GLW22}.
For algorithms and barriers, see~\cite{RSW22,CHLR23,GGNS23,ILW23,CHLR26,KP24}.
\Cref{corollary:strong_avoid_pit_winwin} adds constant-degree polynomials
of~optimal arithmetic circuit complexity to~this list.
In~particular, an~$\cc{FP}^{\NP}$ algorithm for~$\cc{Avoid}$ would
construct such polynomials in~$\cc{FP}^{\NP}$ unconditionally.

\paragraph{Constant-free complexity.}
Constant-free circuits are the model behind the arithmetic alternative
of~\Cref{theorem:unconditional_dichotomy}.
Constant-free versions of~Valiant's classes and their relation to
Boolean complexity have been studied
in~\cite{M03,MP08,K05,B09,KP11}.
Constant-free complexity is~also central to~the $\tau$-conjecture
of~Shub and Smale~\cite{SS95}.
Bhattacharjee, Bl{\"a}ser, Dutta, and Mukherjee~\cite{BBDM26} showed
that this conjecture implies an~exponential lower bound for an~explicit
exponential sum, but only a~subexponential lower bound for the
permanent.
They asked whether it~implies an~exponential lower bound for the
permanent.
\Cref{theorem:oma_from_circuits} gives a~different route to~such
bounds.
The constant-free complexity $\tau(\PERM_N)$, the size of~a~smallest
circuit that computes $\PERM_N$ using only the constants~$\pm1$, is~at
least $\operatorname{asize}_{\Z}(\PERM_N)$.
Hence ruling out perfectly complete input-oblivious Merlin--Arthur
protocols for sparse $3$-\UNSAT{} running in~time $T(n)n^{O(1)}$ gives
$\tau(\PERM_N)\ge T(\Omega(N))$ for infinitely many~$N$.
This loses only a~constant factor in~the dimension of~the permanent.

\subsection{Proof Overview} \label{section:proof_overview}

All our results follow the approach of~Kabanets and Impagliazzo~\cite{KI04} described above: we~guess arithmetic circuits for a~family of~polynomials, verify them using a~downward self-reduction, and then use the verified circuits to~solve a~hard problem.
For the permanent, this approach needs identity tests of~degree~$n$ for checking matrices of~size $n\times n$.
Hence, it~gives nothing for identity testing of~low-degree polynomials.
Our approach replaces the permanent by~polynomials that encode satisfiability, whose degree we~can choose, and which are still downward self-reducible.
Moreover, all these polynomials are projections of~a~single explicit family $\mathcal U\in\cc{VNP}_{\Z}$, so~the assumption that $\mathcal U$ has small circuits gives small circuits for all of~them.
The initial idea of~encoding $\cc{SAT}$ by~low-degree polynomials was inspired by~the polynomial formulations introduced in~\cite{BGKMS25} and used in~\cite{BKMRRS24} as~a~tool for studying barriers to~proving fine-grained lower bounds for various problems.
Below, we~describe the main ideas in~greater detail.

\paragraph{Partition polynomials.}
The polynomials that replace the permanent in~our proofs are called \emph{partition polynomials}.
Recall that the permanent sums, over all permutations, the products of~the matrix entries selected by~the permutation, one entry in~each row.
Similarly, a~partition polynomial sums, over all assignments to~a~set of~Boolean variables, the products of~the table entries selected by~the assignment, one entry in~each table.
Here, a~table should be~thought of~as~a~constraint that sees only some of~the variables and has an~entry for every assignment to~them.
Formally, for a~list $\mathcal S=(S_1,\ldots,S_{d})$ of~subsets of~$[r]$, table~$j \in [d]$ depends on~the variables in~$S_j$ and has an~entry $w_{j,\tau}$ for every local assignment $\tau\in\{0,1\}^{S_j}$, and the partition polynomial is
\[Z_{\mathcal S}(w)=\sum_{\mu\in\{0,1\}^r}\prod_{j=1}^{d}w_{j,\mu|_{S_j}}.\]

Partition polynomials encode constraint satisfaction problems.
If~every entry is~set to~one or~zero, depending on~whether its local assignment is~allowed by~the constraint, then every assignment contributes one if~it~satisfies all constraints and zero otherwise, so~$Z_{\mathcal S}$ counts the solutions.
If, instead, we~set only the forbidden entries to~zero and keep the other entries as~formal variables, then exactly the monomials of~the solutions survive.
Hence, the problem is~unsatisfiable if~and only if~the resulting polynomial is~identically zero.
Thus, unsatisfiability can be~tested using polynomial identity testing.

The degree of~$Z_{\mathcal S}$ equals the number $d$ of~tables.
Hence, by~putting several constraints into one table, we~decrease the degree at~the cost of~increasing the number of~entries.
Every variable has to~be~seen by~some table, so~$d$ tables see at~least $r/d$ variables on~average and have at~least $d2^{r/d}$ entries in~total.
For \Cref{theorem:pit_fixed_degree,corollary:pit_np_size}, we~need a~constant degree~$d$, so~we~can afford only $d$ tables, each of~which sees about $n/d$ of~the $n$ input variables and has about $2^{n/d}$ entries.
The resulting polynomial has degree~$d$ and about $d2^{n/d}$ variables, no~matter how large the circuit is.

Finally, just like the permanent, partition polynomials are downward self-reducible: fixing the Boolean variables one by~one gives
\[Z_i(w)=Z_{i-1}(w)+Z_{i-1}(w^{(i)}),\]
where $Z_i$ is~the sum over the assignments that vanish outside the first $i$ variables, and $w^{(i)}$ is~a~projection of~$w$.
Thus, if~$Z_{\mathcal S}$ has a~small circuit, then a~nondeterministic algorithm can guess circuits for all polynomials $Z_i$ and verify them by~identity tests of~degree~$d$.

\paragraph{The universal polynomial.}
For the permanent, the input is~just a~point at~which we~evaluate one fixed polynomial~$\PERM_n$.
For partition polynomials, however, the input determines the polynomial itself.
Thus, we~need small circuits for many different polynomials, one for every input, with a~common size bound.
To~obtain such a~bound from a~single assumption, we~construct one explicit family that contains all partition polynomials as~projections: then a~small circuit for this family gives small circuits for all partition polynomials.
In~\Cref{section:universal_polynomial}, we~construct such a~family~$\mathcal U=\{\mathcal U_n\}\in\cc{VNP}_{\F}$ for every finite field~$\F$.
Informally, $\mathcal U_n$ has variables that encode the description of~a~$3$-CNF formula $\Psi$ on~$n$ variables, and then $\mathcal U_n$ becomes the polynomial $\sum_\mu\prod_{i:\,\mu_i=1}x_i$, which lists the satisfying assignments $\mu\in\{0,1\}^{n}$ of~$\Psi$ as~monomials.
Therefore, the assumption $\cc{VP}_{\F}=\cc{VNP}_{\F}$ gives small circuits over~$\F$ for all partition polynomials.

\paragraph{Constant-degree PIT.}
To~prove \Cref{theorem:pit_fixed_degree,corollary:pit_np_size}, we~assume that $\mathcal U$ has circuits of~polynomial size, say, every polynomial of~$\mathcal U$ with $t$ variables has a~circuit of~size $O(t^{\kappa})$ for a~constant~$\kappa$, and we~design a~nondeterministic algorithm that decides unsatisfiability of~Boolean circuits of~size $2^{\rho n}$ in~time $O(2^{(1-\rho)n})$.
Together with the small circuits for~$\mathcal U$, such an~algorithm gives $\cc{NE}\not\subseteq\SIZE[2^n/(2n)]$ (\Cref{section:ne_lower_bound}).
Similarly to~Williams~\cite{W13,W14}, Murray and Williams~\cite{MW18}, and Chen~\cite{Che25}, the algorithm certifies truth tables of~high circuit complexity at~infinitely many input lengths.
These tables derandomize Merlin--Arthur protocols in~which Merlin sends a~circuit for~$\mathcal U$, and these protocols suffice to~compute the lexicographically first function on~$n$ inputs that requires circuits of~size larger than $2^n/(2n)$.
In~these protocols, Arthur runs the sum-check protocol of~Lund, Fortnow, Karloff, and Nisan~\cite{LFKN92} for an~arithmetization of~a~$3$-CNF formula on~his own and computes the messages of~the prover from values of~$\mathcal U$, since every partial sum of~this arithmetization is, up~to~an~explicit factor, a~partition polynomial.
The sum-check protocol is~sound against every prover, so~a~wrong circuit cannot make Arthur accept a~satisfiable formula, except with small probability.
To~use $\cc{PIT}$ of~a~constant degree~$d$, we~split the $n$ inputs into $d$ blocks $V_1,\ldots,V_d \subseteq [n]$ and, for a~Boolean function $R\colon\{0,1\}^n\to\{0,1\}$, consider the following polynomial
\[P^{(d)}_R(x)=\sum_{\mu\in\{0,1\}^n}R(\mu)\prod_{j=1}^{d}x_{j,\mu|_{V_j}},\]
which has degree~$d$ and $t=O(2^{n/d})$ variables $x_{j,\tau}$, one for every block $V_j$ and every local assignment $\tau\in\{0,1\}^{V_j}$.
Every satisfying assignment $\mu$ of~$R$ contributes one monomial, which picks in~every block the variable indexed by~the restriction of~$\mu$ to~this block, so~$P^{(d)}_R\equiv0$ if~and only if~$R$ is~unsatisfiable.
Thus, to~decide whether a~Boolean circuit that computes $R$ is~unsatisfiable, it~suffices to~test the identity $P^{(d)}_R\equiv0$ of~degree~$d$ in~$t$ variables.
To~run the $\cc{PIT}$ algorithm, however, we~need an~arithmetic circuit for $P^{(d)}_R$.
This raises two questions: why does a~small arithmetic circuit for $P^{(d)}_R$ exist, and how can we~trust it, given that we~do~not know it~and have to~guess~it?

A~small circuit exists because $P^{(d)}_R$ is~a~specialization of~a~partition polynomial, and hence a~projection of~the universal polynomial.
Indeed, without the factor $R(\mu)$, the polynomial $P^{(d)}_R$ is~just the partition polynomial of~the blocks $V_1,\ldots,V_d$.
To~bring in~$R$, we~take a~Boolean circuit~$C$ that computes~$R$, add a~Boolean variable for every gate, and add a~table for every gate that checks that the value of~the gate is~consistent with the values of~its inputs, together with a~table that checks that the output is~one.
The entries of~these new tables are fixed to~zero or~one, so~they do~not increase the degree: an~assignment to~the inputs and the gates contributes a~nonzero term only if~all gate values are correct and $R(\mu)=1$, and then this term is~exactly $\prod_{j}x_{j,\mu|_{V_j}}$.
Thus, $P^{(d)}_R$ is~a~specialization of~a~partition polynomial with $O(t+\operatorname{size}(C))$ variables and entries, so~it~is~a~projection of~a~member of~$\mathcal U$ with $O(t+\operatorname{size}(C))$ variables, and it~has an~arithmetic circuit of~size about $(t+\operatorname{size}(C))^{\kappa}$, which is~about $t^{\kappa}$ when $\operatorname{size}(C)\le t$.

To~trust a~guessed circuit, we~verify it~using a~downward self-reduction, as~for the permanent.
Suppose we~had a~small branching program for $R$, which computes $R(\mu)$ by~walking from a~start node to~an~accepting or~a~rejecting sink, querying one input at~every node.
Then, such a~self-reduction is~very natural: if~a~node $\alpha$ queries input~$i$ and has successors $\alpha_0$ and $\alpha_1$, then the function $R_\alpha$ computed from~$\alpha$ equals $R_{\alpha_0}$ on~the assignments with $\mu_i=0$ and $R_{\alpha_1}$ on~the assignments with $\mu_i=1$.
For the polynomials $P_\alpha=P^{(d)}_{R_\alpha}$, this gives the recurrence
\[P_\alpha(x)=P_{\alpha_0}(\chi_{i,0}\odot x)+P_{\alpha_1}(\chi_{i,1}\odot x),\]
where the projection $\chi_{i,\sigma}\odot x$ keeps only the monomials of~the assignments with $\mu_i=\sigma$.
Hence, the algorithm guesses circuits for all polynomials $P_\alpha$ and verifies them, starting from the two sinks, by~identity tests of~degree~$d$ on~circuits with $t$ variables and size about~$t^{\kappa}$.
An~algorithm that runs in~time $s^{a}t^{b}$ performs each test in~time $t^{\kappa a+b+o(1)}=2^{(\kappa a+b)n/d+o(n)}$, which is~exponentially faster than exhaustive search if~$\kappa a+b<d$.
In~\Cref{theorem:pit_fixed_degree}, this is~exactly the condition $\kappa<(d-b)/a$.
In~\Cref{corollary:pit_np_size}, we~have $a=o(d)$ and $b=(1-\varepsilon)d$, so, whatever the constant~$\kappa$ is, this condition holds for a~large enough degree~$d$.

For general circuits, we~do~not know such a~recurrence: a~gate computes, for example, the conjunction of~two functions, which corresponds to~the monomial-wise product of~their polynomials, and we~do~not know how to~verify such products by~low-degree identity tests.
Therefore, we~reduce circuits to~branching programs, and here we~use the assumption on~$\mathcal U$ once more.
By~the depth reduction of~Valiant, Skyum, Berkowitz, and Rackoff~\cite{VSBR83}, a~small arithmetic circuit for $P^{(d)}_R$ with a~constant number~$d$ of~blocks gives a~Boolean formula of~small depth for~$R$, since $P^{(d)}_R$ evaluates to~$R(\mu)$ at~the Boolean point that encodes~$\mu$.
As~a~result, Boolean circuits of~size $2^{\rho n}$ have formulas of~size $2^{\zeta n}$, where $\zeta$ tends to~zero with~$\rho$.
Hence, this formula can be~converted into a~branching program and the verification above can proceed.
In~total, it~performs about $2^{(\rho+\zeta)n}$ tests, so~it~runs in~time $2^{((\kappa a+b)/d+\rho+\zeta+o(1))n}$, which is~$O(2^{(1-\rho)n})$ if~$\rho$ is~small enough.

\paragraph{Fast PIT for set-multilinear circuits.}
For \Cref{theorem:unconditional_dichotomy}, we~run the proof of~\Cref{theorem:pit_fixed_degree} over the integers and replace the assumed $\cc{PIT}$ algorithm by~an~unconditional one.
We~notice that we~need to slightly speed-up the \cc{PIT} for \emph{set-multilinear} circuits, i.e. circuits computing a~polynomial~$T$ whose variables are split into $d$~blocks of~at~most $m=2^{\lceil n/d\rceil}$ variables, one block for every~$V_j$, and every monomial of~$T$ takes exactly one variable from every block, and we~need to~solve it~in~$m^{d(1-\varepsilon)}=2^{(1-\varepsilon)n}$ time.
Over the integers, $T\equiv0$ if~and only if~the sum $\|T\|^2$ of~the squares of~its coefficients is~zero, and we~compute this sum efficiently.
Similar sums of~squares have been used for identity testing of~noncommutative circuits~\cite{AJS09,LMP19}.

We~use an~idea similar to~that of~Hyafil~\cite{H79} and of~Valiant, Skyum, Berkowitz, and Rackoff~\cite{VSBR83}, which writes~$T$, given by~a~syntactically set-multilinear circuit with $s$ gates, as~a~sum of~at~most $s$ products of~two set-multilinear polynomials, where every product splits the $d$~blocks into two parts of~at~most $2d/3$ blocks each.
Instead of~expanding these products, we~note that the inner product of~two of~them is~a~sum of~products of~four coefficient arrays, each depending on~only part of~the indices, and, we~can sum out the indices in~a~convenient order.
The better of~two such orders computes the inner product with $O(m^{d-\lceil d/6\rceil})$ operations instead of~about~$m^d$.
Hence $\|T\|^2$ is~computed with $O(s^2m^{5d/6})$ operations, which is~the regime $s^{O(1)}t^{(1-\varepsilon)d}$ of~\Cref{corollary:pit_np_size} with $\varepsilon=1/6$.
The polynomials in~our proof are set-multilinear, but the circuits for them can be~arbitrary, however, a~circuit with $s$ gates that computes a~polynomial~$f$ can be~turned into a~syntactically set-multilinear circuit with $3^{d+1}s$ gates that computes the set-multilinear part of~$f$ (\Cref{section:integer_proof}).
The integers in~this computation are short only if~the guessed circuits have coefficients of~polynomial bit length, and constant-free circuits of~polynomial formal degree have such coefficients.

\paragraph{Hard polynomials from range avoidance.}
For \Cref{corollary:strong_avoid_pit_winwin}, we~build a~circuit that
maps short descriptions of~small arithmetic circuits over~$\Z_q$ to~the
coefficient vectors of~the constant-degree parts of~their polynomials.
Its output is~longer than its input, so~some coefficient vector lies
outside its range, and any such vector describes a~polynomial of~degree
at~most~$d$ that no~small circuit computes.
Under $\cc{PIT}_{\Z}\in\NP$ and $\cc{VP}_{\Z}=\cc{VNP}_{\Z}$, the
Kabanets--Impagliazzo argument gives $\cc{PH}=\NP$, and then the
lexicographically first such vector has a~single-valued $\cc{FNP}$
construction.

\section{Preliminaries}

We~start with the notation used throughout the paper.
All rings are commutative, nonzero and have an~identity element; we~denote a~ring by~$\mathcal R$ and a~finite field by~$\F$.
Besides the integers~$\Z$, we~use the ring $\Z_q\coloneqq\Z/q\Z$ of~integers modulo an~integer $q\ge2$, and the field~$\F_q$ of~order~$q$ for a~prime power~$q$.

We~index the monomials in~variables $x_1,\ldots,x_t$ by~multi-indices $\alpha=(\alpha_1,\ldots,\alpha_t)\in\Z_{\ge0}^t$: we~write $x^\alpha=x_1^{\alpha_1}\cdots x_t^{\alpha_t}$ for the monomial of~degree $|\alpha|=\alpha_1+\cdots+\alpha_t$, and $[x^\alpha]P$ for the coefficient of~$x^\alpha$ in~a~polynomial~$P$.

For a~positive integer~$n$, we~write $[n]=\{1,\ldots,n\}$, and $S_n$ for the set of~permutations of~$[n]$.
All logarithms are base two, and $\widetilde O(\cdot)$ hides polylogarithmic factors.

\subsection{Arithmetic Circuits and Identity Testing}
\label{section:arithmetic_circuits}

An \emph{arithmetic circuit}~$C$ over a ring~$\mathcal R$ and variables $x_1,\ldots,x_t$ is a directed acyclic graph whose gates have in-degree zero or two.
The in-degree-zero gates are labeled by variables or elements of~$\mathcal R$, and the in-degree-two gates are labeled by $+$ or~$\times$.
Unless stated otherwise, a circuit has one designated output gate and computes a polynomial in $\mathcal R[x_1,\ldots,x_t]$.
Polynomial identities are understood formally: two polynomials are identical if all their corresponding coefficients agree.
The degree of a circuit means the total degree of its output polynomial.

\paragraph{Size and encodings.}
We fix binary encodings of the ring elements used as constants.
The \emph{size} of a circuit~$C$ is the number of gates plus the sum of the bit-lengths encoding the constants in the circuit.
We write $\operatorname{asize}_{\mathcal R}(P)$ for the minimum size of an arithmetic circuit over~$\mathcal R$ computing~$P$.
For a size bound~$s(t)$, we write $\cc{ASIZE}_{\mathcal R}[s(t)]$ for polynomial families in $t$ variables that admit circuits of size at most~$s(t)$ for all sufficiently large~$t$; we omit the subscript when the ring is clear.

\paragraph{Polynomial identity testing.}
The problem $\cc{PIT}_{\mathcal R}$ asks whether the polynomial computed by an input arithmetic circuit over~$\mathcal R$ is identically zero.
For a function $f:\mathbb Z_{\ge1}\to\mathbb R_{\ge0}$, \emph{Promise $f(t)$-$\cc{PIT}_{\mathcal R}$} is the restriction to circuits on $t$ variables whose output polynomial has total degree at most~$f(t)$.
Its yes-instances compute the zero polynomial, and its no-instances compute nonzero polynomials satisfying the degree promise.
No correctness requirement is imposed on circuits violating the promise.
For a constant degree bound~$d$, we abbreviate this problem as $d$-$\cc{PIT}_{\mathcal R}$.
Note that the degree bound is~a~promise only on~the output polynomial, and it need not hold at~intermediate gates.

\paragraph{Polynomial families.}
We use the classes $\cc{VP}_{\mathcal R}$ and $\cc{VNP}_{\mathcal R}$ with the circuit-size convention above~\cite{V79a,V82,SY10}.
A family $\{P_n\}_{n\ge1}$, where $P_n$ has $t(n)$ variables, belongs to $\cc{VP}_{\mathcal R}$ if $t(n)$, $\deg P_n$, and $\operatorname{asize}_{\mathcal R}(P_n)$ are polynomially bounded in~$n$.
It belongs to $\cc{VNP}_{\mathcal R}$ if there exist a polynomially bounded function~$m(n)$ and a family $\{G_n\}_{n\ge1}\in\cc{VP}_{\mathcal R}$ such that
\[
    P_n(x)=\sum_{w\in\{0,1\}^{m(n)}}G_n(x,w).
\]

A~circuit over~$\Z$ is~\emph{constant-free} if~its only constants are $1$ and~$-1$.
The \emph{formal degree} of~a~gate is~defined inductively: every leaf, including a~constant, has formal degree one, a~sum gate has the maximum of~the formal degrees of~its children, and a~product gate has their sum.
The formal degree of~a~circuit is~the largest formal degree of~its gates.
A~family $\{P_n\}_{n\ge1}$ of~integer polynomials, where $P_n$ has $t(n)$ variables, belongs to~$\cc{VP}^0_{\Z}$ if~$t(n)$ is~polynomially bounded and $P_n$ is~computed by~a~constant-free circuit whose number of~gates and formal degree are polynomially bounded in~$n$.
It~belongs to~$\cc{VNP}^0_{\Z}$ if~there exist a~polynomially bounded function~$m(n)$ and a~family $\{G_n\}_{n\ge1}\in\cc{VP}^0_{\Z}$ such that $P_n(x)=\sum_{w\in\{0,1\}^{m(n)}}G_n(x,w)$.
Constant-free classes were studied before in~\cite{M03,MP08,K05,B09,KP11}.

A \emph{projection} of a polynomial is obtained by substituting variables or constants for some of its variables.
The projections used in our self-reductions substitute only variables and the constants $0$ and~$1$.

We call an integer polynomial family $\{P_n\}_{n\ge1}$ \emph{coefficient-explicit} if its number of variables and degree are polynomially bounded in~$n$, and a deterministic algorithm, given $1^n$ and a multi-index~$\alpha$ in binary, computes $[x^\alpha]P_n$ in time polynomial in $n$ and the encoding length of~$\alpha$.
In particular, its coefficients have polynomially bounded bit length.
When discussing coefficient access, we abbreviate coefficient-explicit to \emph{explicit}.
This notion is distinct from a nondeterministic construction of an entire coefficient list, defined in~\Cref{section:range_avoid}.

\begin{claim}[\cite{V82,HWY10}]
    Every coefficient-explicit integer polynomial family belongs to $\cc{VNP}_{\Z}$.
\end{claim}

\begin{remark}\label{remark:mod_p}
    For every integer $q\ge2$, reducing the constants of an integer circuit modulo~$q$ preserves its number of gates and computes the coefficientwise reduction of its polynomial in~$\Z_q[x]$.
    This operation need not preserve size as measured by the bit lengths of constant labels.
    Nevertheless, for every fixed finite field~$\F$ of characteristic~$p$,
    \[
        \operatorname{asize}_{\F}(P\bmod p)
        =O_{\F}\bigl(\operatorname{asize}_{\Z}(P)\bigr)
    \]
    for every integer polynomial~$P$, where the coefficients modulo~$p$ are embedded in the prime subfield of~$\F$.
    Consequently, coefficientwise reduction maps families in $\cc{VNP}_{\Z}$ to families in $\cc{VNP}_{\F}$.
\end{remark}

We write $\PERM=\{\PERM_n\}_{n\ge1}$ for the permanent family defined in~\Cref{sec:introduction}.

\begin{theorem}[\cite{V79a}]
    Over every field $\mathbb F$ of characteristic different from two, $\PERM$ is $\cc{VNP}_{\mathbb F}$-complete under polynomially bounded projections allowing field constants.
\end{theorem}

In particular, if the reduction of a coefficient-explicit integer polynomial family requires superpolynomial arithmetic circuits over a fixed finite field~$\F$ of odd characteristic, then $\PERM\notin\cc{VP}_{\F}$.

\paragraph{Truncation and identity testing.}
The following lemma lets us represent the low-degree part of a polynomial without imposing a degree bound on the original circuit.

\begin{lemma}[\cite{S73}]\label{lemma:circuit_truncation}
    Let $A$ be an arithmetic circuit of size~$s$ over a ring~$\mathcal R$, and let $d\ge0$ be an integer.
    One can construct a circuit $\pi_{\le d}(A)$ of size $O\bigl(s(d+1)^2\bigr)$ computing
    \[
        \pi_{\le d}(P)
        \coloneqq
        \sum_{|\alpha|\le d}[x^\alpha]P\,x^\alpha,
    \]
    where $P$ is the polynomial computed by~$A$. Moreover, if $A$ has $g$ gates, then $\pi_{\le d}(A)$ has $O(g(d+1)^2)$ gates.
    The construction takes $\widetilde O\bigl(s(d+1)^2\bigr)$ bit operations and does not perform arithmetic on the ring constants.
    In particular, if $\deg P\le d$, then $\pi_{\le d}(A)$ computes~$P$.
\end{lemma}

We also use the following form of the Schwartz--Zippel lemma.

\begin{lemma}[\cite{S80,Z79,DL78}]\label{lemma:schwartz_zippel}
    Let $f\in\mathbb F[x_1,\ldots,x_t]$ be a nonzero polynomial of total degree at most~$d$ over a field~$\mathbb F$. Let $S\subseteq\F$ be non-empty, and $a_1,\ldots,a_t$ be sampled independently and uniformly at random from $S$. Then
    \[
        \Pr[f(a_1,\ldots,a_t)=0]\le d / |S| \;.
    \]
\end{lemma}

\paragraph{Extension fields.}\label{section:extension_fields}
For a~prime~$p$ and an~integer $k\ge1$, we~represent the field~$\F_{p^k}$ as~$\F_p[u]/(f)$, where $f$ is~the lexicographically first monic irreducible polynomial of~degree~$k$ over~$\F_p$, and we~represent its elements by~their coordinates in~the basis $1,u,\ldots,u^{k-1}$.
Enumerating the $p^k$ monic polynomials of~degree~$k$ and testing their irreducibility~\cite{S05}, we~construct this field in~time polynomial in~$p^k$, and each arithmetic operation in~it~takes time polynomial in~$k\log p$.
If~$k$ divides~$k'$, then an~embedding $\F_{p^k}\hookrightarrow\F_{p^{k'}}$ is~given by~a~root of~the defining polynomial of~$\F_{p^k}$ in~$\F_{p^{k'}}$, which we~find by~exhaustive search in~time polynomial in~$p^{k'}$.
Note that a~circuit over a~field computes the same polynomial over every extension of~this field.

\subsection{Boolean Circuits and Satisfiability}
\label{section:boolean_circuits}

A \emph{Boolean circuit}~$C$ on variables $x_1,\ldots,x_n$ is a directed acyclic graph whose gates have in-degree zero or two.
The in-degree-zero gates are labeled by variables or the constants $0$ and~$1$.
Each in-degree-two gate is labeled by a binary Boolean function, which it applies to its two inputs.
Unless stated otherwise, the circuit has one designated output gate.
Its \emph{size}, denoted by $\operatorname{size}(C)$, is its number of gates.
For circuits with multiple output bits, we specify an ordered list of output gates and include this list in the circuit encoding.

For a class of languages~$\mathcal L$ and a size bound~$s(n)$, we write $\mathcal L\subseteq\cc{SIZE}[s(n)]$ if, for every $L\in\mathcal L$ and all sufficiently large~$n$, the restriction of~$L$ to inputs of length~$n$ is computed by a Boolean circuit of size at most~$s(n)$.
As usual, $\PPoly=\bigcup_{c\ge1}\cc{SIZE}[n^c]$.

A \emph{formula} is a Boolean circuit over $\{\wedge,\vee\}$ whose underlying graph is a tree and whose leaves are labeled by literals or constants.
Its size is its number of leaves.
A \emph{branching program} on $n$ inputs is a directed acyclic graph with a start node and two sinks, labeled \emph{accept} and \emph{reject}.
Every other node~$\alpha$ is labeled by an index $i\in[n]$ and has two outgoing edges leading to nodes $\alpha_0$ and~$\alpha_1$.
On input $\mu\in\{0,1\}^n$, the computation starts at the start node and moves from a node labeled by~$i$ to $\alpha_{\mu_i}$ until it reaches a sink.
The size of a branching program is its number of nodes.

\begin{lemma}[\cite{W87}]\label{lemma:formula_to_bp}
    Every formula of size~$L$ on $n$ inputs can be computed by a branching program of size at most~$L+2$.
\end{lemma}

In $k$-$\cc{SAT}$, the input is a CNF formula with at most $k$ literals per clause, and the task is to determine whether some assignment satisfies all its clauses.
The complementary problem is $k$-$\cc{UNSAT}$.
We remove duplicate clauses, so for fixed~$k$ the binary encoding length is polynomial in the number of variables.

We use the Tseitin transformation in a form that preserves the number of satisfying assignments.

\begin{lemma}[\cite{T83}]\label{lemma:tseitin}
    Every Boolean circuit~$C$ with $n$ inputs can be transformed in polynomial time into a $3$-CNF formula~$F$ with $O(\operatorname{size}(C))$ clauses over the inputs of~$C$ and $O(\operatorname{size}(C))$ auxiliary variables.
    Every auxiliary variable occurs in at least one clause.
    For every $\mu\in\{0,1\}^n$, there is exactly one assignment~$y$ to the auxiliary variables satisfying $F(\mu,y)=1$ if $C(\mu)=1$, and no such assignment otherwise.
\end{lemma}

\subsection{Merlin--Arthur Protocols}
\label{section:merlin_arthur}

The class~$\MA$ consists of languages admitting a proof followed by randomized polynomial-time verification~\cite{B85,BM88}.
On input~$x$, Merlin sends a message~$w$, after which Arthur samples a uniformly random string~$r$ and runs a deterministic verifier~$V(x,w,r)$.
Completeness requires that, if $x\in L$, there exists a message~$w$ satisfying
$\Pr_r[V(x,w,r)=1]\ge2/3$.
Soundness requires that, if $x\notin L$, every message~$w$ satisfies
$\Pr_r[V(x,w,r)=1]\le1/3$.
A protocol runs in time~$T$ if its verification time, message length, and number of random bits are all bounded by~$T$.

A~protocol is~\emph{input-oblivious}~\cite{FSW09,GM15} if, for every input length~$n$, a~single message~$w_n$ satisfies the completeness condition for every $x\in L$ of~length~$n$.
It~is~\emph{perfectly complete} if, moreover, $V(x,w_n,r)=1$ for every such~$x$ and every~$r$.

Our protocols verify claims about sums of~a~polynomial over the Boolean cube by~the sum-check protocol of~Lund, Fortnow, Karloff, and Nisan~\cite{LFKN92}.

\begin{lemma}[\cite{LFKN92}]\label{lemma:sumcheck}
    Let $\F$ be~a~finite field, let $n\ge1$ and $0\le d<|\F|$ be~integers, and let $S\subseteq\F$ be~nonempty.
    Let $H\in\F[y_1,\ldots,y_n]$ have degree at~most~$d$, let $\xi\in\F$, and fix a~set $\Gamma\subseteq\F$ of~$d+1$ distinct points.
    Consider the following protocol for verifying the claim \[\xi=\sum_{\mu\in\{0,1\}^n}H(\mu).\]
    It~has $n$ rounds $j=1,\ldots,n$, and at~the end of~round~$j$, the verifier samples a~uniformly random element $b_j\in S$; we~write $b_{<j}=(b_1,\ldots,b_{j-1})$ for the elements sampled before round~$j$.
    In~round~$j$, the verifier receives from the prover values $v_\gamma\in\F$ for all $\gamma\in\Gamma$; the honest prover sends the values $v_\gamma=g_j(\gamma)$ of~the polynomial \[g_j(z)=\sum_{\mu\in\{0,1\}^{n-j}}H(b_{<j},z,\mu).\]
    The verifier interpolates the polynomial $\widetilde g_j$ of~degree at~most~$d$ with $\widetilde g_j(\gamma)=v_\gamma$ for all $\gamma\in\Gamma$, and rejects unless $\widetilde g_j(0)+\widetilde g_j(1)$ equals~$\xi$ for $j=1$ and $\widetilde g_{j-1}(b_{j-1})$ for $j>1$.
    After round~$n$, the verifier computes $H(b_1,\ldots,b_n)$ and accepts if~and only if~it~equals $\widetilde g_n(b_n)$.
    This protocol has the following properties.
    \begin{itemize}
        \item If~the claim is~true, the honest prover makes the verifier accept with probability one.
        \item If~the claim is~false, every prover makes the verifier accept with probability at~most $nd/|S|$.
    \end{itemize}
\end{lemma}

For the lower bound for~$\cc{NE}$, we~derandomize Merlin--Arthur protocols with truth tables of~high circuit complexity.
For a~string $Y\in\{0,1\}^{2^\ell}$, let $\operatorname{size}(Y)$ be~the minimum size of~a~Boolean circuit with $\ell$ inputs whose truth table is~$Y$.
We~use the generator of~Umans~\cite{Uma03} in~the following form.

\begin{lemma}[{\cite[Theorem~3.2]{Che25}}]\label{lemma:umans}
	There are a~constant $g\ge1$ and a~deterministic algorithm that, given integers $\ell$ and~$m$ with $\ell\le m\le2^{\ell/g}$, a~string $Y\in\{0,1\}^{2^\ell}$, and a~seed $z\in\{0,1\}^{g\ell}$, outputs a~string $G(m,Y,z)\in\{0,1\}^m$ in~time $2^{O(\ell)}$, with the following property.
	If~$\operatorname{size}(Y)>m^g$, then every Boolean circuit~$C$ with $m$ inputs and at~most~$m$ gates satisfies \[\Pr_z[C(G(m,Y,z))=1]\le\Pr_u[C(u)=1]+1/m,\] where $z$ and~$u$ are uniform in~$\{0,1\}^{g\ell}$ and in~$\{0,1\}^m$, respectively.
\end{lemma}

\subsection{Range Avoidance and Nondeterministic Constructions}
\label{section:range_avoid}

Range avoidance provides a framework for studying explicit
construction problems~\cite{K21}.

\begin{definition}[Range avoidance]
    An instance of $\operatorname{Avoid}$ is a Boolean circuit
    $C:\{0,1\}^n\to\{0,1\}^m$ for $m>n$. The task is to find a string $y\in\{0,1\}^m$ outside
    $\operatorname{Range}(C)\coloneqq \{C(x):x\in\{0,1\}^n\}$. We write $\operatorname{LexAvoid}(C)$ for the
    lexicographically smallest such string.
\end{definition}

Since $m>n$, at least one string lies outside the range of~$C$,
so $\operatorname{LexAvoid}(C)$ is defined on all
instances.

We say that a function~$f$ admits a
\emph{single-valued $\cc{FNP}$ construction} if there is a
nondeterministic polynomial-time algorithm that, on every
input~$x$ in the domain of~$f$, has at least one accepting
computation and outputs~$f(x)$ on every accepting computation.
On inputs outside the domain of~$f$, all computations reject.
The accepting computations need not be unique; only their
output must be the same.
This is the standard notion of a function in
$\cc{NPSV}$~\cite{Sel94,HNOS96}.

The following lemma shows that $\cc{PH}=\NP$ suffices for
such a construction of~$\operatorname{LexAvoid}$.

\begin{lemma}\label{lemma:lex_avoid}
    If $\cc{PH}=\NP$, then $\operatorname{LexAvoid}$ has a single-valued $\cc{FNP}$ algorithm.
\end{lemma}

\begin{proof}
    Fix a circuit $C:\{0,1\}^n\to\{0,1\}^m$ with $m>n$.
    A string $y\in\{0,1\}^m$ equals $\operatorname{LexAvoid}(C)$ if and only if both of the following conditions hold:
    \begin{enumerate}
        \item $y$ is outside the range of~$C$: $\forall x\in\{0,1\}^n$, $C(x)\ne y$;
        \item every lexicographically smaller string is in the range of~$C$:
        $\forall z\in\{0,1\}^m$ satisfying $z<_{\mathrm{lex}}y$, $\exists x\in\{0,1\}^n$  such that $C(x)=z$.
    \end{enumerate}
    Their conjunction is a $\Pi_2^{\cc{P}}$ predicate.
    Under $\cc{PH}=\NP$, this predicate belongs to~$\NP$.
    Hence there are a polynomial-time verifier~$V$ and a polynomial~$p$ such that
    \[
        y=\operatorname{LexAvoid}(C)
        \quad\Longleftrightarrow\quad
        \exists w\in\{0,1\}^{\le p(|\langle C\rangle|)},
        \quad V(C,y,w)=1.
    \]
    A nondeterministic algorithm guesses $y$ and~$w$, runs the verifier, and outputs~$y$ if the verifier accepts; otherwise it rejects.
    The string $\operatorname{LexAvoid}(C)$ exists and has an accepting certificate, so at least one branch accepts.
    No other string has an accepting certificate, so every accepting branch outputs $\operatorname{LexAvoid}(C)$.
\end{proof}

\section{Deterministic PIT for Constant-Degree Polynomials}
\label{section:low_degree_pit}

In~this section, we prove \Cref{theorem:low_degree_pit_upper},
which improves the deterministic running time $O_d(s\,t^d)$
of~the naive algorithm for $d$-$\cc{PIT}_{\F}$ by~a~factor
of~$t$.
Throughout, $d\ge1$ is fixed.

\ThmLowDegreePITUpper*

Our algorithm evaluates the given circuit at~points with few
nonzero coordinates.
We first choose a~field $\mathbb K\supseteq\F$ with at~least
$d+1$ elements.
If~$|\F|\ge d+1$, we take $\mathbb K=\F$.
Otherwise, $\F$ is a~finite field of~size $q\le d$, and we take
an~extension of~degree
$m=\lceil\log_q(d+1)\rceil$.
The given circuit can be evaluated over~$\mathbb K$, and extending
the field does not change whether its polynomial is identically zero.

Fix a~set $E\subseteq\mathbb K$ of~size $d+1$ containing zero.
For an~integer $0\le r\le d$, let $H_r\subseteq E^t$ be~the set of~points with at~most $r$ nonzero coordinates.
Then
\[
    |H_r|
    = \sum_{j=0}^{\min\{r,t\}}\binom{t}{j}d^j
    = O_d(t^r).
\]
The following lemma shows that $H_r$ detects every polynomial
of~degree at~most $d$ containing a~monomial involving at~most
$r$ variables.

\begin{lemma}\label{lemma:few_nonzero_points}
	Let $0\le r\le d$, and let
	$g\in\mathbb K[x_1,\ldots,x_t]$ be a~polynomial of~degree
	at~most~$d$ containing a~monomial involving at~most $r$ variables.
	Then $g(a)\ne0$ for some $a\in H_r$.
\end{lemma}
\begin{proof}
	Let $U$ be the set of~variables occurring in~such a~monomial,
	so~$|U|\le r$.
	Let $\mathcal{G}_U$ be the polynomial in~the variables~$U$ obtained
	from~$g$ by~setting all variables outside~$U$ to~zero, so~$\mathcal{G}_U\not\equiv0$.
	Moreover, $\deg \mathcal{G}_U\le d$.
	By~\Cref{lemma:schwartz_zippel}, a~uniformly random point
	$e\in E^U$ satisfies $\mathcal{G}_U(e)=0$ with probability at~most
	$d/(d+1)<1$.
	Thus, $\mathcal{G}_U(e)\ne0$ for some $e\in E^U$.
	Extending~$e$ by~zero outside~$U$ gives a~point $a\in H_r$
	with $g(a)=\mathcal{G}_U(e)\ne0$.
\end{proof}

Every monomial of~a~polynomial of~degree at~most $d$ involves
at~most $d$ variables.
Thus, \Cref{lemma:few_nonzero_points} gives a~straightforward
algorithm: evaluate the circuit at~all points of~$H_d$ and accept
if~and only if~all computed values are zero.
Its running time is $O_d(s\,|H_d|)=O_d(s\,t^d)$.

To~save a~factor of~$t$, we evaluate on~$H_{d-1}$ instead.
By~\Cref{lemma:few_nonzero_points}, the only nonzero polynomials
of~degree at~most $d$ that can vanish on~this set are homogeneous
multilinear polynomials of~degree~$d$.
Such a~polynomial has a~nonzero first partial derivative,
which has degree $d-1$ and is therefore detected by~$H_{d-1}$.
The following result of~Baur and Strassen~\cite{BS83} lets us
evaluate all first partial derivatives together with the original
polynomial at~a~constant-factor overhead.
Here, $\partial_i f$ denotes the formal partial derivative
of~$f$ with respect to~$x_i$.

\begin{lemma}[\cite{BS83}]\label{lemma:baur_strassen}
	Let $\F$ be a~field.
	Given an~arithmetic circuit over~$\F$ of~size $s$ with $t$ input
	variables that computes a~polynomial $f$, one can construct
	in~time $O(s)$ an~arithmetic circuit over~$\F$ of~size $O(s)$
	with $t+1$ output gates computing
	$f,\partial_1 f,\ldots,\partial_t f$.
\end{lemma}

\begin{proof}[Proof of~\Cref{theorem:low_degree_pit_upper}]
	Let $f$ be the polynomial computed by~the given circuit.
	The algorithm constructs the circuit
	from~\Cref{lemma:baur_strassen}, evaluates all its outputs $f,\partial_1f,\ldots,\partial_tf$ at~every point of~$H_{d-1}$, and accepts if~and only if~all these values are zero.

	If~$f\equiv0$, then all its partial derivatives are zero
	as~well, so~the algorithm accepts.
	Suppose that $f\not\equiv0$.
	If~$f$ contains a~monomial involving at~most $d-1$ variables,
	then \Cref{lemma:few_nonzero_points} gives a~point
	$a\in H_{d-1}$ with $f(a)\ne0$.

	Otherwise, since $\deg f\le d$, every monomial of~$f$ is
	a~product of~$d$ distinct variables.
	Choose a~variable $x_i$ occurring in~one of~these monomials.
	Differentiation with respect to~$x_i$ removes $x_i$ from each
	monomial containing it and leaves its coefficient unchanged.
	Distinct such monomials remain distinct, so~$\partial_i f$
	is a~nonzero polynomial of~degree $d-1$.
	By~\Cref{lemma:few_nonzero_points}, there is a~point
	$a\in H_{d-1}$ with $\partial_i f(a)\ne0$.
	Thus, in~both cases the algorithm rejects.

	Finally, the circuit from~\Cref{lemma:baur_strassen} has size
	$O(s)$, so~evaluating all its outputs at~one point takes
	$O(s)$ arithmetic operations in~$\mathbb K$, and hence
	$O_d(s)$ time.
	We enumerate~$H_{d-1}$ by~listing the nonzero coordinates
	and their values, which takes $O_d(|H_{d-1}|)$ time.
	Initializing the input assignment takes $O(t)=O(s)$ time.
	Between consecutive points, at~most $2(d-1)$ coordinates
	need to~be updated.
	Since $|H_{d-1}|=O_d(t^{d-1})$, the total running time is
	$O_d(s\,t^{d-1})$.
\end{proof}

\section{Partition Polynomials and the Universal Polynomial}\label{section:universal_polynomial}

In~this section, we~formally define partition polynomials and construct the universal polynomial~$\mathcal U$.

Let $r$ and $d$ be~positive integers, and let $\mathcal S=(S_1,\ldots,S_{d})$ be~a~list of~subsets of~$[r]$ with $S_1\cup\cdots\cup S_{d}=[r]$.
We~think of~every $j\in[d]$ as~a~table that depends on~the Boolean variables indexed by~$S_j$ and has an~\emph{entry} $w_{j,\tau}$ for every local assignment $\tau\in\{0,1\}^{S_j}$.
Treating the entries as~formal variables, we~define the \emph{partition polynomial} of~$\mathcal S$ as
\[Z_{\mathcal S}(w)=\sum_{\mu\in\{0,1\}^r}\prod_{j=1}^{d}w_{j,\mu|_{S_j}}.\]
In~other words, every assignment $\mu$ selects one entry in~each table, and $Z_{\mathcal S}$ sums the products of~the selected entries.
The polynomial $Z_{\mathcal S}$ has $t=\sum_{j}2^{|S_j|}$ variables, one for every entry, and is~homogeneous of~degree~$d$.
Since the sets $S_j$ cover~$[r]$, distinct Boolean assignments select distinct tuples of~table entries and therefore contribute distinct monomials.
Thus, every nonzero coefficient of~$Z_{\mathcal S}$ is~one.

The name comes from the partition functions of~weighted constraint satisfaction problems.
For a~vector $a\in\{0,1\}^t$ with one bit $a_{j,\tau}$ for each of~the $t$ entries, we~say that an~assignment $\mu\in\{0,1\}^r$ to~the $r$ Boolean variables \emph{satisfies} $(\mathcal S,a)$ if~$a_{j,\mu|_{S_j}}=1$ for every $j\in[d]$.
Thus, $(\mathcal S,a)$ is~a~constraint satisfaction problem in~which table~$j$ lists the allowed local assignments on~$S_j$, and $Z_{\mathcal S}(a)$ is~its number of~satisfying assignments.
For example, a~$3$-CNF formula $F$ gives one table for every clause, which depends on~the variables of~this clause, together with the vector $a_F$ that marks the local assignments satisfying the clause; then $Z_{\mathcal S}(a_F)$ is~the number of~satisfying assignments of~$F$.

The following lemma expresses the unsatisfiability of~$(\mathcal S,a)$ as~a~polynomial identity over every ring.
For $a\in\{0,1\}^t$ and a~tuple of~variables $x=(x_1,\ldots,x_t)$, we~write $a\odot x=(a_1x_1,\ldots,a_tx_t)$.

\begin{lemma}\label{lemma:masking}
	Let $P\in\Z[x_1,\ldots,x_t]$ have coefficients in~$\{0,1\}$, and let $a\in\{0,1\}^t$.
	Then, over every ring, $P(a\odot x)\equiv0$ if~and only if~$P(a)=0$ over~$\Z$.
\end{lemma}
\begin{proof}
	Every monomial $M$ of~$P$ satisfies $M(a\odot x)=M(a)\cdot M(x)$, where $M(a)\in\{0,1\}$.
	Hence, $P(a\odot x)$ is~the sum of~the monomials $M$ of~$P$ with $M(a)=1$.
	This sum is~identically zero if~and only if~there are no~such monomials, that~is, if~and only if~$P(a)=0$ over~$\Z$.
\end{proof}

In~particular, $Z_{\mathcal S}(a\odot w)\equiv0$ if~and only if~$(\mathcal S,a)$ has no~satisfying assignment.
Next, we~describe the downward self-reduction of~partition polynomials, which fixes the Boolean variables one by~one.

\begin{definition}
Let $\mathcal S=(S_1,\ldots,S_{d})$ be~a~list of~subsets of~$[r]$ with $S_1\cup\cdots\cup S_{d}=[r]$.
For $0\le i\le r$, define
\[
Z_i(w):=
\sum_{\substack{\mu\in\{0,1\}^r\\
                \mu_k=0\text{ for every }k>i}}
\prod_{j=1}^{d} w_{j,\mu|_{S_j}}.
\]
Thus, $Z_i$ sums over assignments whose first $i$ bits
are unrestricted and whose remaining bits are zero.

For each $i\in[r]$, define the substitution $w\mapsto w^{(i)}$ by
\[
w^{(i)}_{j,\tau}:=
\begin{cases}
w_{j,\tau^{\,i\leftarrow 1}}, & \text{if }i\in S_j,\\
w_{j,\tau}, & \text{if }i\notin S_j,
\end{cases}
\]
where $\tau^{\,i\leftarrow 1}$ agrees with $\tau$ except
that its value at $i$ is set to one.
\end{definition}

\begin{lemma}\label{lemma:partition_self_reduction}
The partial partition polynomials satisfy
\[
Z_0(w)=\prod_{j=1}^{d}w_{j,0^{S_j}},
\qquad
Z_r(w)=Z_{\mathcal S}(w),
\]
and, for every $i\in[r]$,
\[
Z_i(w)=Z_{i-1}(w)+Z_{i-1}(w^{(i)}).
\]
Moreover, every $Z_i$ is a projection of $Z_{\mathcal S}$
obtained by setting some table variables to zero.
\end{lemma}

\begin{proof}
	The formulas for $Z_0$ and $Z_r$ follow immediately from the definition.
	For $i\in[r]$, split the assignments defining $Z_i$ according to~the value of~$\mu_i$.
	Those with $\mu_i=0$ contribute $Z_{i-1}(w)$.
	Changing the $i$th bit from zero to~one induces the substitution $w\mapsto w^{(i)}$, so~those with $\mu_i=1$ contribute $Z_{i-1}(w^{(i)})$.

	Finally, obtain $Z_i$ from $Z_{\mathcal S}$ by~setting $w_{j,\tau}=0$ whenever $\tau$ assigns one to~an~index in~$S_j\setminus[i]$.
	Since the sets $S_j$ cover~$[r]$, this removes exactly the monomials whose assignments have a~nonzero bit outside~$[i]$.
\end{proof}

For the rest of~this section, fix a~prime~$p$.
We~now construct an~integer polynomial family $\mathcal U=(\mathcal U_n)_{n\ge3}$, which may depend on~$p$; we~suppress this dependence in~the notation.
The construction has two steps.
First, we~construct a~polynomial $\mathcal U_n$ with $O(n)$ variables that, after substituting an~encoding of~a~$3$-CNF formula~$\Psi$ on~$n$ variables with at~most $3n$ clauses, becomes the generating polynomial of~its satisfying assignments, \[G_\Psi(x)=\sum_{\mu\in\{0,1\}^n:\,\Psi(\mu)=1}\ \prod_{i:\,\mu_i=1}x_i.\]
Distinct satisfying assignments contribute distinct monomials with coefficient one, so, in~every characteristic, $G_\Psi\equiv0$ if~and only if~$\Psi$ is~unsatisfiable.
Second, we~show that every partition polynomial is~a~projection of~such a~generating polynomial for a~suitable formula~$\Psi$.

To~encode the $O(n)$ literal positions using only $O(n)$ variables, we~represent each literal by~one element of~the extension field $\F_Q$ of~degree $k=\lceil\log_p(2n)\rceil$ over~$\F_p$, so~that $Q=p^k\ge2n$.
Encoding each literal by~its bits would instead require $\Theta(n\log n)$ variables.
We~first construct literal indicators over~$\F_Q$.
We~then express their coefficients in~a~power basis $1,z_n,\ldots,z_n^{k-1}$ over~$\F_p$, replace $z_n$ by~a~formal variable~$z$, and lift the resulting prime-field coefficients to~integers.
This produces integer polynomials whose specialization at~$z=z_n$, in~characteristic~$p$, recovers the indicators.

\begin{lemma}\label{lemma:packed_sat_polynomial}
	Fix a~prime~$p$.
	For every $n\ge3$, there is~a~polynomial $\mathcal U_n(z,a,x)$ with integer coefficients and the following properties.
	\begin{enumerate}
		\item It~has $10n+1$ variables, namely $z$, $a=(a_{j,s})_{j\in[3n],\,s\in[3]}$, and $x=(x_1,\ldots,x_n)$, and degree $O(n^2)$.
		Its coefficients have bit length $O(n\log n)$. Given a~monomial, its coefficient can be~computed in~time polynomial in~$n$ and the encoding length of~the monomial.
		Moreover, \[\mathcal U_n(z,a,x)=\sum_{\mu\in\{0,1\}^n}A_n(z,a,x,\mu)\] for an~integer polynomial~$A_n$ of~degree $n^{O(1)}$ that has a~circuit with $n^{O(1)}$ gates and constants in~$\{-1,0,\ldots,p-1\}$, constructed in~time polynomial in~$n$.
		\item Let $Q=p^{\lceil\log_p(2n)\rceil}$.
		There is~an~element $z_n\in\F_Q$ such that, for every $3$-CNF formula $\Psi$ on~$n$ variables with at~most $3n$ clauses, one can compute in~polynomial time a~vector $a_\Psi\in\F_Q^{9n}$ such that, over~$\F_Q$,
		\[\mathcal U_n(z_n,a_\Psi,x)=G_\Psi(x).\]
	\end{enumerate}
\end{lemma}

By~the first property and the definition of~$\cc{VNP}_{\Z}$, we~have $\mathcal U\in\cc{VNP}_{\Z}$, and hence, for every finite field~$\F$ of~characteristic~$p$, the reduction of~$\mathcal U$ modulo~$p$ belongs to~$\cc{VNP}_{\F}$ by~\Cref{remark:mod_p}.

\begin{proof}
	Let $k=\lceil\log_p(2n)\rceil$, so~that $Q=p^k\ge2n$.
	We~represent $\F_Q=\F_p[u]/(f_n)$ as~in~\Cref{section:extension_fields}, in~time polynomial in~$Q<2pn$, and let $z_n$ be~the class of~$u$.
	Every element of~$\F_Q$ is~uniquely written as~$\sum_{v<k}c_vz_n^v$ with $c_v\in\F_p$.
	For $0\le i<2n$, let $i=\sum_{v<k}i_vp^v$ in~base~$p$, and let $\operatorname{enc}(i)=\sum_{v<k}i_vz_n^v$.
	We~encode the positive literal of~the $i$th variable by~$\operatorname{enc}(i-1)$ and the negative one by~$\operatorname{enc}(n+i-1)$.

	Next, we~lift the indicators of~the encodings to~integer polynomials.
	For $0\le i<2n$, expand $1-(a-\operatorname{enc}(i))^{Q-1}$ as~a~polynomial in~$a$ over~$\F_Q$, write each of~its coefficients as~$\sum_{v<k}c_vz_n^v$ with $c_v\in\{0,\ldots,p-1\}$, and then replace $z_n$ by~$z$ and regard every $c_v$ as~an~integer.
	The result is~a~polynomial $\Delta_i(z,a)$ with coefficients in~$\{0,\ldots,p-1\}$ such that $\deg_a\Delta_i\le Q-1$ and $\deg_z\Delta_i\le k-1$.
	All these polynomials can be~constructed in~time polynomial in~$n$.
	By~construction, $\Delta_i(z_n,a)=1-(a-\operatorname{enc}(i))^{Q-1}$ over~$\F_Q$, so, for every $a\in\F_Q$, the value $\Delta_i(z_n,a)$ is~one if~$a=\operatorname{enc}(i)$ and zero otherwise.

	For $\mu\in\{0,1\}^n$, let
	\[\operatorname{val}_\mu(z,a)=\sum_{i=1}^{n}\bigl(\mu_i\Delta_{i-1}(z,a)+(1-\mu_i)\Delta_{n+i-1}(z,a)\bigr).\]
	If~$a$ is~the encoding of~a~literal, then $\operatorname{val}_\mu(z_n,a)$ is~the value of~this literal under~$\mu$.
	We~define
	\[\mathcal U_n(z,a,x)=\sum_{\mu\in\{0,1\}^n}\left[\prod_{j=1}^{3n}\left(1-\prod_{s=1}^{3}\bigl(1-\operatorname{val}_\mu(z,a_{j,s})\bigr)\right)\right]\prod_{i:\,\mu_i=1}x_i.\]
	Since $Q<2pn$ and $k=O(\log n)$, the degree of~every $\operatorname{val}_\mu$ is~$O(n)$, so~the degree of~$\mathcal U_n$ is~$O(n^2)$.
	Replacing $\mu$ in~the summand by~variables~$y$ and $\prod_{i:\,\mu_i=1}x_i$ by~$\prod_{i=1}^n(1-y_i+y_ix_i)$ gives the polynomial~$A_n(z,a,x,y)$ of~the first property: computing every~$\Delta_i$ from its at~most $Qk$ monomials gives a~circuit for~$A_n$ with $n^{O(1)}$ gates and constants in~$\{-1,0,\ldots,p-1\}$.

    Each $\Delta_i$ has at most $Qk$ monomials and coefficients bounded by $p-1$. Since $p$ is fixed and $Qk=n^{O(1)}$, each clause factor has polynomially many monomials and polynomially bounded coefficients. Expanding a product of $3n$ such factors gives coefficients of magnitude at most $n^{O(n)}$. Summing over the $2^n$ assignments therefore gives coefficients of magnitude at most $2^n n^{O(n)}=2^{O(n\log n)}$, and hence bit length $O(n\log n)$.

    The coefficient of~$z^e a^\alpha x^\beta$ is~zero unless
	$\beta\in\{0,1\}^n$; otherwise, only the summand indexed
	by~$\mu=\beta$ contributes.
	Since the clause factors use disjoint groups of~$a$-variables,
	we~extract the prescribed powers of~these variables from
	each factor separately.
	This gives $3n$ univariate polynomials in~$z$, each of~degree
	at~most $3(k-1)$ and computable in~polynomial time.
	We~multiply them and extract the coefficient of~$z^e$.
	The product has degree $O(nk)$, and the preceding expansion
	argument shows that every intermediate coefficient has
	bit length $O(n\log n)$.
	Thus, the computation uses polynomially many arithmetic
	operations on integers of~polynomial bit length and takes
	polynomial time.

	For the second property, let $\Psi$ be~a~$3$-CNF formula on~$n$ variables with at~most $3n$ clauses.
	If~$\Psi$ contains an~empty clause, first replace $\Psi$ by~the conjunction of~a~variable and its negation, viewed as~a~formula on~the same $n$ variables.
	Repeat literals within each short clause until it~has exactly three literals, and append tautological clauses until there are exactly $3n$ clauses.
	These operations preserve the set of~satisfying assignments.
	Let $a_\Psi$ assign to~$a_{j,s}$ the encoding of~the $s$th literal of~the $j$th clause.
	Then, for every $\mu$, the factor of~the $j$th clause in~$\mathcal U_n(z_n,a_\Psi,x)$ is~one if~$\mu$ satisfies this clause and zero otherwise.
	Hence, the expression in~the square brackets is~one if~$\Psi(\mu)=1$ and zero otherwise, which gives the claimed identity.

\end{proof}

Next, we~show that every partition polynomial is~a~projection of~$\mathcal U_n$ over a~suitable extension field.
Recall that the variables~$a$ of~$\mathcal U_n$ can hold any $3$-CNF formula~$\Psi$ on~$n$ variables with at most $3n$ clauses.
Namely, by~\Cref{lemma:packed_sat_polynomial}, setting $z$ and~$a$ to~the constants $z_n$ and~$a_\Psi$ from~$\F_Q$ turns $\mathcal U_n$ into the generating polynomial of~the satisfying assignments of~$\Psi$.
Hence, it~suffices to~construct a~formula~$\Psi$ whose generating polynomial becomes $Z_{\mathcal S}$ after we~replace every variable $x_i$ by~an~entry $w_{j,\tau}$ or~by~one.
We~choose $\Psi$ so~that its satisfying assignments are in~one-to-one correspondence with the assignments $\mu\in\{0,1\}^r$ and it~has, for every entry $w_{j,\tau}$, a~variable $v_{j,\tau}$ that is~true exactly when table~$j$ selects this entry, that~is, when $\mu|_{S_j}=\tau$.
We~replace the variable $x_i$ of~every $v_{j,\tau}$ by~the entry $w_{j,\tau}$, and all other variables $x_i$ by~one.

Defining $v_{j,\tau}$ directly would require a~clause with $|S_j|+1$ literals, which may exceed the allowed width of~three. Instead, for each table~$j$, we~use a~complete binary
decision tree that reads the variables of~$S_j$ one by~one. Each node has a~variable that is~true exactly when the
assignment agrees with the path from the root to~that node. We~force the root variable to~be~true and define each non-root variable from its parent and the next input bit using three clauses of~width at~most three. The leaves are the variables $v_{j,\tau}$. Since each tree has $2^{|S_j|}$ leaves and fewer than twice as~many nodes, the trees together use $O(t)$ variables
and clauses.

\begin{lemma}\label{lemma:weighted_table_projection}
	Let $p$ be~a~prime, let $\mathcal S=(S_1,\ldots,S_{d})$ be~a~list of~subsets of~$[r]$ with $S_1\cup\cdots\cup S_{d}=[r]$, let $t=\sum_{j}2^{|S_j|}$ be~the total number of~entries, and let $n\ge2(r+t)$ be~an~integer.
	Put $Q=p^{\lceil\log_p(2n)\rceil}$.
	Then $Z_{\mathcal S}$ is~a~projection of~$\mathcal U_n$ over~$\F_Q$.
	Namely, there is~a~substitution that replaces every variable $x_i$ of~$\mathcal U_n$ by~an~entry $w_{j,\tau}$ or~by~one, and all other variables of~$\mathcal U_n$ by~constants from~$\F_Q$, such that the resulting polynomial over~$\F_Q$ is~$Z_{\mathcal S}$.
	Moreover, this substitution can be~computed in~time polynomial in~$n$.
\end{lemma}
\begin{proof}
	We~construct a~$3$-CNF formula $\Psi$ on~$n$ variables as~follows.
	It~has variables $y_1,\ldots,y_r$, which represent an~assignment $\mu\in\{0,1\}^r$, and, for every $i\in[r]$, a~variable $\bar y_i$, which is~forced to~equal $\neg y_i$ by~the clauses $(y_i\vee\bar y_i)$ and $(\neg y_i\vee\neg\bar y_i)$.
	For every table~$j$, we~simulate the selection of~the entry $w_{j,\mu|_{S_j}}$ by~a~decision tree.
	Namely, we~order the variables of~$S_j$ and, for every binary string $\tau$ of~length at~most $|S_j|$, introduce a~variable $v_{j,\tau}$ that is~true exactly when the first $|\tau|$ variables of~$S_j$ are assigned~$\tau$.
	This is~enforced by~the clause $(v_{j,\varnothing})$ and, if~$y_i$ is~the $(|\tau|+1)$st variable in~the chosen
order on~$S_j$, by~the constraints
	\[v_{j,\tau0}=v_{j,\tau}\wedge\bar y_i,\qquad v_{j,\tau1}=v_{j,\tau}\wedge y_i,\]
	where each constraint $v=v'\wedge\lambda$ is~written as~the three clauses $(\neg v\vee v')$, $(\neg v\vee\lambda)$, and $(v\vee\neg v'\vee\neg\lambda)$.
	The leaves $v_{j,\tau}$ with $|\tau|=|S_j|$ correspond to~the entries of~table~$j$.
	For example, if~$S_j=\{1,2\}$, then the assignment $y_1=1$, $y_2=0$ makes true exactly the variables $v_{j,\varnothing}$, $v_{j,1}$, and $v_{j,10}$ of~this tree.
	Finally, we~pad $\Psi$ to~exactly $n$ variables by~adding dummy variables forced to~zero. By~construction, every assignment to~$y_1,\ldots,y_r$ extends uniquely to~a~satisfying assignment of~$\Psi$.

	To~apply \Cref{lemma:packed_sat_polynomial}, we~check the variable and clause counts.
	The tree of~table~$j$ has $2^{|S_j|+1}-1$ variables, so, before the padding, $\Psi$ has $2r+2t-d\le n$ variables.
	The formula has $2r$ clauses for the variables $\bar y_i$, $3\cdot2^{|S_j|+1}-5$ clauses for the tree of~table~$j$, and one clause for each of~the $n-2r-2t+d$ dummy variables.
	Hence, it~has $n+4t-4d\le3n$ clauses, since $2t\le n$.

	Therefore, over~$\F_Q$,
	\[\mathcal U_n(z_n,a_\Psi,x)=\sum_{\mu'\in\{0,1\}^n:\,\Psi(\mu')=1}\ \prod_{i:\,\mu'_i=1}x_i.\]
	We~then replace each variable $x_i$ that corresponds to~a~leaf $v_{j,\tau}$ by~$w_{j,\tau}$, and every other variable $x_i$ by~one.
	The unique satisfying assignment extending~$\mu$ makes exactly one leaf variable true in~each table, namely $v_{j,\mu|_{S_j}}$.
	Its monomial therefore becomes $\prod_jw_{j,\mu|_{S_j}}$, so~the right-hand side becomes $Z_{\mathcal S}(w)$.
	All steps of~this construction take time polynomial in~$n$.
\end{proof}

Since the projection is~an~identity of~polynomials over~$\F_Q$, it~remains valid over every field that contains~$\F_Q$.
We~obtain circuits over a~fixed finite field~$\F$ by~simulating the computation over a~common extension of~$\F$ and $\F_Q$ in~a~basis over~$\F$.

\begin{corollary}\label{corollary:partition_circuits}
	Let $\F$ be~a~finite field of~characteristic~$p$, let $\mathcal S=(S_1,\ldots,S_{d})$ be~a~list of~subsets of~$[r]$ with $S_1\cup\cdots\cup S_{d}=[r]$, let $t=\sum_{j}2^{|S_j|}$ be~the total number of~entries, and let $n\ge2(r+t)$ be~an~integer.
	Let $Z$ be~obtained from $Z_{\mathcal S}$ by~replacing some entries by~constants from~$\F$.
	Then $\operatorname{asize}_{\F}(Z)\le\operatorname{asize}_{\F}(\mathcal U_n)\cdot\log^{O(1)}n$.
\end{corollary}
\begin{proof}
	 Applying the substitution from~\Cref{lemma:weighted_table_projection} to~a~minimum circuit for $\mathcal U_n$ over~$\F$ and replacing the chosen entries by~constants gives a~circuit with the same number of~gates that computes~$Z$ over the smallest field~$\F'$ that contains $\F$ and~$\F_Q$.
	Put $k'=[\F':\F]\le\lceil\log_p(2n)\rceil$, and choose a~basis $e_1,\ldots,e_{k'}$ of~$\F'$ over~$\F$ with $e_1=1$.
	We~simulate the circuit over~$\F$ by~computing, for every gate, the coordinates of~its polynomial in~this basis.
	Addition is~performed coordinatewise.
	To~simulate a~multiplication gate, we~compute the $(k')^2$ products of~a~coordinate of~one factor and a~coordinate of~the other, and then each of~the $k'$ coordinates of~the product as~a~fixed linear combination of~these $(k')^2$ products.
	This uses $O((k')^3)$ gates per gate of~the original circuit.
	Since $Z$ has coefficients in~$\F$ and $e_1=1$, its first coordinate is~$Z$ and all its other coordinates are zero.
	The first output coordinate therefore gives the required circuit over~$\F$, with only a~$\log^{O(1)}n$ overhead.
\end{proof}

\section{Fast Oblivious Merlin--Arthur Protocols from~Small Circuits}\label{section:oma}

In~this section, we~prove \Cref{theorem:oma_from_circuits}, which turns small circuits into fast input-oblivious Merlin--Arthur protocols for sparse $3$-\UNSAT{}.

We~first define the Boolean language from~\Cref{theorem:oma_from_circuits}, using the universal family~$\mathcal U$ constructed for the characteristic two.
For $N\ge3$, let $t_N=10N+1$ be~the number of~variables of~$\mathcal U_N$, put $k_N=\lceil\log(2N)\rceil$, and let $\mathbb K_N$ be~the field with $2^{2k_N}$ elements, represented as~in~\Cref{section:extension_fields} by~$2k_N$ coordinates over~$\F_2$, so~that addition in~$\mathbb K_N$ is~bitwise XOR.
Let $L_\oplus$ be~the language of~the strings $1^N0\omega\operatorname{bin}(j)$ such that $N\ge3$, $\omega\in\mathbb K_N^{t_N}$, $0\le j<2k_N$, and coordinate~$j$ of~$\mathcal U_N(\omega)$ is~one, where $\omega$ is~encoded by~$2k_Nt_N$ bits and $\operatorname{bin}(j)$ is~the binary encoding of~$j$ with $\lceil\log(2k_N)\rceil$ bits.
These strings have length $M_N=\Theta(N\log N)$.

\begin{lemma}\label{lemma:oplus_language}
	The language $L_\oplus$ belongs to~$\oplus\cc P\cap\cc{DTIME}[2^{O(M/\log M)}]$, where $M$ is~the input length.
\end{lemma}
\begin{proof}
	By~\Cref{lemma:packed_sat_polynomial}, $\mathcal U_N(\omega)=\sum_{\mu\in\{0,1\}^N}A_N(\omega,\mu)$, where $A_N(\omega,\mu)$ can be~computed from its circuit in~time polynomial in~$N$.
	On~an~input $1^N0\omega\operatorname{bin}(j)$, a~nondeterministic machine guesses $\mu\in\{0,1\}^N$ and accepts if~and only if~coordinate~$j$ of~$A_N(\omega,\mu)$ is~one.
	Since addition in~$\mathbb K_N$ is~bitwise XOR, the number of~accepting computations is~odd if~and only if~coordinate~$j$ of~$\mathcal U_N(\omega)$ is~one, so~$L_\oplus\in\oplus\cc P$.
	Enumerating all~$\mu$ decides $L_\oplus$ in~time $2^NN^{O(1)}=2^{O(M/\log M)}$, since $N=\Theta(M_N/\log M_N)$.
\end{proof}

We~now restate \Cref{theorem:oma_from_circuits}.

\ThmOMAFromCircuits*

In~all three cases, Merlin sends circuits for $N=O(n)$, which depend only on~$n$.
For the permanent, Arthur checks these circuits with the downward self-reduction of~the permanent, as~in~\cite{KI04}, and then computes the number of~satisfying assignments of~the formula modulo a~large prime (\Cref{section:oma_permanent}).
For the universal polynomial and for~$L_\oplus$, Arthur does not check the circuit: he~runs the sum-check protocol on~his own and computes the messages of~the prover from values of~the circuit, and this protocol is~sound against every prover (\Cref{section:oma_sumcheck}).

\subsection{Certifying the Permanent}\label{section:oma_permanent}

Fix $N\ge1$ and regard $\PERM_k$, for $k\le N$, as~a~polynomial in~the entries of~an~$N\times N$ matrix~$X$ that depends only on~its top-left $k\times k$ submatrix.
Then \[\PERM_1(X)=X_{1,1},\qquad\PERM_k(X)=\sum_{j=1}^{k}X_{1,j}\PERM_{k-1}(X^{(k,j)})\quad(2\le k\le N),\] where $X^{(k,j)}$ is~obtained by~placing the minor formed from the first $k$ rows and columns of~$X$, with the first row and the $j$th column deleted, in~the top-left corner of~an~otherwise zero $N\times N$ matrix.
Starting from $\PERM_1$, these identities uniquely determine all the polynomials $\PERM_k$.
Arthur can therefore test candidate circuits by~checking these identities at~random points, after truncating the circuits to~the required degrees.

\begin{lemma}\label{lemma:perm_oma}
	Let $N,s\ge1$ be~integers, and let $p\ge3(N+1)$ be~a~prime.
	There is~a~randomized algorithm that, given an~$N\times N$ matrix~$A$ over~$\F_p$ and circuits $C_2,\ldots,C_N$ over~$\F_p$ with at~most $s$ gates each, runs in~time $N^{O(1)}s\log^{O(1)}(sp)$ and has the following properties.
	\begin{itemize}
		\item If~$C_k$ computes $\PERM_k$ for every~$k$, then it~outputs $\PERM(A)\bmod p$ with probability one.
		\item For all circuits $C_2,\ldots,C_N$, with probability at~least $2/3$, it~either rejects or~outputs $\PERM(A)\bmod p$.
	\end{itemize}
\end{lemma}
\begin{proof}
	Set $C'_1(X)=X_{1,1}$ and, for $2\le k\le N$, construct $C'_k=\pi_{\le k}(C_k)$ by~\Cref{lemma:circuit_truncation}.
	For every $k\ge2$, the algorithm tests the identity \[C'_k(X)=\sum_{j=1}^{k}X_{1,j}C'_{k-1}(X^{(k,j)})\] by~evaluating the difference of~its two sides, which has degree at~most~$k$ and a~circuit with $O(N^3s)$ gates, at~an~independent uniformly random point of~$\F_p^{N^2}$.
	It~rejects if~some difference is~nonzero, and otherwise outputs $C'_N(A)$.
	The construction and all evaluations take time $N^{O(1)}s\log^{O(1)}(sp)$.

	If~every $C_k$ computes $\PERM_k$, then truncation does not change these polynomials, all differences are zero, and the algorithm outputs $\PERM(A)\bmod p$ with probability one.
	For arbitrary circuits, if~$C'_k\equiv\PERM_k$ for every~$k$, then the output is~correct.
	Otherwise, let $k$ be~the least index with $C'_k\not\equiv\PERM_k$; then $k\ge2$.
	By~the choice of~$k$ and the downward self-reduction, the difference for this~$k$ equals $C'_k-\PERM_k$, which is~a~nonzero polynomial of~degree at~most~$k$.
	By~\Cref{lemma:schwartz_zippel}, it~vanishes at~a~uniformly random point with probability at~most $k/p\le N/(3(N+1))<1/3$, so~the algorithm rejects with probability at~least $2/3$.
\end{proof}

To~apply this algorithm to~sparse $3$-\UNSAT{}, we~use the reduction of~Ben-Dor and Halevi~\cite{BH93}.

\begin{lemma}[\cite{BH93}]\label{lemma:sat_to_permanent}
	Given a~$3$-CNF formula~$F$ with $n$ variables and $m$ clauses, one can construct in~polynomial time a~matrix~$A_F$ of~dimension $O(n+m)$ with entries in~$\{-1,0,1,2,3\}$ and a~positive integer $\kappa_F\le2^{O(m)}$ such that \[\PERM(A_F)=\kappa_F\cdot\#F,\] where $\#F$ is~the number of~satisfying assignments of~$F$.
\end{lemma}

\begin{proof}[Proof of~\Cref{theorem:oma_from_circuits} for the permanent]
	By~\Cref{lemma:sat_to_permanent}, every sparse formula~$F$ gives a~matrix~$A_F$ of~dimension $N=D_1n$ and a~factor $\kappa_F\le2^{bn}$, for constants $D_1$ and~$b$ depending only on~$c$.
	Merlin sends a~prime $p\in(2^{(b+1)n},2^{(b+1)n+1})$ and the reductions modulo~$p$ of~minimum integer circuits for $\PERM_2,\ldots,\PERM_N$, which have at~most $s(N)+O(1)$ gates by~\Cref{remark:mod_p} and depend only on~$n$.
	Arthur checks the message and accepts if~and only if~the algorithm of~\Cref{lemma:perm_oma} outputs zero on~$A_F\bmod p$.
	Since $0\le\PERM(A_F)=\kappa_F\cdot\#F<p$, this value is~zero if~and only if~$F$ is~unsatisfiable, so~\Cref{lemma:perm_oma} gives perfect completeness and soundness error $1/3$.
	The message has length $s(N)n^{O(1)}$, since $\log s(N)\le N$ and $\log p=O(n)$, and the protocol runs in~time $s(D_1n)n^{O(1)}$.
\end{proof}

\subsection{Sum-Check with an~Evaluation Oracle}\label{section:oma_sumcheck}

In~the proof for the permanent, Arthur computes $\kappa_F\cdot\#F$ modulo a~prime $p>\kappa_F2^n$ chosen by~Merlin, so~that this value is~zero if~and only if~$F$ is~unsatisfiable.
The universal polynomial, however, is~constructed for a~fixed characteristic~$p$, and the number of~satisfying assignments of~a~satisfiable formula may be~divisible by~$p$.
We~therefore replace this number by~the generating polynomial~$G_F$ of~\Cref{section:universal_polynomial}, which is~identically zero if~and only if~$F$ is~unsatisfiable, in~every characteristic.
Arthur tests this identity at~a~random point and verifies the resulting claim by~the sum-check protocol of~\Cref{lemma:sumcheck}.
Every partial sum needed by~the honest prover is, up~to~an~explicit factor, a~partition polynomial, which Arthur evaluates with one query to~$\mathcal U_N$ for $N=O(n)$.
Since the sum-check protocol is~sound against every prover, the following lemma holds for every oracle.

\begin{lemma}\label{lemma:oracle_sumcheck}
    Fix an~integer $c\ge1$ and a~prime~$p$.
    Let $n\ge1$ and $N\ge(6+16c)n$ be~integers, and let $t$ be~the number of~variables of~$\mathcal U_N$.
    Let $\F$ be~a~finite field of~characteristic~$p$ with at~least $4N^2$ elements and containing~$\F_Q$, where $Q=p^{\lceil\log_p(2N)\rceil}$.
    There is~a~randomized algorithm~$\mathcal A$ that, given a~$3$-CNF formula~$F$ on~$n$ variables with at~most $cn$ clauses and oracle access to~a~function $\mathcal O\colon\F^t\to\F$, has the following properties.
    \begin{itemize}
        \item If~$F$ is~unsatisfiable and $\mathcal O$ agrees with $\mathcal U_N$ on~$\F^t$, then $\mathcal A$ accepts with probability one.
        \item If~$F$ is~satisfiable, then, for every function~$\mathcal O$, the algorithm $\mathcal A$ accepts with probability at~most $1/16$.
    \end{itemize}
    Moreover, $\mathcal A$ makes at~most $n(N+1)$ oracle queries and uses additional time polynomial in~$N$ and $\log|\F|$.
\end{lemma}
\begin{proof}
    We~arithmetize~$F$, then evaluate its partial sums with the oracle, and simulate the sum-check protocol.

    \paragraph{Arithmetization.}
    Let $F$ have $m\le cn$ clauses.
    For a~literal~$\lambda$ on~a~variable~$y_i$, put $\lambda(y)=y_i$ if~the literal is~positive, and $\lambda(y)=1-y_i$ otherwise.
    Define \[H_F(y,x)=\prod_{C\in F}\Bigl(1-\prod_{\lambda\in C}(1-\lambda(y))\Bigr)\prod_{i=1}^n(1-y_i+y_ix_i).\]
    For a~Boolean assignment~$\mu$, the first product is~the indicator that $\mu$ satisfies~$F$, and the second product is~$\prod_{i:\,\mu_i=1}x_i$.
    The degree of~$H_F$ is~at~most $3m+2n\le N$.
    For $0\le j\le n$ and $\beta=(\beta_1,\ldots,\beta_j)\in\F^j$, define the partial sum over the last $n-j$ variables \[R_j(\beta,x)=\sum_{\mu\in\{0,1\}^{n-j}}H_F(\beta,\mu,x).\]
    Thus $R_0=G_F$ and $R_n=H_F$.

    \paragraph{Evaluating partial sums with the oracle.}
    Fix $j$ and a~point $(\beta,x)$ with $\beta\in\F^j$ and $x\in\F^n$.
    For each clause~$C$, form the table of~values of~$1-\prod_{\lambda\in C}(1-\lambda(y))$ on~the variables among $y_{j+1},\ldots,y_n$ occurring in~$C$, with $y_1,\ldots,y_j$ fixed to~$\beta_1,\ldots,\beta_j$.
    For each $i>j$, also form the table on~$y_i$ with entries $1$ and~$x_i$.
    The partition polynomial of~these tables, evaluated at~their entries and multiplied by~$\prod_{i\le j}(1-\beta_i+\beta_ix_i)$, equals $R_j(\beta,x)$.

    The tables have at~most $8m+2(n-j)\le(8c+2)n$ entries in~total and involve at~most $n-j\le n$ Boolean variables.
    Since $2\bigl(n-j+(8c+2)n\bigr)\le(6+16c)n\le N$, \Cref{lemma:weighted_table_projection} gives a~point $\omega\in\F^t$, computable in~time polynomial in~$N$ and $\log|\F|$, such that \[R_j(\beta,x)=\mathcal U_N(\omega)\prod_{i\le j}(1-\beta_i+\beta_ix_i).\]
    Define $\widetilde R_j(\beta,x)$ by~replacing $\mathcal U_N(\omega)$ with the oracle answer $\mathcal O(\omega)$ in~this expression.
    Each value of~$\widetilde R_j$ therefore requires at~most one oracle query.

    \paragraph{Simulating sum-check.}
    Fix an~ordering of~$\F$, and let $S$ consist of~its first $2^{\lfloor\log|\F|\rfloor}$ elements.
    Then $|S|\ge|\F|/2$, and a~uniform element of~$S$ can be~sampled using $\lfloor\log|\F|\rfloor$ random bits.
    Also fix a~set $\Gamma\subseteq\F$ of~$N+1$ distinct interpolation points.
    The algorithm samples $x\in S^n$ uniformly and runs the verifier of~\Cref{lemma:sumcheck} for the polynomial $H(y)=H_F(y,x)$, the set~$S$, and the claim \[\sum_{\mu\in\{0,1\}^n}H(\mu)=0.\]
    In~round~$j$, it~uses the values $\widetilde R_j(b_{<j},\gamma,x)$ for $\gamma\in\Gamma$ as~the prover's message, which is~fixed before $b_j$ is~sampled.
    It~computes the value $H(b_1,\ldots,b_n)$ for the final check by~evaluating~$H_F$ directly.

    If~$\mathcal O$ agrees with $\mathcal U_N$, then $\widetilde R_j=R_j$ for every~$j$, so~the simulated prover is~honest.
    If~$F$ is~unsatisfiable, then $R_0(x)=G_F(x)=0$ for every~$x$, and the completeness of~the sum-check protocol gives acceptance with probability one.

    Now suppose that $F$ is~satisfiable, and fix an~arbitrary function~$\mathcal O$.
    The polynomial $G_F$ is~nonzero and has degree at~most~$n$, so~\Cref{lemma:schwartz_zippel} gives $\Pr_{x\in S^n}[G_F(x)=0]\le n/|S|$.
    For every fixed~$x$ with $G_F(x)\ne0$, the oracle determines a~prover for a~false claim, and each of~its messages depends only on~$x$ and on~the previously sampled elements, so~\Cref{lemma:sumcheck} bounds the acceptance probability by~$nN/|S|$.
    The overall acceptance probability is~thus at~most
	\[\frac{n(N+1)}{|S|}\le\frac{2n(N+1)}{4N^2}\le\frac{n}{N}\le\frac1{16}.\qedhere\]
\end{proof}

\begin{proof}[Proof of~\Cref{theorem:oma_from_circuits} for the universal polynomial]
	Let $p$ be~the characteristic of~$\F$, write $|\F|=p^a$, put $N=(6+16c)n$ and $Q=p^{\lceil\log_p(2N)\rceil}$, and let $\mathbb K$ be~the field with $Q^{2a}$ elements.

	Merlin sends a~circuit~$C$ over~$\F$ with at~most $s(N)$ gates; the honest Merlin sends a~minimum circuit for~$\mathcal U_N$, which depends only on~$n$.
	Arthur runs the algorithm of~\Cref{lemma:oracle_sumcheck} with $\mathbb K$ in~place of~$\F$ on~the input formula, answers every oracle query~$\omega$ by~evaluating~$C$ at~$\omega$ over~$\mathbb K$, and accepts if~and only if~this algorithm accepts.
	If~$C$ computes~$\mathcal U_N$, then it~computes $\mathcal U_N$ also over~$\mathbb K$, so~Arthur accepts every unsatisfiable formula with probability one.
	For every message~$C$, Arthur accepts every satisfiable formula with probability at~most $1/16$.
	Arthur evaluates~$C$ at~most $n(N+1)$ times over a~field with polynomially many elements, so~the protocol runs in~time $s((6+16c)n)n^{O(1)}$.
\end{proof}

\begin{proof}[Proof of~\Cref{theorem:oma_from_circuits} for the Boolean language]
	Put $N=(6+16c)n$, and recall that $\mathcal U$ is~now the universal family for the characteristic two.
	The field $\mathbb K_N$ has $2^{2k_N}\ge4N^2$ elements and contains $\F_Q$ for $Q=2^{k_N}$.

	Merlin sends a~Boolean circuit~$C$ with $M_N$ inputs and at~most $s(N)$ gates; the honest message is~a~minimum circuit for~$L_\oplus$ on~inputs of~length~$M_N$.
	Arthur defines a~function $\mathcal O\colon\mathbb K_N^{t_N}\to\mathbb K_N$ by~letting coordinate~$j$ of~$\mathcal O(\omega)$ be~$C(1^N0\omega\operatorname{bin}(j))$ for $0\le j<2k_N$.
	He~then runs the algorithm of~\Cref{lemma:oracle_sumcheck} with $p=2$ and $\mathbb K_N$ in~place of~$\F$ on~the input formula, answers every oracle query with $2k_N$ evaluations of~$C$, and accepts if~and only if~this algorithm accepts.
	If~$C$ decides~$L_\oplus$ on~inputs of~length~$M_N$, then $\mathcal O$ agrees with $\mathcal U_N$ on~$\mathbb K_N^{t_N}$, so~Arthur accepts every unsatisfiable formula with probability one.
	For every message~$C$, the function~$\mathcal O$ is~fixed before Arthur samples his random bits, so~he accepts every satisfiable formula with probability at~most $1/16$.
	The protocol makes $O(n^2\log n)$ evaluations of~$C$, so~it~runs in~time $s((6+16c)n)n^{O(1)}$.
\end{proof}

\section{Constant-Degree PIT and Faster Satisfiability Algorithms}\label{section:pit_np_size}

In~this section, we~assume that $\operatorname{asize}_{\F}(\mathcal U_N)\le N^{\kappa}$ for a~constant~$\kappa$, and, as~outlined in~\Cref{section:proof_overview}, we~turn a~fast nondeterministic algorithm for $d$-$\cc{PIT}_{\F}$ into a~nondeterministic algorithm that decides unsatisfiability of~Boolean circuits with $n$ inputs and size at~most $2^{\rho n}$ in~time $O(2^{(1-\rho)n})$.
Together with the circuits for~$\mathcal U$, such an~algorithm gives the lower bound for~$\cc{NE}$ in~the first alternative of~\Cref{theorem:pit_fixed_degree,corollary:pit_np_size}, whose proofs we~complete in~\Cref{section:ne_lower_bound}.
The construction has four steps.
\begin{enumerate}
	\item If~$R$ has a~Boolean circuit~$C$, then $P^{(d)}_R$, which has $t=\Theta(d2^{n/d})$ variables, has an~arithmetic circuit of~size $(t+\operatorname{size}(C))^{\kappa}\log^{O(1)}(t+\operatorname{size}(C))$ (\Cref{lemma:uniform_assignment_bound}).
	\item We~use guessed circuits for the assignment polynomials and the assumed $d$-$\cc{PIT}$ algorithm to~decide unsatisfiability of~branching programs of~size $2^{\lambda n}$ in~nondeterministic time $O(2^{(1-\lambda)n})$, for some constant $\lambda>0$ (\Cref{lemma:bp_verifier}).
	\item Using only the small-circuit assumption for~$\mathcal U$, we~show that, for every constant $\zeta>0$, there is~a~constant $\eta>0$ such that every Boolean circuit of~size $2^{\eta n}$ has an~equivalent formula of~size at~most $2^{\zeta n}$, for sufficiently large~$n$ (\Cref{lemma:formula_compression}).
	\item To~decide unsatisfiability of~a~circuit~$C$ of~size $2^{\rho n}$, we~guess, for every gate~$\alpha$ of~$C$, a~formula $F_\alpha$ of~size at~most $2^{\zeta n}$ that computes the value of~$\alpha$ as~a~function of~the $n$ inputs of~$C$.
	We~check consistency at~each gate and unsatisfiability at~the output.
	Each check is~the unsatisfiability of~a~formula of~size $O(2^{\zeta n})$, which we~convert into a~branching program and verify by~step~2 (\Cref{lemma:circuit_unsat}).
\end{enumerate}

\subsection{Assignment Polynomials}

We~first define assignment polynomials and show that they inherit small circuits from~$\mathcal U$.
Fix integers $n\ge d\ge1$, and split $[n]$ into $d$ consecutive blocks $V_1,\ldots,V_{d}$, each of~size $\lfloor n/d\rfloor$ or~$\lceil n/d\rceil$.
For every $j\in[d]$ and every local assignment $\tau\in\{0,1\}^{V_j}$, we~introduce a~variable $x_{j,\tau}$.
Hence, the number of~variables is
\[t=\sum_{j=1}^{d}2^{|V_j|},\qquad\text{where}\qquad d2^{n/d-1}\le t\le d2^{n/d+1}.\]
For a~Boolean function $R\colon\{0,1\}^n\to\{0,1\}$, we~define its \emph{assignment polynomial} as
\[P^{(d)}_R(x)=\sum_{\mu\in\{0,1\}^n}R(\mu)\prod_{j=1}^{d}x_{j,\mu|_{V_j}}.\]
In~other words, $P^{(d)}_R$ is~obtained from the partition polynomial of~the blocks $V_1,\ldots,V_{d}$ by~removing the monomials of~the assignments that do~not satisfy~$R$.
Distinct satisfying assignments contribute distinct monomials of~degree~$d$, each with coefficient one.
Hence, over every field, $P^{(d)}_R\equiv0$ if~and only if~$R$ has no~satisfying assignment.

We~first write $P^{(d)}_R$ as~a~partition polynomial of~degree~$d$ with some entries fixed.

\begin{lemma}\label{lemma:assignment_tables}
	Let $n\ge d\ge1$, let $R\colon\{0,1\}^n\to\{0,1\}$ be~computed by~a~Boolean circuit~$C$, and let $F$ be~the $3$-CNF formula given by~\Cref{lemma:tseitin} for~$C$, with auxiliary variables~$y$.
	Consider the partition polynomial~$Z$ whose Boolean variables are the inputs~$\mu$ and the auxiliary variables~$y$, with a~table for every block~$V_j$ and a~table for every clause of~$F$, which depends on~the variables of~this clause.
	Then $P^{(d)}_R$ is~obtained from~$Z$ by~replacing every entry~$\tau$ of~the table of~$V_j$ by~$x_{j,\tau}$, and every entry of~a~clause table by~one if~its local assignment satisfies the clause and by~zero otherwise.
\end{lemma}
\begin{proof}
	After this replacement, every assignment $(\mu,y)$ contributes the term $F(\mu,y)\prod_jx_{j,\mu|_{V_j}}$, and $\sum_yF(\mu,y)=R(\mu)$ for every $\mu\in\{0,1\}^n$ by~\Cref{lemma:tseitin}.
	Hence the result is~\[\sum_{\mu\in\{0,1\}^n}\sum_yF(\mu,y)\prod_{j=1}^{d}x_{j,\mu|_{V_j}}=\sum_{\mu\in\{0,1\}^n}R(\mu)\prod_{j=1}^{d}x_{j,\mu|_{V_j}}=P^{(d)}_R(x).\qedhere\]
\end{proof}

Hence, if~$\mathcal U$ has small circuits, then so~does the assignment polynomial of~every function with small Boolean circuits.

\begin{lemma}\label{lemma:uniform_assignment_bound}
	Let $\F$ be~a~finite field, and suppose that $\operatorname{asize}_{\F}(\mathcal U_N)\le N^{\kappa}$ for a~constant $\kappa\ge1$ and all sufficiently large~$N$.
	Then, for all integers $n\ge d\ge1$ and every function $R\colon\{0,1\}^n\to\{0,1\}$ computed by~a~Boolean circuit~$C$,
	\[\operatorname{asize}_{\F}\bigl(P^{(d)}_R\bigr)\le\bigl(t+\operatorname{size}(C)\bigr)^{\kappa}\log^{O(1)}\bigl(t+\operatorname{size}(C)\bigr),\]
	where $t$ is~the number of~variables of~$P^{(d)}_R$.
\end{lemma}
\begin{proof}
	The partition polynomial~$Z$ of~\Cref{lemma:assignment_tables} has $n+O(\operatorname{size}(C))$ Boolean variables and $t+O(\operatorname{size}(C))$ entries.
	Since $n\le t$, \Cref{corollary:partition_circuits} gives
	\[\operatorname{asize}_{\F}(P^{(d)}_R)\le\operatorname{asize}_{\F}(\mathcal U_N)\cdot\log^{O(1)}N\]
	for some sufficiently large $N=O(t+\operatorname{size}(C))$.
	The assumed bound on~$\operatorname{asize}_{\F}(\mathcal U_N)$ gives the result.
\end{proof}

\subsection{Verifying Branching Programs}\label{section:verifying_bp}

For a~node $\alpha$ of~a~branching program~$G$, let $R_\alpha\colon\{0,1\}^n\to\{0,1\}$ be~the function with $R_\alpha(\mu)=1$ if~the computation on~input~$\mu$ started at~$\alpha$ reaches the accepting sink, and $R_\alpha(\mu)=0$ otherwise.
Then every $R_\alpha$ can be~computed by~a~Boolean circuit of~size $O(n+\operatorname{size}(G))$.

We~now describe the self-reduction that we~use for verification.
Let $P_\alpha=P^{(d)}_{R_\alpha}$ for every node~$\alpha$.
Since the blocks $V_1,\ldots,V_d$ depend only on~$n$ and~$d$, all these polynomials have the same $t=\sum_{j=1}^{d}2^{|V_j|}$ variables $x_{j,\tau}$.
For $i\in[n]$ and $\sigma\in\{0,1\}$, let $\chi_{i,\sigma}\in\{0,1\}^{t}$ be~the vector with $(\chi_{i,\sigma})_{j,\tau}=0$ if~and only if~$i\in V_j$ and $\tau$ assigns $1-\sigma$ to~input~$i$.
Then the substitution $x\mapsto\chi_{i,\sigma}\odot x$ keeps the monomial $\prod_jx_{j,\mu|_{V_j}}$ of~every assignment $\mu$ with $\mu_i=\sigma$ and sets the monomials of~all other assignments to~zero.
Hence, for every node $\alpha$ labelled by~$i$,
\[P_\alpha(x)=P_{\alpha_0}(\chi_{i,0}\odot x)+P_{\alpha_1}(\chi_{i,1}\odot x),\]
while $P_{\mathrm{reject}}=0$ and $P_{\mathrm{accept}}(x)=\prod_{j=1}^{d}\sum_{\tau\in\{0,1\}^{V_j}}x_{j,\tau}$.
In~words, $P_\alpha$ is~the sum of~the polynomials of~its two successors, restricted to~the assignments with $\mu_i=0$ and $\mu_i=1$, respectively.

The verification algorithm guesses a~circuit~$C_\alpha$ for every polynomial~$P_\alpha$ and checks the recurrence at~every node: for a~node~$\alpha$ labelled by~$i$, it~tests the identity \[C_\alpha(x)=C_{\alpha_0}(\chi_{i,0}\odot x)+C_{\alpha_1}(\chi_{i,1}\odot x),\] that~is, that the difference of~its two sides is~zero, and it~also tests that $C_{\mathrm{start}}$ computes zero.
A~guessed circuit, however, may compute a~polynomial of~large degree, while our identity tests apply only to~polynomials of~degree at~most~$d$.
Therefore, we~first delete the monomials of~degree larger than~$d$ by~the truncation~$\pi_{\le d}$, and we~test only that the truncated difference is~zero.
The argument uses only two properties of~the truncation, and we~state it~for every map~$\pi$ with these properties, since in~\Cref{section:integer_proof} we~apply it~with another map.
First, $\pi$ is~a~linear map on~polynomials in~the variables~$x$ that keeps every monomial of~the~$P_\alpha$, that~is, every monomial that takes one variable from every block.
Second, $\pi$ commutes with the substitutions $x\mapsto\chi\odot x$, that~is, $\pi(f(\chi\odot x))=(\pi f)(\chi\odot x)$ for every polynomial~$f$ and every $\chi\in\{0,1\}^t$.
The truncation has both properties, since the~$P_\alpha$ have degree~$d$ and such a~substitution deletes some monomials of~$f$ and keeps the others unchanged.
Finally, let $C_{\mathrm{reject}}=0$ and $C_{\mathrm{accept}}=\prod_{j=1}^{d}\sum_{\tau\in\{0,1\}^{V_j}}x_{j,\tau}$ be~fixed circuits for the polynomials of~the sinks.

\begin{lemma}\label{lemma:bp_from_tests}
	Let $n\ge d\ge1$, let $G$ be~a~branching program on~$n$ inputs, let $\pi$ be~as~above, and let $\mathcal C$ be~a~set of~circuits in~the variables~$x$ that contains $C_{\mathrm{reject}}$ and~$C_{\mathrm{accept}}$.
	Suppose that membership in~$\mathcal C$ can be~decided in~time~$T$, and let $A$ be~a~nondeterministic algorithm that, given $i\in[n]$ and circuits $C,C_0,C_1\in\mathcal C$ computing polynomials $Q,Q_0,Q_1$, runs in~time~$T$ and accepts only if \[\pi\bigl(Q(x)-Q_0(\chi_{i,0}\odot x)-Q_1(\chi_{i,1}\odot x)\bigr)=0.\]
	Suppose also that, if~$G$ is~unsatisfiable, then there are circuits $C^*_\alpha\in\mathcal C$ computing~$P_\alpha$, with $C^*_\alpha=C_\alpha$ for the two sinks, such that $A$ has an~accepting computation on~$(i,C^*_\alpha,C^*_{\alpha_0},C^*_{\alpha_1})$ for every node~$\alpha$ labelled by~$i$ and on~$(1,C^*_{\mathrm{start}},C_{\mathrm{reject}},C_{\mathrm{reject}})$.
	Then unsatisfiability of~$G$ can be~decided in~nondeterministic time $O((\operatorname{size}(G)+1)T)$.
\end{lemma}
\begin{proof}
	The algorithm guesses a~circuit $C_\alpha\in\mathcal C$ for every node~$\alpha$ that is~not a~sink, runs~$A$ on~$(i,C_\alpha,C_{\alpha_0},C_{\alpha_1})$ for every node~$\alpha$ labelled by~$i$ and on~$(1,C_{\mathrm{start}},C_{\mathrm{reject}},C_{\mathrm{reject}})$, and accepts if~and only if~all runs accept.
	Suppose that all runs accept, and let $Q_\alpha$ be~$\pi$ of~the polynomial computed by~$C_\alpha$ for every node~$\alpha$.
	Since $\pi$ is~linear and commutes with substitutions $x\mapsto\chi\odot x$, $Q_\alpha(x)=Q_{\alpha_0}(\chi_{i,0}\odot x)+Q_{\alpha_1}(\chi_{i,1}\odot x)$ for every node~$\alpha$ labelled by~$i$, and $Q_\alpha=P_\alpha$ for the sinks, since $\pi$ keeps the monomials of~$P_{\mathrm{accept}}$.
	By~the recurrence for the polynomials~$P_\alpha$ and topological induction from the sinks, $Q_\alpha=P_\alpha$ for every node~$\alpha$, so~$P_{\mathrm{start}}=Q_{\mathrm{start}}=0$ by~the last run, and $G$ is~unsatisfiable.
	Conversely, if~$G$ is~unsatisfiable, then the algorithm can guess $C_\alpha=C^*_\alpha$, and then all runs have accepting computations.
	The algorithm makes at~most $\operatorname{size}(G)$ membership tests and $\operatorname{size}(G)+1$ runs of~$A$.
\end{proof}

Over a~finite field, we~apply this lemma with the assumed $\cc{PIT}$ algorithm.

\begin{lemma}\label{lemma:bp_verifier}
	Let $\F$ be~a~finite field, and suppose that $\operatorname{asize}_{\F}(\mathcal U_N)\le N^{\kappa}$ for a~constant $\kappa\ge1$ and all sufficiently large~$N$.
	Suppose also that $d$-$\cc{PIT}_{\F}\in\NTIME[s^{a}t^{b}]$ for an~integer~$d$ and constants $a\ge1$ and $b\ge0$ with $\kappa a+b<d$.
	Then there is~a~constant $\lambda>0$ such that unsatisfiability of~branching programs on~$n$ inputs of~size at~most $2^{\lambda n}$ can be~decided in~nondeterministic time $O(2^{(1-\lambda)n})$.
\end{lemma}
\begin{proof}
	Choose a~constant $\lambda>0$ with $\lambda<1/d$ and $2\lambda<1-(\kappa a+b)/d$, which exists because $\kappa a+b<d$.
	Let $G$ be~a~branching program on~$n$ inputs of~size at~most $2^{\lambda n}$, where $n\ge d$ is~sufficiently large.
	Since $\lambda<1/d$, every $R_\alpha$ can be~computed by~a~Boolean circuit of~size $O(n+2^{\lambda n})\le t$, so, by~\Cref{lemma:uniform_assignment_bound}, every $P_\alpha$ has a~circuit over~$\F$ with at~most $S=t^{\kappa}\log^{O(1)}t$ gates.
	We~apply \Cref{lemma:bp_from_tests} with $\pi=\pi_{\le d}$, the set~$\mathcal C$ of~circuits over~$\F$ in~the variables~$x$ with at~most~$S$ gates, which contains $C_{\mathrm{reject}}$ and~$C_{\mathrm{accept}}$, and the algorithm~$A$ that writes a~circuit for $Q(x)-Q_0(\chi_{i,0}\odot x)-Q_1(\chi_{i,1}\odot x)$, truncates it~to~degree~$d$ by~\Cref{lemma:circuit_truncation}, and runs the $\cc{PIT}$ algorithm on~the result.
	The truncated circuit computes a~polynomial of~degree at~most~$d$, so~$A$ accepts only if~$\pi_{\le d}$ of~the difference is~zero, and it~has an~accepting computation if~the difference is~zero, as~for the circuits computing the~$P_\alpha$.
	This circuit has $t$ variables and $O(d^2S)$ gates, so~$A$ runs in~time $t^{\kappa a+b}\log^{O(1)}t=2^{(\kappa a+b)n/d+O(\log n)}$, since $t\le d2^{n/d+1}$, and, since $\kappa\le\kappa a+b$, this bound also covers the membership tests.
	Hence the total running time is~$2^{(\lambda+(\kappa a+b)/d)n+O(\log n)}$, which is~$O(2^{(1-\lambda)n})$ by~the choice of~$\lambda$.
\end{proof}

\subsection{Verifying Boolean Circuits}\label{section:verifying_circuits}

To~decide unsatisfiability of~a~Boolean circuit, we~guess a~small formula for every gate, convert the formulas that check the gates into branching programs, and verify them by~\Cref{lemma:bp_verifier}.

We~use the depth reduction of~Valiant, Skyum, Berkowitz, and Rackoff~\cite{VSBR83}.
\begin{theorem}[\cite{VSBR83}]\label{theorem:vsbr}
	Over every field~$\mathbb F$, a~polynomial of~degree at~most $d\ge1$ computed by~an~arithmetic circuit of~size $s\ge1$ is~also computed by~an~arithmetic circuit of~fan-in two and depth $O(\log(2d)\log(2sd))$.
\end{theorem}

The next lemma shows that small circuits for~$\mathcal U$ turn Boolean circuits of~size $2^{\eta n}$ into formulas of~size $2^{\zeta n}$ with $\zeta$ arbitrarily small.
This step uses only the small-circuit assumption for~$\mathcal U$, with no~PIT assumption.

\begin{lemma}\label{lemma:formula_compression}
	Let $\F$ be~a~finite field, and suppose that $\operatorname{asize}_{\F}(\mathcal U_N)\le N^{\kappa}$ for a~constant $\kappa\ge1$ and all sufficiently large~$N$.
	Then, for every constant $\zeta>0$, there is~a~constant $\eta>0$ such that, for all sufficiently large~$n$, every Boolean function on~$n$ inputs computed by~a~Boolean circuit of~size at~most $2^{\eta n}$ is~also computed by~a~formula of~size at~most $2^{\zeta n}$.
\end{lemma}
\begin{proof}
	We~encode the elements of~$\F$ by~binary strings of~length $\lceil\log|\F|\rceil$.
	Since $\F$ is~fixed, addition and multiplication on~these encodings are computed by~Boolean circuits over~$\{\wedge,\vee,\neg\}$ of~constant depth and fan-in at~most two.
	Let $\ell\ge1$ be~an~integer, which we~choose at~the end of~the proof, and put $\eta=1/(8\ell)$.

	Let $R$ be~computed by~a~Boolean circuit~$C$ of~size at~most $2^{\eta n}$, where $n\ge\ell$ is~sufficiently large, and let $t$ be~the number of~variables of~$P^{(\ell)}_R$.
	Since $\operatorname{size}(C)\le t$, \Cref{lemma:uniform_assignment_bound} gives an~arithmetic circuit over~$\F$ of~size $t^{\kappa}\log^{O(1)}t$ that computes $P^{(\ell)}_R$.
	By~\Cref{theorem:vsbr}, this polynomial of~degree at~most~$\ell$ is~also computed by~an~arithmetic circuit $\widehat C$ over~$\F$ of~fan-in two and depth $O(\log(2\ell)\log t)$.
    Let $V_1,\ldots,V_\ell$ be~the blocks defining $P_R^{(\ell)}$.
    For $\mu\in\{0,1\}^n$, define $e(\mu)\in\{0,1\}^t$ by
        \[
        e(\mu)_{j,\tau}=
        \begin{cases}
        1 & \text{if }\mu|_{V_j}=\tau,\\
        0 & \text{otherwise}.
        \end{cases}
        \]
	Then the monomial $\prod_jx_{j,\mu'|_{V_j}}$ of~an~assignment $\mu'$ equals one at~$e(\mu)$ if~$\mu'=\mu$ and zero otherwise, so~$P^{(\ell)}_R(e(\mu))=R(\mu)$.

	We~now simulate $\widehat C$ on~the point $e(\mu)$ by~a~Boolean circuit whose inputs are the bits of~$\mu$.
	Every coordinate of~$e(\mu)$ is~a~conjunction of~at~most $\lceil n/\ell\rceil$ literals, so~its encoding is~computed in~depth $O(\log n)$.
	Every gate of~$\widehat C$ is~replaced by~a~circuit of~constant depth that computes the encoding of~its value from the encodings of~the values of~its inputs, and at~the output we~test whether the value equals one.
	The resulting Boolean circuit has fan-in at~most two, computes~$R$, and has depth
	\[\Delta=O(\log(2\ell)\log t+\log n).\]
	Unfolding it~into a~tree and pushing the negations to~the leaves by~De~Morgan's laws, we~obtain a~formula of~size at~most $2^{\Delta}$ that computes~$R$.

	Finally, we~bound $\Delta$ and choose~$\ell$.
	Since $t\le\ell2^{n/\ell+1}$ and $\ell\le n$, we~have $\log t\le n/\ell+O(\log n)$ and $\log(2\ell)\le\log(2n)$, so, for a~constant $c_0$ independent of~$\ell$,
	\[\Delta\le c_0\frac{\log(2\ell)}{\ell}\,n+O(\log^2n).\]
	Choose $\ell$ large enough that $c_0\log(2\ell)/\ell\le\zeta/2$.
	Then $\Delta\le\zeta n$ for all sufficiently large~$n$.
\end{proof}

Finally, we~combine the two steps into a~fast nondeterministic algorithm for deciding unsatisfiability of~Boolean circuits.

\begin{lemma}\label{lemma:circuit_unsat}
	Suppose that, for a~constant $\lambda\in(0,1/2]$, unsatisfiability of~branching programs on~$n$ inputs of~size at~most $2^{\lambda n}$ can be~decided in~nondeterministic time $O(2^{(1-\lambda)n})$.
	Suppose also that, for every constant $\zeta>0$, there is~a~constant $\eta>0$ such that, for all sufficiently large~$n$, every Boolean function on~$n$ inputs computed by~a~Boolean circuit of~size at~most $2^{\eta n}$ is~also computed by~a~formula of~size at~most $2^{\zeta n}$.
	Then there is~a~constant $\rho\in(0,1/2)$ such that unsatisfiability of~Boolean circuits with $n$ inputs and size at~most $2^{\rho n}$ can be~decided in~nondeterministic time $O(2^{(1-\rho)n})$.
\end{lemma}
\begin{proof}
	Let $\eta$ be~the constant from the second assumption for $\zeta=\lambda/4$, and fix a~constant $\rho$ with $0<\rho<\min\{\eta,\lambda/2\}$.
	Let $C$ be~a~Boolean circuit with $n$ inputs and size at~most $2^{\rho n}$, where $n$ is~sufficiently large.

	The algorithm guesses, for every gate $\alpha$ of~$C$, a~formula $F_\alpha$ of~size at~most $2^{\zeta n}$ on~the $n$ inputs, which should compute the value of~$\alpha$.
	It~then checks that the following formulas are unsatisfiable: $F_{\mathrm{out}}$ for the output gate, and, for every gate~$\alpha$, the formula $E_\alpha=F_\alpha\oplus f(F_{\alpha_0},F_{\alpha_1})$ if~$\alpha$ applies a~binary function $f$ to~its inputs $\alpha_0$ and~$\alpha_1$, or~$E_\alpha=F_\alpha\oplus\ell$ if~$\alpha$ is~labelled by~a~variable or~a~constant~$\ell$, so~every check formula has size $O(2^{\zeta n})$.
	The algorithm converts each check formula into a~branching program of~size $O(2^{\zeta n})$ by~\Cref{lemma:formula_to_bp}.
	Since $\zeta<\lambda$, the algorithm can check its unsatisfiability by~the first assumption.

	If~all checks pass, then, by~induction along~$C$, every $F_\alpha$ computes the value of~$\alpha$, so~$C$ is~unsatisfiable, since $F_{\mathrm{out}}$ is.
	Conversely, if~$C$ is~unsatisfiable, then the second assumption supplies correct formulas for all gate functions, since each gate function has a~circuit of~size at~most $\operatorname{size}(C)\le2^{\rho n}\le2^{\eta n}$.
	The algorithm can guess these formulas, and all checks then have accepting computations.
	It~makes $O(2^{\rho n})$ calls to~the algorithm from~the first assumption, each taking time $O(2^{(1-\lambda)n})$, and spends time $2^{(\rho+\zeta)n}\poly(n)$ on~everything else.
	Its total running time is~therefore $O(2^{(1-\rho)n})$, since $2\rho<\lambda\le1/2$ and $\zeta=\lambda/4$.
\end{proof}

\section{A~Nearly Maximal Lower Bound for~\texorpdfstring{$\cc{NE}$}{NE} and Constant-Degree PIT}\label{section:ne_lower_bound}

In~this section, we~show that small circuits for the universal polynomial together with a~slightly faster nondeterministic algorithm for unsatisfiability of~Boolean circuits give a~nearly maximal circuit lower bound for~$\cc{NE}$ (\Cref{section:ne_bound}).
Combined with the satisfiability algorithm of~\Cref{section:pit_np_size}, this lower bound proves \Cref{theorem:pit_fixed_degree,corollary:pit_np_size} (\Cref{section:pit_fixed_degree_proof}).

\subsection{A~Lower Bound for~\texorpdfstring{$\cc{NE}$}{NE} from~Satisfiability Algorithms}\label{section:ne_bound}

We~first prove the lower bound for~$\cc{NE}=\NTIME[2^{O(n)}]$ in~the first alternative of~\Cref{theorem:unconditional_dichotomy,theorem:pit_fixed_degree}, and hence of~\Cref{corollary:pit_np_size}.

\begin{theorem}\label{theorem:ne_lower_bound}
	Let $\F$ be~a~finite field, and let $\mathcal U$ be~the universal family constructed for the characteristic of~$\F$.
	Suppose that $\operatorname{asize}_{\F}(\mathcal U_N)\le N^{\kappa}$ for a~constant $\kappa\ge1$ and all sufficiently large~$N$.
	Suppose also that unsatisfiability of~Boolean circuits with $n$ inputs and size at~most $2^{\rho n}$ can be~decided in~nondeterministic time $O(2^{(1-\rho)n})$ for a~constant $\rho\in(0,1/2)$.
	Then $\cc{NE}\not\subseteq\SIZE[2^n/(2n)]$.
\end{theorem}

Let $F_n\in\{0,1\}^{2^n}$ be~the lexicographically first truth table of~a~function on~$n$ inputs that requires circuits of~size larger than $2^n/(2n)$.
For all sufficiently large~$n$, such a~function exists by~the counting argument of~Shannon~\cite{Sha49}.
We~construct a~nondeterministic algorithm that runs in~time $2^{O(n)}$ and computes~$F_n$ at~infinitely many lengths~$n$.
It~guesses a~string~$T$ and has to~check that $T=F_n$, which is~a~$\Pi_2^{\cc P}$ condition on~$T$.
For this, we~use the Merlin--Arthur protocols of~\Cref{section:oma}, in~which Merlin sends a~circuit for~$\mathcal U$ (\Cref{theorem:oma_from_circuits}).
To~remove Arthur's randomness, we~use truth tables of~high circuit complexity, which the satisfiability algorithm certifies at~infinitely many lengths, similarly to~the proofs of~Murray and Williams~\cite{MW18} and of~Chen~\cite{Che25}: our algorithm guesses such a~table and runs Arthur on~all outputs of~the generator of~\Cref{lemma:umans} built from~it.

We~now show that small circuits for~$\mathcal U$ give such protocols for~$\Sigma_2^{\cc P}$.

\begin{lemma}\label{lemma:sigma2_protocol}
	Let $\F$ be~a~finite field, and let $\mathcal U$ be~the universal family constructed for the characteristic of~$\F$.
	Suppose that $\operatorname{asize}_{\F}(\mathcal U_N)\le N^{\kappa}$ for a~constant $\kappa\ge1$ and all sufficiently large~$N$.
	Then, for every $L\in\Sigma_2^{\cc P}$, there is~a~deterministic polynomial-time algorithm $V(x,w,r)\in\{0,1\}$, where $w$ and~$r$ have length polynomial in~$|x|$, such that, for every~$x$,
	\begin{enumerate}
		\item if~$x\in L$, then some~$w$ satisfies $V(x,w,r)=1$ for every~$r$, and
		\item if~$x\notin L$, then every~$w$ satisfies $\Pr_r[V(x,w,r)=1]\le1/3$.
	\end{enumerate}
	Moreover, $\Sigma_2^{\cc P}=\Pi_2^{\cc P}$.
\end{lemma}
\begin{proof}
	Let $L\in\Sigma_2^{\cc P}$, so~that $x\in L$ if~and only if~$R(x,y,z)$ holds for some~$y$ and every~$z$, where $y$ and~$z$ have length polynomial in~$|x|$ and $R$ is~decided in~polynomial time.
	For fixed $x$ and~$y$, \Cref{lemma:tseitin}, applied to~a~circuit that computes $\neg R(x,y,z)$ from~$z$, gives a~$3$-CNF formula $\Psi_{x,y}$ that is~unsatisfiable if~and only if~$R(x,y,z)$ holds for every~$z$.
	We~may assume that $\Psi_{x,y}$ has at~most as~many clauses as~variables, and that its number $n'$ of~variables depends only on~$|x|$.
	By~\Cref{theorem:oma_from_circuits} with $c=1$ and $s(N)=N^{\kappa}$, there are an~algorithm $A(\Psi,w',r)$ that runs in~time polynomial in~$n'$ and, for every~$n'$, a~message~$w_{n'}$ with the following properties for every $3$-CNF formula~$\Psi$ on~$n'$ variables with at~most $n'$ clauses: if~$\Psi$ is~unsatisfiable, then $A(\Psi,w_{n'},r)=1$ for every~$r$, and if~$\Psi$ is~satisfiable, then every~$w'$ satisfies $\Pr_r[A(\Psi,w',r)=1]\le1/3$.
	The algorithm~$V$ reads~$w$ as~a~string~$y$ and a~message~$w'$, and outputs $A(\Psi_{x,y},w',r)$.
	If~$x\in L$, let $y$ be~such that $R(x,y,z)$ holds for every~$z$; then $\Psi_{x,y}$ is~unsatisfiable, and the message $w=(y,w_{n'})$ satisfies $V(x,w,r)=1$ for every~$r$.
	If~$x\notin L$, then $\Psi_{x,y}$ is~satisfiable for every~$y$, so~every~$w$ satisfies $\Pr_r[V(x,w,r)=1]\le1/3$.

	For the second claim, let $k$ be~the length of~$w$.
	Then $x\in L$ if~and only if, for all $r_1,\ldots,r_k$, some~$w$ satisfies $V(x,w,r_i)=1$ for every $i\in[k]$.
	Indeed, if~$x\in L$, the message~$w$ of~the first claim works for all $r_1,\ldots,r_k$.
	If~$x\notin L$, then a~fixed~$w$ satisfies all $k$ conditions with probability at~most $3^{-k}$, so~with probability at~least $1-2^k3^{-k}>0$ no~$w$ satisfies them.
	This is~a~$\Pi_2^{\cc P}$ condition, so~$\Sigma_2^{\cc P}\subseteq\Pi_2^{\cc P}$, and taking complements gives $\Pi_2^{\cc P}\subseteq\Sigma_2^{\cc P}$.
\end{proof}

Next, we~certify truth tables of~high circuit complexity.
The following lemma is~a~theorem of~Chen and Ren~\cite[Theorem~D.3]{CR20} (Appendix~E.1 of~the full version).\footnote{They state it~for circuit depth, but their proof works for circuit size verbatim.}

\begin{lemma}[{\cite[Theorem~D.3]{CR20}}]\label{lemma:hard_tables}
	Suppose that unsatisfiability of~Boolean circuits with $n$ inputs and size at~most $2^{\rho n}$ can be~decided in~nondeterministic time $O(2^{(1-\rho)n})$ for a~constant $\rho\in(0,1/2)$.
	Then there are a~constant $\delta\in(0,1)$ and a~deterministic algorithm~$W$ that, given $1^\ell$ and $Y\in\{0,1\}^{2^\ell}$, runs in~time $2^{O(\ell)}$, with the following property.
	For infinitely many~$\ell$, the algorithm~$W$ accepts $(1^\ell,Y)$ for some $Y\in\{0,1\}^{2^\ell}$, and every such~$Y$ satisfies $\operatorname{size}(Y)>2^{\delta\ell}$.
\end{lemma}

Finally, we~prove \Cref{theorem:ne_lower_bound}: we~define the hard function, verify it~by~the protocol of~\Cref{lemma:sigma2_protocol}, and derandomize this protocol with the truth tables of~\Cref{lemma:hard_tables}.

\begin{proof}[Proof of~\Cref{theorem:ne_lower_bound}]
	A~string $T\in\{0,1\}^{2^n}$ equals~$F_n$ if~and only if~no~circuit with at~most $2^n/(2n)$ gates computes~$T$, and every lexicographically smaller string in~$\{0,1\}^{2^n}$ is~computed by~such a~circuit.
	This is~a~$\Pi_2^{\cc P}$ condition on~$T$.
	By~\Cref{lemma:sigma2_protocol}, there is~a~deterministic algorithm~$V(T,w,r)$, where $w$ and~$r$ have length polynomial in~$|T|$, with the following properties for every $T\in\{0,1\}^{2^n}$: if~$T=F_n$, then some~$w$ satisfies $V(T,w,r)=1$ for every~$r$, and if~$T\ne F_n$, then every~$w$ satisfies $\Pr_r[V(T,w,r)=1]\le1/3$.
	On~strings~$T$ of~length~$2^n$, it~runs in~time $2^{O(n)}$.
	Hence there is~a~function~$m(n)$ with $2^n\le m(n)\le2^{O(n)}$ such that, for all fixed $T$ and~$w$, the function that maps~$r$ to~$V(T,w,r)$ is~computed by~a~circuit with at~most $m(n)$ inputs and at~most $m(n)$ gates.

	Let $\delta$ and~$W$ be~given by~\Cref{lemma:hard_tables}, let $g$ be~the constant of~\Cref{lemma:umans}, and fix a~constant integer $c\ge1$ such that $2^{\delta cn}\ge m(n)^g$ for all~$n$.
	By~\Cref{lemma:hard_tables}, the set~$I$ of~the lengths~$\ell$ such that $W$ accepts $(1^\ell,Y)$ for some~$Y$ and every such~$Y$ satisfies $\operatorname{size}(Y)>2^{\delta\ell}$ is~infinite.
	Some $j\in\{0,\ldots,c-1\}$ is~congruent modulo~$c$ to~infinitely many elements of~$I$, and we~put $\ell(n)=cn+j$.
	For large~$n$, we~have $\ell(n)\le2^n\le m(n)\le2^{\delta\ell(n)/g}\le2^{\ell(n)/g}$, so~$\ell(n)$ and~$m(n)$ satisfy the conditions of~\Cref{lemma:umans}.

	On~an~input $x\in\{0,1\}^n$, the nondeterministic algorithm for~the language $L$ puts $\ell=\ell(n)$, guesses $T\in\{0,1\}^{2^n}$, $Y\in\{0,1\}^{2^\ell}$, and~$w$, and accepts if~and only if~$W$ accepts $(1^\ell,Y)$, $V(T,w,G(m(n),Y,z))=1$ for every seed $z\in\{0,1\}^{g\ell}$, and the bit of~$T$ indexed by~$x$ is~one; for small~$n$, it~rejects.
	It~runs in~time $2^{O(\ell)}2^{O(n)}=2^{O(n)}$, so~$L\in\cc{NE}$.

	Let $n$ be~large and such that $\ell(n)\in I$.
	Then $W$ accepts some~$Y$, and every such~$Y$ satisfies $\operatorname{size}(Y)>2^{\delta\ell(n)}\ge m(n)^g$.
	If~$T=F_n$, then the message~$w$ given by~\Cref{lemma:sigma2_protocol} satisfies $V(T,w,r)=1$ for every~$r$, and in~particular for every output of~the generator.
	If~$T\ne F_n$, then every~$w$ satisfies $\Pr_r[V(T,w,r)=1]\le1/3$, so, by~\Cref{lemma:umans}, at~most a~$1/3+1/m(n)<1$ fraction of~the seeds~$z$ satisfy $V(T,w,G(m(n),Y,z))=1$, and the algorithm rejects.
	Hence, on~inputs of~length~$n$, the algorithm accepts~$x$ if~and only if~the bit of~$F_n$ indexed by~$x$ is~one.
	Thus, for infinitely many~$n$, the language~$L$ requires circuits of~size larger than $2^n/(2n)$ on~inputs of~length~$n$, and $\cc{NE}\not\subseteq\SIZE[2^n/(2n)]$.
\end{proof}

\subsection{Proof of~the Results on~Constant-Degree PIT}\label{section:pit_fixed_degree_proof}

We~now prove \Cref{theorem:pit_fixed_degree} by~combining \Cref{lemma:bp_verifier,lemma:formula_compression,lemma:circuit_unsat} with \Cref{theorem:ne_lower_bound}, and then derive \Cref{corollary:pit_np_size} from it.

\ThmPITFixedDegree*
\begin{proof}[Proof of~\Cref{theorem:pit_fixed_degree}]
	We~prove the theorem with $P_N=\mathcal U_N$.
	This family is~explicit, has $10N+1$ variables by~\Cref{lemma:packed_sat_polynomial}, and depends only on~the characteristic of~$\F$.
	If~$\mathcal U$ does not satisfy the second alternative, then $\operatorname{asize}_{\F}(\mathcal U_N)\le N^{\kappa}$ for some constant $1<\kappa<(d-b)/a$ and all sufficiently large~$N$.
	Since $\kappa a+b<d$, \Cref{lemma:bp_verifier,lemma:formula_compression} give the assumptions of~\Cref{lemma:circuit_unsat}, which gives a~constant $\rho\in(0,1/2)$ such that unsatisfiability of~Boolean circuits with $n$ inputs and size at~most $2^{\rho n}$ can be~decided in~nondeterministic time $O(2^{(1-\rho)n})$.
	Hence \Cref{theorem:ne_lower_bound} gives $\cc{NE}\not\subseteq\SIZE[2^n/(2n)]$, which is~the first alternative.
\end{proof}

\CrlyPITNPSize*
\begin{proof}[Proof of~\Cref{corollary:pit_np_size}]
	Suppose that both alternatives fail.
	By~\Cref{lemma:packed_sat_polynomial}, the image of~$\mathcal U$ over~$\F$ belongs to~$\cc{VNP}_{\F}=\cc{VP}_{\F}$.
	Choose a~constant $\kappa>1$ such that $\operatorname{asize}_{\F}(\mathcal U_N)\le N^{\kappa}$ for all sufficiently large~$N$.
	Choose a~sufficiently large degree~$d$ from the infinite set in~the hypothesis such that $\kappa a(d)<\varepsilon d$, and put $a=a(d)$ and $b=(1-\varepsilon)d$.
	Then $\kappa a+b<d$, so~\Cref{lemma:bp_verifier,lemma:formula_compression} give the assumptions of~\Cref{lemma:circuit_unsat}, which gives a~nondeterministic algorithm that decides unsatisfiability of~Boolean circuits with $n$ inputs and size at~most $2^{\rho n}$ in~time $O(2^{(1-\rho)n})$ for a~constant $\rho\in(0,1/2)$, and \Cref{theorem:ne_lower_bound} gives $\cc{NE}\not\subseteq\SIZE[2^n/(2n)]$.
	This contradicts the assumption that the first alternative fails.
\end{proof}

\section{An~Unconditional Win--Win Result}\label{section:win_win}

In~this section, we~prove our main result.

\ThmUnconditionalDichotomy*

As~outlined in~\Cref{section:proof_overview}, we~run the proof of~\Cref{theorem:pit_fixed_degree} over the integers to~obtain the satisfiability algorithm of~\Cref{theorem:ne_lower_bound} unconditionally: we~replace $\mathcal U$ by~a~partition polynomial in~$\cc{VNP}^0_{\Z}$ (\Cref{section:constant_free}) and the assumed $\cc{PIT}$ algorithm by~an~unconditional one (\Cref{section:zero_test}), verify branching programs with them (\Cref{section:integer_proof}), and prove \Cref{theorem:unconditional_dichotomy} in~\Cref{section:integer_proof_main}.

\subsection{Constant-Free Circuits and a~Universal Partition Polynomial}\label{section:constant_free}

Every gate~$\alpha$ of~a~circuit computes a~polynomial~$f_\alpha$, and we~write $f_C$ for the polynomial of~the output gate of~a~circuit~$C$.
For a~polynomial $f\in\Z[x]$, let $\|f\|_1$ be~the sum of~the absolute values of~its coefficients.
We~start by~observing that constant-free circuits of~small formal degree compute polynomials with small coefficients.

\begin{proposition}\label{lemma:constant_free_coefficients}
	Let $C$ be~a~constant-free circuit with $s$ gates and formal degree at~most~$d$.
	Then every gate~$\alpha$ of~$C$ satisfies $\|f_\alpha\|_1\le2^{sd}$.
\end{proposition}

We~use the partition polynomials of~\Cref{section:universal_polynomial} for a~fixed list of~tables.
Recall that a~list $\mathcal S=(S_1,\ldots,S_k)$ of~subsets of~$[r]$ with $S_1\cup\cdots\cup S_k=[r]$ describes $k$~tables: table~$j$ depends on~the Boolean variables indexed by~$S_j$ and has an~entry~$w_{j,\tau}$, a~formal variable, for every local assignment $\tau\in\{0,1\}^{S_j}$.
Every assignment $\mu\in\{0,1\}^r$ selects the entry $w_{j,\mu|_{S_j}}$ in~every table~$j$, and the partition polynomial~$Z_{\mathcal S}$ is~the sum, over all~$\mu$, of~the products of~the selected entries.
For $N\ge3$, let the list~$\mathcal T_N$ consist of~the following sets:
\begin{itemize}
	\item every set $\{i_1,i_2,i_3\}\subseteq[N]$ of~three indices, whose table will hold a~clause of~a~$3$-CNF formula on~these three variables;
	\item every interval $\{i,i+1,\ldots,i'\}\subseteq[N]$ of~length at~most $\log N$, whose table will hold a~block of~variables of~an~assignment polynomial.
\end{itemize}
A~set that is~both, such as $\{1,2,3\}$, gives two different tables.
Writing $\mathcal T_N=(S_1,\ldots,S_k)$, we~obtain \[Z_{\mathcal T_N}(w)=\sum_{\mu\in\{0,1\}^N}\prod_{j=1}^{k}w_{j,\mu|_{S_j}},\] a~polynomial with $\sum_j2^{|S_j|}=O(N^3)$ variables that is~homogeneous of~degree $k=O(N^3)$.
Unlike the projections of~$\mathcal U$, the projections of~$Z_{\mathcal T_N}$ below use only the constants $0$ and~$1$.

\begin{lemma}\label{lemma:partition_membership}
	The family $\{Z_{\mathcal T_N}\}_{N\ge3}$ belongs to~$\cc{VNP}^0_{\Z}$.
\end{lemma}
\begin{proof}
	For a~table $S_j$ of~$\mathcal T_N$, consider the constant-free polynomial \[A_j(w,y)=\sum_{\tau\in\{0,1\}^{S_j}}w_{j,\tau}\prod_{i\in S_j:\tau_i=1}y_i\prod_{i\in S_j:\tau_i=0}(1-y_i).\]
	For $y\in\{0,1\}^N$, the product over~$i$ is~one for $\tau=y|_{S_j}$ and zero otherwise, so~$A_j(w,y)=w_{j,y|_{S_j}}$, and $Z_{\mathcal T_N}(w)=\sum_{y\in\{0,1\}^N}\prod_jA_j(w,y)$.
	Every $A_j$ has $2^{|S_j|}\le\max\{8,N\}$ terms with $O(\log N)$ factors each, and there are $O(N^3)$ tables.
	Hence the product of~all $A_j$ has a~constant-free circuit with $N^{O(1)}$ gates and formal degree $N^{O(1)}$.
\end{proof}

The next lemma obtains assignment polynomials from~$Z_{\mathcal T_N}$, as~\Cref{lemma:uniform_assignment_bound} obtains them from~$\mathcal U$.

\begin{lemma}\label{lemma:integer_projection}
	Let $n\ge d\ge1$, let $R\colon\{0,1\}^n\to\{0,1\}$ be~computed by~a~Boolean circuit~$C$, and let $F$ be~the $3$-CNF formula given by~\Cref{lemma:tseitin} for~$C$.
	Let $N\ge2^{\lceil n/d\rceil}$ be~larger than $n$ plus the number of~auxiliary variables of~$F$.
	Then $P^{(d)}_R$ is~obtained from~$Z_{\mathcal T_N}$ by~replacing every entry by~$0$, by~$1$, or~by~a~variable $x_{j,\tau}$.
\end{lemma}
\begin{proof}
	Let $y$ be~the auxiliary variables of~$F$.
	By~\Cref{lemma:tseitin}, $P^{(d)}_R=\sum_{\mu,y}F(\mu,y)\prod_jx_{j,\mu|_{V_j}}$.
	To~obtain this polynomial from~$Z_{\mathcal T_N}$, we~let its Boolean variables $z_1,\ldots,z_N$ hold $\mu$ and then~$y$, and we~add the clause $\neg z_i$ for every remaining position~$i$.
	Every block $V_j$ is~an~interval of~length at~most $\lceil n/d\rceil\le\log N$, so~$\mathcal T_N$ has a~table for it, and we~replace its entry at~$\tau$ by~$x_{j,\tau}$.
	We~assign every clause to~a~triple of~$\mathcal T_N$ that contains its variables, replace the entry of~a~triple at~$\tau$ by~one if~$\tau$ satisfies all clauses assigned to~it and by~zero otherwise, and replace all other entries by~one.
	Then an~assignment $z\in\{0,1\}^N$ satisfies all clauses exactly when $z=(\mu,y,0)$ with $F(\mu,y)=1$, and it~then contributes $\prod_jx_{j,\mu|_{V_j}}$, while every other assignment contributes zero.
\end{proof}

\subsection{PIT for Set-Multilinear Circuits}\label{section:zero_test}

In~this subsection, we~give the unconditional $\cc{PIT}$ algorithm that replaces the assumed one in~the proof of~\Cref{theorem:pit_fixed_degree}.
The polynomials that it~tests are set-multilinear: their variables are split into $d$~blocks, and every monomial takes exactly one variable from every block, as~the monomials of~the assignment polynomials~$P^{(d)}_R$ do.
Such a~polynomial with blocks of~$m$~variables has up~to~$m^d$ monomials, and we~decide whether it~is~zero in~time about $m^{d-\lceil d/6\rceil}$, using the fact that, over the integers, a~polynomial is~zero if~and only if~the sum of~the squares of~its coefficients is~zero.

Let the variables be~split into disjoint blocks $X_1,\ldots,X_d$ of~at~most $m$ variables each.
A~monomial is~\emph{set-multilinear of~type} $B\subseteq[d]$ if~it~is~a~product of~exactly one variable from every block $X_j$ with $j\in B$, and of~no~other variables.
A~polynomial $f\in\Z[X_1,\ldots,X_d]$ is~set-multilinear of~type~$B$ if~all its monomials are; for such~$f$, we~write $f[u]$, for $u\in[m]^B$, for the coefficient of~the monomial that takes the $u_j$th variable of~$X_j$, setting it~to~zero when this variable does not exist.
For polynomials $f$ and~$f'$ of~type~$[d]$, let \[\langle f,f'\rangle=\sum_{u\in[m]^d}f[u]\,f'[u],\qquad\|f\|^2=\langle f,f\rangle.\]
Over the integers, $\|f\|^2=0$ if~and only if~$f=0$.

A~circuit is~\emph{syntactically set-multilinear} if~its gates~$\alpha$ can be~assigned types $B(\alpha)\subseteq[d]$, from the leaves up, as~follows.
A~leaf labeled by~a~variable of~$X_j$ has type~$\{j\}$, a~leaf labeled by~a~nonzero integer has type~$\varnothing$, and a~leaf labeled by~zero may have any type, since the zero polynomial is~set-multilinear of~every type.
For a~sum gate, both children must have the same type, and the gate has this type.
For a~product gate, the children must have disjoint types, and the gate has their union as~its type.
By~induction, every gate~$\alpha$ computes a~polynomial that is~set-multilinear of~type $B(\alpha)$.

If~the types of~all gates come from a~single tree of~partitions of~$[d]$, the circuit is~called \emph{pure}, and Raz and Shpilka~\cite{RS05} test pure circuits for zero in~deterministic polynomial time, using the partial derivative matrices of~Nisan~\cite{N91}.
The circuits that we~test are not pure, since different product gates may split the same type in~different ways.

\ThmZeroTest*

The algorithm computes $\|f_\Phi\|^2$: it~writes $f_\Phi$ as~a~sum of~products of~two set-multilinear polynomials of~lower degree, and computes the inner products of~these products.
Arvind, Chatterjee, Datta, and Mukhopadhyay~\cite{ACDM18} used the same approach, computing the sum of~the squares of~the coefficients as~a~sum of~inner products of~pairs of~summands, to~beat brute force for depth-$3$ circuits.

\paragraph{The heavy decomposition.}
The next lemma is~the set-multilinear form of~the decomposition behind the depth reduction of~Hyafil~\cite{H79} and of~Valiant, Skyum, Berkowitz, and Rackoff~\cite{VSBR83}.
Let $\Phi$ be~a~syntactically set-multilinear circuit whose output gate has type~$[d]$, with $d\ge2$.
Call a~gate \emph{heavy} if~its type has more than $2d/3$ blocks, and let $\mathcal F$ be~the set of~heavy product gates whose children are not heavy.
Note that a~product gate has at~most one heavy child, since two disjoint subsets of~$[d]$ cannot both have more than $2d/3$ elements.
Let $\widetilde\Phi$ be~obtained from~$\Phi$ by~replacing every gate $\beta\in\mathcal F$ by~a~new variable~$y_\beta$.
The gates that are not heavy compute the same polynomials in~$\Phi$ and~$\widetilde\Phi$, which do~not depend on~the new variables.
Every heavy gate~$\alpha$ of~$\widetilde\Phi$ computes a~polynomial that is~linear in~the new variables, \[\sum_{\beta\in\mathcal F}[\alpha:\beta]\,y_\beta,\] and we~write $[\alpha:\beta]$ for the coefficient of~$y_\beta$ in~it, which is~a~polynomial in~the original variables.
Intuitively, $[\alpha:\beta]$ is~the sum, over all paths from~$\alpha$ down~to~$\beta$, of~the products of~the polynomials computed by~the children that leave the path at~its product gates.
It~is~zero unless $B(\beta)\subseteq B(\alpha)$, in~which case it~is~set-multilinear of~type $B(\alpha)\setminus B(\beta)$.
All these claims follow by~induction from the leaves~up: a~gate $\beta\in\mathcal F$ computes~$y_\beta$, so~$[\beta:\beta]=1$ and $[\beta:\beta']=0$ for $\beta'\ne\beta$; both children of~a~heavy sum gate~$\alpha$ are heavy, and $[\alpha:\beta]=[\alpha_0:\beta]+[\alpha_1:\beta]$; and a~heavy product gate $\alpha\notin\mathcal F$ has a~heavy child~$\alpha_0$ and a~child~$\alpha_1$ that is~not heavy, and $[\alpha:\beta]=[\alpha_0:\beta]\,f_{\alpha_1}$, where $f_{\alpha_1}$ has type $B(\alpha)\setminus B(\alpha_0)$.

\begin{lemma}\label{lemma:balanced}
	Let $\alpha_{\mathrm{out}}$ be~the output gate of~$\Phi$.
	For every $\beta\in\mathcal F$, let $\beta_0$ and~$\beta_1$ be~the children of~$\beta$, ordered so~that $|B(\beta_0)|\ge|B(\beta_1)|$.
	Then \[f_\Phi=\sum_{\beta\in\mathcal F}f_{\beta_0}\cdot\left(f_{\beta_1}[\alpha_{\mathrm{out}}:\beta]\right),\] where $f_{\beta_0}$ is~set-multilinear of~type $S_\beta=B(\beta_0)$ with $d/3<|S_\beta|\le2d/3$, and $f_{\beta_1}[\alpha_{\mathrm{out}}:\beta]$ is~set-multilinear of~type $[d]\setminus S_\beta$.
\end{lemma}
\begin{proof}
	Substituting $f_\beta$ for every~$y_\beta$ in~the polynomial of~a~gate of~$\widetilde\Phi$ gives the polynomial of~this gate in~$\Phi$.
	The output gate is~heavy, because $d>2d/3$, so~$f_\Phi=\sum_{\beta\in\mathcal F}f_\beta\,[\alpha_{\mathrm{out}}:\beta]$.
	For $\beta\in\mathcal F$, we~have $f_\beta=f_{\beta_0}f_{\beta_1}$.
	The child $\beta_0$ is~not heavy, and its type has at~least $|B(\beta)|/2>d/3$ blocks.
	Finally, $f_{\beta_1}[\alpha_{\mathrm{out}}:\beta]$ has type $B(\beta_1)\cup([d]\setminus B(\beta))=[d]\setminus S_\beta$.
\end{proof}

\paragraph{Inner products.}
We~now compute the inner product of~two terms of~this decomposition.
For two polynomials of~type~$[d]$ given by~their coefficients, computing the inner product clearly takes about $m^d$ operations, since each of~them has up~to~$m^d$ coefficients.
However, it~turns out that $\langle gh,g'h'\rangle$ can be~computed faster from the coefficients of~$g$, $h$, $g'$, and~$h'$, without writing down the products.

\begin{lemma}\label{lemma:cycle}
	Let $S,S'\subseteq[d]$ have $d/3<|S|,|S'|\le2d/3$.
	Let $g,h,g',h'$ be~set-multilinear of~types $S$, $[d]\setminus S$, $S'$, and $[d]\setminus S'$, respectively, given by~their coefficients.
	Then $\langle gh,g'h'\rangle$ can be~computed with $O(m^{d-\lceil d/6\rceil})$ arithmetic operations, and every intermediate value is~a~sum of~at~most $m^d$ products of~at~most four coefficients of~$g$, $h$, $g'$, and~$h'$.
\end{lemma}
\begin{proof}
	For $X\subseteq[d]$, we~write $\overline X=[d]\setminus X$, and $u_X$ for the restriction of~$u\in[m]^d$ to~$X$.
	Then \[\langle gh,g'h'\rangle=\sum_{u\in[m]^d}g[u_S]\,h[u_{\overline S}]\,g'[u_{S'}]\,h'[u_{\overline{S'}}].\]
	The sets $S\cap S'$, $S\setminus S'$, $S'\setminus S$, and~$\overline{S\cup S'}$ partition~$[d]$, and we~write every $u\in[m]^d$ as~$u=(a,b,b',c)$, where \[a=u_{S\cap S'},\qquad b=u_{S\setminus S'},\qquad b'=u_{S'\setminus S},\qquad c=u_{\overline{S\cup S'}}.\]
	Then $u_S=(a,b)$, $u_{\overline S}=(b',c)$, $u_{S'}=(a,b')$, and $u_{\overline{S'}}=(b,c)$, so \[\langle gh,g'h'\rangle=\sum_{a,b,b',c}g[a,b]\,h'[b,c]\,h[b',c]\,g'[a,b'].\]
	So, we~can rewrite it~in~the following form:
	\begin{align*}
		\langle gh,g'h'\rangle&=\sum_{a,c}A[a,c]\,A'[a,c],\text{ where} \\
		A[a,c]&=\sum_bg[a,b]\,h'[b,c], \\
		A'[a,c]&=\sum_{b'}g'[a,b']\,h[b',c].
	\end{align*}
	The array~$A$ has a~term for every $(a,b,c)$, so~it~is~computed with $O(m^{d-|S'\setminus S|})$ operations; similarly, $A'$ is~computed with $O(m^{d-|S\setminus S'|})$ operations, and the final sum with fewer operations.
	Symmetrically, summing first over~$a$ and over~$c$, that is, computing $\sum_ag[a,b]\,g'[a,b']$ and $\sum_ch'[b,c]\,h[b',c]$ for all $(b,b')$, takes $O(m^{d-|\overline{S\cup S'}|}+m^{d-|S\cap S'|})$ operations.
	In~both orders, every intermediate value is~a~sum of~at~most $m^d$ products of~at~most four coefficients.

	The better of~the two orders uses $O(m^{d-\mu})$ operations, where \[\mu=\max\{\min(|S\setminus S'|,|S'\setminus S|),\min(|S\cap S'|,|\overline{S\cup S'}|)\}.\]
	We~claim that $\mu\ge d/6$; otherwise, one of~$|S\setminus S'|$ and~$|S'\setminus S|$ and one of~$|S\cap S'|$ and~$|\overline{S\cup S'}|$ are smaller than $d/6$, and the sum of~these two sizes is~one of~$|S|$, $|\overline{S'}|$, $|S'|$, and~$|\overline S|$.
	Each of~these is~at~least $d/3$, which is~a~contradiction.
\end{proof}

We~now prove \Cref{theorem:zero_test}, which we~first restate.

\ThmZeroTest*

\begin{proof}[Proof of~\Cref{theorem:zero_test}]
	The algorithm reads~$\Phi$ and rejects if~a~constant of~$\Phi$ has absolute value larger than~$2^L$.
	Let $M=2^{(L+1)(s+1)}$; we~show below that, if~every gate~$\alpha$ of~$\Phi$ satisfies $\|f_\alpha\|_1\le2^L$, then every integer that the algorithm computes has absolute value at~most $s^2m^dM^4$, which has $O(sL+d\log(2m))$ bits.
	So~the algorithm stops and rejects as~soon as~it~computes a~larger integer.

	The algorithm computes the coefficients of~every gate of~$\Phi$ that is~not heavy, then $[\alpha:\beta]$ for all heavy gates~$\alpha$ and all $\beta\in\mathcal F$, then $h_\beta=f_{\beta_1}[\alpha_{\mathrm{out}}:\beta]$ for all $\beta\in\mathcal F$, and finally
	\[\|f_\Phi\|^2=\sum_{\beta,\beta'\in\mathcal F}\langle f_{\beta_0}h_\beta,f_{\beta'_0}h_{\beta'}\rangle\]
	by~\Cref{lemma:balanced,lemma:cycle}.
	It~accepts if~and only if~$\|f_\Phi\|^2=0$.

	Now, we~bound the numbers that the algorithm uses.
	Suppose that $\|f_\alpha\|_1\le2^L$ for every gate~$\alpha$ of~$\Phi$.
	Then the constants of~$\Phi$, which are computed by~leaves, pass the first check.
	By~induction along the recurrences for $[\alpha:\beta]$, we~have $\|[\alpha:\beta]\|_1\le2^{(L+1)s_\alpha}$, where $s_\alpha$ is~the number of~gates of~the subcircuit of~$\alpha$.
	Hence $\|f_{\beta_0}\|_1$ and~$\|h_\beta\|_1$ are at~most~$M$.
	Every integer computed before the inner products is~a~coefficient of~a~gate of~$\Phi$, of~some $[\alpha:\beta]$, or~of~some~$h_\beta$, since a~product of~two polynomials of~disjoint types has coefficients that are products of~two coefficients.
	By~\Cref{lemma:cycle}, every integer computed afterwards has absolute value at~most $s^2m^dM^4$.

	Then, every gate of~$\Phi$ that is~not heavy has a~type of~at~most $\lfloor2d/3\rfloor$ blocks, which is~bounded by~$d - \lceil d/6\rceil$, hence its coefficients are computed with $O(m^{d-\lceil d/6\rceil})$ operations.
	Every $[\alpha:\beta]$ has a~type of~fewer than $d/3$ blocks, and every $h_\beta$ of~fewer than $2d/3$ blocks, so~each of~these at~most $2s^2$ values is~also computed with $O(m^{d-\lceil d/6\rceil})$ operations.
	Therefore, by~\Cref{lemma:cycle}, all $|\mathcal F|^2$ inner products use $O(s^2m^{d-\lceil d/6\rceil})$ operations.
	Hence the algorithm runs in~time $m^{d-\lceil d/6\rceil}\cdot(dsL\log(2m))^{O(1)}$ on~every input.
\end{proof}

\subsection{Verifying Branching Programs}\label{section:integer_proof}

The circuits tested in~\Cref{lemma:bp_verifier} are not set-multilinear, so~we~first make them set-multilinear, similarly to~the truncation of~\Cref{lemma:circuit_truncation}.
For a~polynomial~$f$ and $B\subseteq[d]$, let $\pi_B(f)$ be~the sum of~the monomials of~$f$ that are set-multilinear of~type~$B$, with their coefficients.
For all polynomials $f$ and~$f'$, we~have $\pi_B(f+f')=\pi_B(f)+\pi_B(f')$ and \[\pi_B(ff')=\sum_{B'\subseteq B}\pi_{B'}(f)\,\pi_{B\setminus B'}(f'),\] because a~product of~two monomials is~set-multilinear of~type~$B$ exactly when the first one is~set-multilinear of~some type $B'\subseteq B$ and the second one of~type $B\setminus B'$.
Hence, from a~circuit~$C$ with $s$ gates we~obtain a~syntactically set-multilinear circuit $\operatorname{SM}(C)$ with a~gate $(\alpha,B)$ computing $\pi_B(f_\alpha)$ for every gate~$\alpha$ of~$C$ and every $B\subseteq[d]$, where, for a~product gate~$\alpha$, the gate $(\alpha,B)$ adds up the $2^{|B|}$ products one by~one.
Thus $\operatorname{SM}(C)$ has at~most $3^{d+1}s$ gates, and its output gate $(\alpha_{\mathrm{out}},[d])$ computes $\pi_{[d]}(f_C)$.
If~every gate~$\alpha$ of~$C$ satisfies $\|f_\alpha\|_1\le2^L$, then every gate of~$\operatorname{SM}(C)$ computes a~polynomial~$f$ with $\|f\|_1\le2^{2L}$, since $\|\pi_B(f_\alpha)\|_1\le\|f_\alpha\|_1$, and every partial sum of~the products for $(\alpha,B)$ has $\|\cdot\|_1$ at~most $\sum_{B'}\|\pi_{B'}(f_{\alpha_0})\|_1\|\pi_{B\setminus B'}(f_{\alpha_1})\|_1\le\|f_{\alpha_0}\|_1\|f_{\alpha_1}\|_1$.

We~now apply \Cref{lemma:bp_from_tests} with $\pi=\pi_{[d]}$, and with \Cref{theorem:zero_test} in~place of~the assumed $\cc{PIT}$ algorithm.
Recall that $Z_{\mathcal T_N}$ is~the partition polynomial of~\Cref{section:constant_free} with a~table for every possible clause and every possible block on~$N$ Boolean variables, and that, by~\Cref{lemma:integer_projection}, every assignment polynomial~$P^{(d)}_R$ is~obtained from it~by~replacing its entries by~$0$, $1$, and variables.
Hence small circuits for~$Z_{\mathcal T_N}$ give small circuits for all these polynomials, which is~what \Cref{lemma:bp_from_tests} needs.

\begin{lemma}\label{lemma:integer_bp_verifier}
	Suppose that there is~an~integer $c\ge1$ such that, for all sufficiently large~$N$, the polynomial $Z_{\mathcal T_N}$ has a~circuit over~$\Z$ with at~most $N^c$ gates in~which every gate computes a~polynomial~$f$ with $\|f\|_1\le2^{N^c}$.
	Then there is~a~constant $\lambda>0$ such that unsatisfiability of~branching programs on~$n$ inputs of~size at~most $2^{\lambda n}$ can be~decided in~nondeterministic time $O(2^{(1-\lambda)n})$.
\end{lemma}
\begin{proof}
	Let $d$ be~an~integer, chosen at~the end of~the proof depending only on~$c$, and let $\lambda=1/(2d)$.
	Let $G$ be~a~branching program on~$n$ inputs of~size at~most $2^{\lambda n}$, where $n$ is~sufficiently large.
	We~use the blocks $V_1,\ldots,V_d$ and the polynomials $P_\alpha$ of~\Cref{section:verifying_bp}, whose $t$ variables are split into $d$~blocks of~at~most $m=2^{\lceil n/d\rceil}$ variables $x_{j,\tau}$, one block for every~$V_j$.

	Every $R_\alpha$ has a~Boolean circuit of~size $O(n+2^{\lambda n})=O(m)$, since $\lambda<1/d$.
	Hence, by~\Cref{lemma:integer_projection}, there is~a~constant integer~$K$ such that, for $N=Km$, every $P_\alpha$ is~obtained from~$Z_{\mathcal T_N}$ by~replacing every entry by~$0$, by~$1$, or~by~a~variable.
	Such a~replacement does not increase $\|\cdot\|_1$, so, by~the hypothesis, every $P_\alpha$ has a~circuit over~$\Z$ in~the variables~$x$ with at~most $N^c$ gates in~which every gate computes a~polynomial~$f$ with $\|f\|_1\le2^L$, where $L=N^c$.

	We~apply \Cref{lemma:bp_from_tests} with $\pi=\pi_{[d]}$, the set~$\mathcal C$ of~circuits over~$\Z$ in~the variables~$x$ with at~most $N^c$ gates whose constants have absolute value at~most $2^L$, together with $C_{\mathrm{reject}}$ and~$C_{\mathrm{accept}}$, and the algorithm~$A$ that writes a~circuit~$D$ for $Q(x)-Q_0(\chi_{i,0}\odot x)-Q_1(\chi_{i,1}\odot x)$ and runs the algorithm of~\Cref{theorem:zero_test} on~$\operatorname{SM}(D)$.
	By~\Cref{theorem:zero_test}, $A$ accepts only if~$\pi_{[d]}(f_D)=0$.
	For the circuits computing the polynomials~$P_\alpha$ described above, every such difference circuit computes zero, and their gates compute polynomials~$f$ with $\|f\|_1\le3\cdot2^L\le2^{L+2}$, since a~substitution $x\mapsto\chi\odot x$ does not increase $\|\cdot\|_1$ and the gates of~$C_{\mathrm{accept}}$ have $\|f\|_1\le2^n\le2^L$.
	Hence every gate of~their set-multilinear circuits has $\|f\|_1\le2^{2L+4}$, and $A$ accepts by~\Cref{theorem:zero_test}.

	Every circuit given to~$A$ has $O(m^c)$ gates, since $t\le dm$, and so~does its set-multilinear circuit, which $A$ writes in~time $m^{O(c)}$.
	By~\Cref{theorem:zero_test}, whose constants are absolute, $A$ runs in~time $m^{d-\lceil d/6\rceil+O(c)}n^{O(1)}$, where the constant in~$O(c)$ does not depend on~$d$.
	Hence we~can choose $d$ such that this exponent of~$m$ is~at~most $11d/12$.
	Then, since $m\le2^{n/d+1}$, $A$ runs in~time $O(2^{11n/12}n^{O(1)})$.
	By~\Cref{lemma:bp_from_tests}, the total time is~$O(2^{(\lambda+11/12)n}n^{O(1)})$, which is~$O(2^{(1-\lambda)n})$, since $2\lambda=1/d<1/12$.
\end{proof}

Next, we~show that the second alternative of~\Cref{theorem:unconditional_dichotomy} follows from $\cc{VP}_{\F}\ne\cc{VNP}_{\F}$ for any single field~$\F$.
In~the proof of~the theorem, we~use the first part of~the following proposition, which needs only small integer circuits for~$Z_{\mathcal T_N}$.

\begin{proposition}\label{proposition:vp0_to_vpf}
\begin{enumerate}
	\item If~$Z_{\mathcal T_N}$ has a~circuit over~$\Z$ with $N^{O(1)}$ gates for all sufficiently large~$N$, then $\cc{VP}_{\F}=\cc{VNP}_{\F}$ for every finite field~$\F$.
	\item If~$\cc{VP}^0_{\Z}=\cc{VNP}^0_{\Z}$, then $\cc{VP}_{\F}=\cc{VNP}_{\F}$ for every field~$\F$.
\end{enumerate}
\end{proposition}
\begin{proof}
	Let $R\colon\{0,1\}^{q\times q}\to\{0,1\}$ be~the function of~the variables $e_{i,i'}$ that is~one exactly when the pairs $(i,i')$ with $e_{i,i'}=1$ form a~permutation of~$[q]$ with one cycle.
	It~has a~Boolean circuit of~size $q^{O(1)}$, which checks that every row and every column of~$e$ contains exactly one one and that the orbit of~$1$ under this permutation has $q$~elements.
	The Hamiltonian cycle polynomial \[\mathrm{HC}_q(x)=\sum_{e\in\{0,1\}^{q\times q}}R(e)\prod_{e_{i,i'}=1}x_{i,i'}=\sum_\sigma\prod_{i=1}^qx_{i,\sigma(i)},\] where $\sigma$ ranges over the permutations of~$[q]$ with one cycle, is~obtained from the assignment polynomial $P^{(q^2)}_R$, whose $q^2$ blocks are single variables, by~replacing its variable at~$e_{i,i'}=0$ by~one and its variable at~$e_{i,i'}=1$ by~$x_{i,i'}$.
	Hence, by~\Cref{lemma:integer_projection} with $d=n=q^2$, it~is~obtained from~$Z_{\mathcal T_N}$ for some $N=q^{O(1)}$ by~replacing every entry by~$0$, by~$1$, or~by~a~variable.
	In~the first part, reducing the constants of~the circuits for~$Z_{\mathcal T_N}$ modulo the characteristic of~$\F$ gives circuits over~$\F$ for~$\mathrm{HC}_q$ with $q^{O(1)}$ gates by~\Cref{remark:mod_p}.
	In~the second part, $Z_{\mathcal T_N}$ has constant-free circuits of~polynomial size by~\Cref{lemma:partition_membership}, and the resulting circuits for~$\mathrm{HC}_q$ can be~read over every field~$\F$.
	In~both cases, $\{\mathrm{HC}_q\}\in\cc{VP}_{\F}$, and, since this family is~$\cc{VNP}_{\F}$-complete over every field~\cite{V79a}, we~obtain $\cc{VP}_{\F}=\cc{VNP}_{\F}$.
\end{proof}

\subsection{Proof of~the Win--Win Result}\label{section:integer_proof_main}

Finally, we~combine \Cref{theorem:ne_lower_bound} with the satisfiability algorithm of~\Cref{lemma:integer_bp_verifier} to~prove \Cref{theorem:unconditional_dichotomy}.
We~prove the following stronger statement, in~which the second alternative is~replaced by~the existence of~small integer circuits for~$Z_{\mathcal T_N}$ whose gates compute polynomials with coefficients of~polynomial bit length.

\begin{theorem}\label{theorem:integer_dichotomy_strong}
	Suppose that there is~an~integer $c\ge1$ such that, for all sufficiently large~$N$, the polynomial $Z_{\mathcal T_N}$ has a~circuit over~$\Z$ with at~most $N^c$ gates in~which every gate computes a~polynomial~$f$ with $\|f\|_1\le2^{N^c}$.
	Then $\cc{NE}\not\subseteq\SIZE[2^n/(2n)]$.
\end{theorem}
\begin{proof}
	By~\Cref{proposition:vp0_to_vpf}, $\cc{VP}_{\F_2}=\cc{VNP}_{\F_2}$.
	By~\Cref{lemma:packed_sat_polynomial}, the universal family~$\mathcal U$ constructed for $p=2$ belongs to~$\cc{VNP}_{\F_2}=\cc{VP}_{\F_2}$, so~$\operatorname{asize}_{\F_2}(\mathcal U_N)\le N^{\kappa}$ for a~constant~$\kappa$ and all sufficiently large~$N$.
	\Cref{lemma:integer_bp_verifier} and \Cref{lemma:formula_compression} for $\F=\F_2$ give the assumptions of~\Cref{lemma:circuit_unsat}, which gives a~constant $\rho\in(0,1/2)$ such that unsatisfiability of~Boolean circuits with $n$ inputs and size at~most $2^{\rho n}$ can be~decided in~nondeterministic time $O(2^{(1-\rho)n})$.
	Hence \Cref{theorem:ne_lower_bound} for $\F=\F_2$ gives $\cc{NE}\not\subseteq\SIZE[2^n/(2n)]$.
\end{proof}

\begin{proof}[Proof of~\Cref{theorem:unconditional_dichotomy}]
	Suppose that $\cc{VP}^0_{\Z}=\cc{VNP}^0_{\Z}$.
	By~\Cref{lemma:partition_membership}, the family $\{Z_{\mathcal T_N}\}$ has constant-free circuits with at~most $N^c$ gates and formal degree at~most $N^c$ for some integer $c\ge1$ and all sufficiently large~$N$.
	By~\Cref{lemma:constant_free_coefficients}, every gate of~these circuits computes a~polynomial~$f$ with $\|f\|_1\le2^{N^{2c}}$, so~\Cref{theorem:integer_dichotomy_strong} gives $\cc{NE}\not\subseteq\SIZE[2^n/(2n)]$.
\end{proof}

\section{Constructing Hard Polynomials via Range Avoidance}
\label{section:strong_avoidance}

We show that $\cc{PIT}_{\Z}\in\NP$ implies either
$\cc{VP}_{\Z}\ne\cc{VNP}_{\Z}$ or a single-valued nondeterministic
polynomial-time construction, for every fixed $d\ge1$, of polynomials
in $t$ variables of degree at most~$d$ and arithmetic circuit complexity
$\Theta_d(t^d)$.
The proof has two parts. First, we give an unconditional reduction that
turns any solution to a suitable range-avoidance instance into a hard
polynomial. Second, under $\cc{VP}_{\Z}=\cc{VNP}_{\Z}$, the
Kabanets--Impagliazzo argument~\cite{KI04} gives $\cc{PH}=\NP$.
By~\Cref{lemma:lex_avoid}, this collapse supplies a single-valued
algorithm for solving the range-avoidance instances.
\CrlyStrongAvoidPITWinWin*

The reduction encodes small arithmetic circuits by strings shorter than
their coefficient vectors. A string outside the range of the resulting
map therefore describes a polynomial that no such circuit computes.
We carry out this construction over $\Z_q$ and then lift the coefficients
to~$\Z$. Choosing $q$ comparable to the number of monomials gives a
lower bound linear in that number.

\begin{lemma}
\label{lemma:hard_poly_to_avoid}
For every fixed integer $d\ge 1$, there is a constant $c>0$
with the following property. For all sufficiently large $t$, let
\[
    M=\binom{t+d}{d},\qquad
    q=2^{\lceil\log_2(M+t)\rceil},\qquad
    L=M\log_2 q,\qquad
    s=\lfloor cM\rfloor.
\]
The following properties hold, with construction and
reconstruction performed deterministically in time
polynomial in~$t$:
\begin{enumerate}
    \item \emph{Construction.}
    Given $1^t$, construct a Boolean circuit
    \[
        C_t:\{0,1\}^{L-1}\longrightarrow\{0,1\}^{L}
    \]
    of size polynomial in $t$ whose range contains the encodings
    of the degree-at-most-$d$ coefficient vectors of all
    arithmetic circuits over $\Z_q$ with at most $s$ gates.

    \item \emph{Reconstruction.}
    Given $1^t$ and $y\in\{0,1\}^{L}$, output the coefficient
    list of a polynomial
    \[
        G_{t,y}\in\Z[x_1,\ldots,x_t]
    \]
    of degree at most $d$, with coefficients in
    $\{0,\ldots,q-1\}$.
    The algorithm can also construct an arithmetic circuit
    over $\Z$ of size $O_d(t^d)$ computing $G_{t,y}$.

    \item \emph{Hardness.}
    If $y\notin\operatorname{Range}(C_t)$, then every arithmetic
    circuit over $\Z$ computing $G_{t,y}$ has more than $s$ gates.
    In particular,
    \[
        \operatorname{asize}_{\Z}(G_{t,y})
        =\Theta_d(t^d).
    \]
\end{enumerate}
\end{lemma}

\begin{proof}
Fix an ordering of the $M$ monomials of total degree at most $d$
in $x_1,\ldots,x_t$. Since $q$ is a power of two, each coefficient
in $\Z_q$ can be represented by exactly $\log_2 q$ bits.
Concatenating these representations in the fixed monomial order
gives a bijection between coefficient vectors in $(\Z_q)^M$
and binary strings of length $L=M\log_2 q$.

We first describe the circuit encodings used in the construction.
Every arithmetic circuit over $\Z_q$ with at most $s$ gates
has an encoding of length
\[
    O\bigl(s\log(t+s+q)\bigr)=O(s\log q).
\]
The encoding records the type of each gate, the indices of its
predecessors, any variable or constant labels, and the designated
output gate. The circuit has at most $s$ gates, and its constants
may be arbitrary elements of $\Z_q$.

For a sufficiently small choice of $c>0$ and all sufficiently
large $t$, these encodings can be padded to a common length
$\ell<L$.
For a polynomial $P$, write $[x^\alpha]P$ for the coefficient
of the monomial $x^\alpha$ in $P$, and let $\pi_{\le d}$ denote
the truncation from~\Cref{lemma:circuit_truncation}.

Define $\Phi_t:\{0,1\}^{\ell}\to(\Z_q)^M$ by
\[
    \Phi_t(e)=
    \begin{cases}
        \bigl([x^\alpha]\pi_{\le d}(A_e)\bigr)_{|\alpha|\le d},
        & \text{if $e$ encodes an arithmetic circuit $A_e$
          with at most $s$ gates},\\[2mm]
        0^M,
        & \text{otherwise}.
    \end{cases}
\]
The coordinates are listed in the fixed monomial order.
For a valid encoding $e$, the vector $\Phi_t(e)$ consists
of the coefficients of all monomials of total degree at most $d$
in the polynomial computed by $A_e$.

The map $\Phi_t$ is~computable in~time polynomial in~$t$: as~in~\Cref{lemma:circuit_truncation}, we~keep at~every gate the coefficients of~degree at~most~$d$, which are elements of~$\Z_q$ with $O_d(\log t)$ bits.

Viewing the output of $\Phi_t$ as an $L$-bit string, define
\[
    C_t(z)=\Phi_t(z_1,\ldots,z_\ell)
    \qquad
    \text{for }z\in\{0,1\}^{L-1}.
\]
The circuit ignores the remaining input bits. It can be
constructed uniformly in time polynomial in $t$, and
\[
    \operatorname{Range}(C_t)
    =
    \operatorname{Range}(\Phi_t).
\]
In particular, its range contains the truncated coefficient
vector of every arithmetic circuit over $\Z_q$ with at most
$s$ gates.

For reconstruction, partition $y\in\{0,1\}^{L}$ into
$M$ blocks of $\log_2 q$ bits. Interpret these blocks as
integers $a_\alpha\in\{0,\ldots,q-1\}$, and output
\[
    G_{t,y}(x)
    =
    \sum_{|\alpha|\le d} a_\alpha x^\alpha.
\]
This gives its coefficient list in polynomial time.
To construct an arithmetic circuit for $G_{t,y}$, compute
all monomials of degree at most $d$, generate the integers
$0,\ldots,q-1$ by successive additions starting from $0$ and $1$,
and form the indicated sum. Sharing these monomials and
integer constants gives a circuit with $O_d(M+q)=O_d(t^d)$
gates and only the constant labels $0$ and $1$.
Its size, including the encoding lengths of constant labels,
is therefore $O_d(t^d)$.

Finally, suppose that $y\notin\operatorname{Range}(C_t)$
but that an arithmetic circuit $A$ over $\Z$ with at most
$s$ gates computes $G_{t,y}$. By~\Cref{remark:mod_p}, reducing the constants of~$A$ modulo~$q$ gives a~circuit over~$\Z_q$ with at~most $s$ gates computing $G_{t,y}\bmod q$.
Because $G_{t,y}$ has degree at most $d$, the truncated
coefficient vector of this circuit is exactly the vector
encoded by $y$. By construction, this places $y$ in
$\operatorname{Range}(C_t)$, a contradiction.

Thus every integer arithmetic circuit computing $G_{t,y}$
has more than $s=\Omega_d(t^d)$ gates. The reconstruction
procedure also produces an integer arithmetic circuit for
$G_{t,y}$ of size $O_d(M+q)=O_d(t^d)$. Therefore,
\[
    \operatorname{asize}_{\Z}(G_{t,y})=\Theta_d(t^d).
\]

\end{proof}

The lemma uses ordinary range avoidance with one more output bit than
input bits. Its deterministic reconstruction also preserves
single-valuedness: a unique accepting output~$y$ determines a unique
coefficient list.

\begin{proof}[Proof of~\Cref{corollary:strong_avoid_pit_winwin}]
Assume $\cc{PIT}_{\Z}\in\NP$.
If $\cc{VP}_{\Z}\ne\cc{VNP}_{\Z}$, the first alternative holds.
Otherwise, $\PERM$ has polynomial-size arithmetic circuits over~$\Z$.
Together with the PIT assumption, the Kabanets--Impagliazzo
argument~\cite{KI04} puts the graph of the permanent on $0$--$1$
matrices in~$\NP$.
Consequently, a nondeterministic polynomial-time machine can simulate
a $\cc{P}^{\#\cc{P}}$ computation by guessing and certifying the exact
answers to its permanent queries.
Thus $\cc{P}^{\#\cc{P}}\subseteq\NP$, and Toda's theorem~\cite{T91}
gives
\[
    \cc{PH}\subseteq\cc{P}^{\#\cc{P}}\subseteq\NP.
\]
Hence $\cc{PH}=\NP$.
By~\Cref{lemma:lex_avoid}, $\operatorname{LexAvoid}$ has a single-valued
$\cc{FNP}$ algorithm.

Fix~$d\ge1$.
On input~$1^t$, construct~$C_t$ from~\Cref{lemma:hard_poly_to_avoid},
compute $y_t=\operatorname{LexAvoid}(C_t)$ using this algorithm,
and output the coefficient list of $G_t=G_{t,y_t}$.
The deterministic steps take polynomial time, and the middle step
has at least one accepting branch, all of whose accepting outputs
equal~$y_t$.
The resulting construction is therefore single-valued.

Since $y_t\notin\operatorname{Range}(C_t)$, the lemma gives
$\deg G_t\le d$, coefficients of bit length $O_d(\log t)$, and
\[
    \operatorname{asize}_{\Z}(G_t)=\Theta_d(t^d)
\]
for all sufficiently large~$t$.
Set $G_t=0$ at the finitely many remaining values of~$t$.
The upper bound places $\{G_t\}_{t\ge1}$ in~$\cc{VP}_{\Z}$,
completing the second alternative.
\end{proof}

\section{A~Kabanets--Impagliazzo Theorem over Finite Fields}\label{section:KI_finite_fields}

In~this section, we~prove \Cref{theorem:KI_finite_fields}.
The proof is~similar to~the argument of~Kabanets and Impagliazzo~\cite{KI04}, with partition polynomials in~place of~the permanent.
In~the proof, we~heavily use the notation of~\Cref{section:universal_polynomial}.

\ThmKIFiniteFields*
\begin{proof}
	By contradiction, suppose that $\cc{VP}_{\F}=\cc{VNP}_{\F}$ and $\NEXP\subseteq\PPoly$.
	Let $\mathcal U$ be~the universal family constructed for the characteristic of~$\F$.
	By~\Cref{lemma:packed_sat_polynomial}, the image of~$\mathcal U$ over~$\F$ belongs to~$\cc{VNP}_{\F}=\cc{VP}_{\F}$, so~$\operatorname{asize}_{\F}(\mathcal U_N)\le N^{\kappa}$ for a~constant~$\kappa$ and all sufficiently large~$N$.
	We~first show that $3$-$\cc{UNSAT}\in\cc{NSUBEXP}$.

	Let $\Psi$ be~a~$3$-CNF formula with $m$ clauses on~$n$ variables.
	As~in~\Cref{section:universal_polynomial}, $\Psi$ gives a~list~$\mathcal S$ with one table for every clause, which depends on~the variables of~this clause, and the vector $a_\Psi\in\{0,1\}^t$ that marks the local assignments satisfying the clause, where $t\le8m$ is~the number of~variables.
	Since $Z_{\mathcal S}(a_\Psi)$ is~the number of~satisfying assignments of~$\Psi$ and all nonzero coefficients of~$Z_{\mathcal S}$ are one, \Cref{lemma:masking} gives that $Z_{\mathcal S}(a_\Psi\odot w)\equiv0$ over~$\F$, where $w=(w_{j,\tau})$ are the entries of~$Z_{\mathcal S}$, if~and only if~$\Psi$ is~unsatisfiable.
	By~\Cref{lemma:partition_self_reduction}, every $Z_i$ is~obtained from $Z_{\mathcal S}$ by~setting some entries to~zero, so~\Cref{corollary:partition_circuits} with $N=2(n+t)$ gives $\operatorname{asize}_{\F}(Z_i)\le N^{\kappa}\log^{O(1)}N$ for every $i\in[n]$.
	Since $n\le3m$ and $t\le8m$, this bound is~polynomial in~the length~$\ell$ of~$\Psi$.

	The algorithm guesses circuits $C_1,\ldots,C_n$ over~$\F$ of~size $N^{\kappa}\log^{O(1)}N$ on~the $t$ entries, which should compute $Z_1,\ldots,Z_n$, and replaces every $C_i$ by~$\pi_{\le m}(C_i)$ by~\Cref{lemma:circuit_truncation}.
	Let $C_0$ be~a~circuit for the product $Z_0(w)=\prod_{j=1}^{m}w_{j,0^{S_j}}$.
	The algorithm tests the identities $C_i(w)=C_{i-1}(w)+C_{i-1}(w^{(i)})$ for every $i\in[n]$ and the identity $C_n(a_\Psi\odot w)=0$ by~running the $\cc{PIT}$ algorithm on~the difference of~the two sides, and accepts if~and only if~all tests pass.
	Every tested circuit has $t$ variables and degree at~most~$m$ where $m\le t$.
	Therefore, the algorithm runs in~nondeterministic time $2^{\ell^{\varepsilon}}$ for every $\varepsilon>0$ and sufficiently large~$\ell$.

	Note that if~all tests pass, then, by~induction on~$i$ and \Cref{lemma:partition_self_reduction}, every $C_i$ computes~$Z_i$, so~$Z_{\mathcal S}(a_\Psi\odot w)=C_n(a_\Psi\odot w)\equiv0$, and $\Psi$ is~unsatisfiable.
	Conversely, if~$\Psi$ is~unsatisfiable, then the algorithm can guess circuits of~the above size that compute $Z_1,\ldots,Z_n$.
	Since every $Z_i$ has degree at~most~$m$, the truncation keeps them correct, so~all tested identities hold, and the algorithm has an~accepting computation.
	Therefore, $3$-$\cc{UNSAT}\in\cc{NSUBEXP}$.

	Finally, we~derive the contradiction.
	Since $3$-$\cc{UNSAT}$ is~$\coNP$-complete, $\coNP\subseteq\cc{NSUBEXP}$.
	Hence $\Sigma_2^{\cc P}\subseteq\cc{NSUBEXP}$: for a~language in~$\Sigma_2^{\cc P}$ of~the form $\{x:\exists y\,\forall z\;R(x,y,z)\}$, the algorithm guesses~$y$ and decides the $\coNP$ condition $\forall z\;R(x,y,z)$ in~$\cc{NSUBEXP}$.
	Since $\NEXP\subseteq\PPoly$, we~have $\NEXP=\MA$~\cite{IKW02}, and $\MA\subseteq\Sigma_2^{\cc P}$.
	Thus $\NEXP\subseteq\cc{NSUBEXP}$, which contradicts the nondeterministic time hierarchy theorem~\cite{C73,Z83}.
\end{proof}

\section{AI~Methodology}\label{section:ai_methodology}

We~follow the recommendations of~the working group report~\cite{AITCS26} and~use its template for~this section.

\begin{flushleft}
	\renewcommand{\arraystretch}{1.3}
	\begin{tabular}{@{}l@{\qquad}c@{}}
		\textbf{Activity} & \textbf{AI used} \\
		Asking the research question & No \\
		Coming up with the approach & Yes \\
		Proof development & Yes \\
		Writing and exposition & Yes \\
		Checking for bugs & Yes \\
		Other supporting tasks & Yes \\
	\end{tabular}
\end{flushleft}

\paragraph{Details (elaborating on~the points above).}
For a~few years, the authors worked on~the idea that derandomizing $\cc{PIT}$ even for polynomials of~low degree implies circuit lower bounds.
Our initial idea was to~use the polynomial formulations of~\cite{BGKMS25,BKMRRS24}.
Following this idea, we~obtained a~weaker version of~\Cref{corollary:pit_np_size} and a~slightly weaker version of~\Cref{corollary:strong_avoid_pit_winwin}.
Starting from these results and our techniques, the model GPT~6~Astra improved them to~the versions stated in~\Cref{sec:introduction}.
It~also suggested \Cref{theorem:low_degree_pit_upper}, and it~found and proved \Cref{theorem:pit_fixed_degree}.
Then, over many rounds of~interactive discussion with Claude Opus~5.5, we~tried to~design a~fast $\cc{PIT}$ algorithm for restricted classes of~arithmetic circuits; this work led to~\Cref{theorem:zero_test} and, eventually, to~\Cref{theorem:unconditional_dichotomy}.
Thus, the tool materially affected these statements and their proofs in~\Cref{section:low_degree_pit,section:pit_np_size,section:win_win,section:strong_avoidance}.
Nevertheless, the whole text of~the paper was written by~the authors: we~carefully rewrote all proofs suggested by~the model, simplified and cleaned up the proofs, and carefully worked out the presentation of~all results.
The authors verified the correctness and originality of~all content, including references.
We~also used Claude Opus~5.5 to~help with finding typos and errors and with editing the exposition.
Below, we~state our initial results and compare them with the final ones.

\paragraph{Constant-degree $\cc{PIT}$.}
Our version of~\Cref{corollary:pit_np_size} assumed that $D$-$\cc{PIT}\in\NTIME[s^{o(D^{1/3})}]$ for every constant degree $D\ge1$.
It~concluded that either $\cc{E}^{\NP}$ requires circuits of~size $\omega(n)$, or, for every constant $\gamma\ge1$, there is~an~explicit family of~constant-degree polynomials on~$t$ variables over~$\Z$ requiring arithmetic circuits of~size $\Omega(t^{\gamma})$.
GPT~6~Astra derived \Cref{corollary:pit_np_size} from \Cref{theorem:pit_fixed_degree}, a~clever fixed-degree statement that it~found and proved based on~our techniques.

\paragraph{Satisfiability algorithms and Boolean lower bounds.}
Earlier, we~also wanted to~use the results of~Murray and Williams~\cite{MW18} and Williams~\cite{W13,W14}, which turn nontrivial satisfiability algorithms into circuit lower bounds, but we~did not manage to~apply them since we~had no~polynomial formulation for \cc{Circuit}-\cc{SAT}.
GPT~6~Astra applied them in~a~clever way via a~reduction to~branching programs.
To~decide unsatisfiability of~a~Boolean circuit, its algorithm guesses a~formula for every gate of~the circuit, converts each check into a~branching program, and verifies the branching programs by~$\cc{PIT}$ (see \Cref{section:verifying_bp,section:verifying_circuits}).
This gives the Boolean alternatives of~\Cref{theorem:pit_fixed_degree,corollary:pit_np_size}.

\paragraph{Range avoidance.}
We~proved \Cref{corollary:strong_avoid_pit_winwin} ourselves in~the following form.
If~$\cc{PIT}\in\NP$, then either $\cc{VP}\ne\cc{VNP}$, or~$\NP=\coNP$, in~which case $\cc{StrongAvoid}\in\cc{FNP}$, then $\cc{E}^{\NP}$ requires circuits of~size $2^n/(2n)$~\cite{K21}, and, for every fixed $D\ge1$, there is~an~$\cc{FNP}$-constructible family of~polynomials over~$\Z$ of~degree at~most~$D$ in~$\cc{VP}$ requiring arithmetic circuits of~size $\Omega_D(t^D/\log t)$.
GPT~6~Astra suggested an~elegant way to~shave the logarithmic factor (see \Cref{lemma:hard_poly_to_avoid}), which gives the tight bound $\Theta_d(t^d)$ in~\Cref{corollary:strong_avoid_pit_winwin}.

\paragraph{Unconditional win--win result over the integers.}
\Cref{theorem:unconditional_dichotomy} was obtained while we~tried to~derive unconditional circuit lower bounds by~creating fast \cc{PIT} and~plugging it~into~\Cref{theorem:pit_fixed_degree}; the proof was found by~Claude Opus~5.5, based on~the techniques we~had in~this paper.

\bibliographystyle{alpha}
\bibliography{refs}

\newcommand{\etalchar}[1]{$^{#1}$}
\begin{thebibliography}{FPdJdV15}

\bibitem[Aar06]{A06}
Scott Aaronson.
\newblock {Oracles Are Subtle But Not Malicious}.
\newblock In {\em CCC}, 2006.

\bibitem[ABKM09]{ABKM09}
Eric Allender, Peter B{\"{u}}rgisser, Johan Kjeldgaard{-}Pedersen, and
  Peter~Bro Miltersen.
\newblock On the complexity of numerical analysis.
\newblock {\em {SIAM} Journal on Computing}, 2009.

\bibitem[ACDM18]{ACDM18}
Vikraman Arvind, Abhranil Chatterjee, Rajit Datta, and Partha Mukhopadhyay.
\newblock Beating brute force for polynomial identity testing of general
  depth-3 circuits.
\newblock {\em Electron. Colloquium Comput. Complex.}, {TR18-111}, 2018.

\bibitem[Agr05]{A05}
Manindra Agrawal.
\newblock {Proving Lower Bounds Via Pseudo-random Generators}.
\newblock In {\em FSTTCS}, 2005.

\bibitem[AIT26]{AITCS26}
{AI and TCS: The Next Six Months. Working Group Report}.
\newblock Simons Institute for the Theory of Computing,
  \url{https://simons.berkeley.edu/AITCSreport}, 2026.

\bibitem[AJS09]{AJS09}
Vikraman Arvind, Pushkar~S. Joglekar, and Srikanth Srinivasan.
\newblock Arithmetic circuits and the {H}adamard product of polynomials.
\newblock In {\em {FSTTCS}}, 2009.

\bibitem[AKT25]{AKT25}
Robert Andrews, Deepanshu Kush, and Roei Tell.
\newblock {Polynomial-Time {PIT} from (Almost) Necessary Assumptions}.
\newblock In {\em {STOC}}, 2025.

\bibitem[{Ant}26]{Ant26}
{Anthropic}.
\newblock {A $2^{o(n)}$ Time Algorithm for {3-SAT}}.
\newblock \url{https://github.com/anthropics/formal-math/tree/main/eth}, 2026.

\bibitem[AvM11]{AM11}
Scott Aaronson and Dieter van Melkebeek.
\newblock {On Circuit Lower Bounds from Derandomization}.
\newblock {\em Theory of Computing}, 2011.

\bibitem[AW26]{AVW26}
Josh Alman and Virginia~Vassilevska Williams.
\newblock {Truly Subquadratic {3SUM} and Truly Subcubic {APSP} via Triangles in
  Sparse Lopsided Graphs}, 2026.

\bibitem[Bab85]{B85}
L{\'{a}}szl{\'{o}} Babai.
\newblock {Trading Group Theory for Randomness}.
\newblock In {\em {STOC}}, 1985.

\bibitem[BBDM26]{BBDM26}
Somnath Bhattacharjee, Markus Bl{\"{a}}ser, Pranjal Dutta, and Saswata
  Mukherjee.
\newblock Exponential lower bound via exponential sums.
\newblock {\em CoRR}, abs/2601.00387, 2026.

\bibitem[BGK{\etalchar{+}}25]{BGKMS25}
Tatiana Belova, Alexander Golovnev, Alexander~S. Kulikov, Ivan Mihajlin, and
  Denil Sharipov.
\newblock {Polynomial Formulations as a Barrier for Reduction-Based Hardness
  Proofs}.
\newblock {\em {ACM} Transactions on Algorithms}, 2025.

\bibitem[BH93]{BH93}
Amir Ben{-}Dor and Shai Halevi.
\newblock {Zero-One Permanent is {\#P}-Complete, A Simpler Proof}.
\newblock In {\em {ISTCS}}, 1993.

\bibitem[BKM{\etalchar{+}}24]{BKMRRS24}
Tatiana Belova, Alexander~S. Kulikov, Ivan Mihajlin, Olga Ratseeva, Grigory
  Reznikov, and Denil Sharipov.
\newblock {Computations with polynomial evaluation oracle: Ruling out
  superlinear SETH-based lower bounds}.
\newblock In {\em {SODA}}, 2024.

\bibitem[BM88]{BM88}
L{\'{a}}szl{\'{o}} Babai and Shlomo Moran.
\newblock {Arthur-Merlin Games: {A} Randomized Proof System, and a Hierarchy of
  Complexity Classes}.
\newblock {\em Journal of Computer and System Sciences}, 1988.

\bibitem[BS83]{BS83}
Walter Baur and Volker Strassen.
\newblock {The Complexity of Partial Derivatives}.
\newblock {\em Theoretical Computer Science}, 1983.

\bibitem[B{\"u}r00]{Burgisser00}
Peter B{\"u}rgisser.
\newblock {Cook's versus Valiant's Hypothesis}.
\newblock {\em Theoretical Computer Science}, 2000.

\bibitem[B{\"{u}}r09]{B09}
Peter B{\"{u}}rgisser.
\newblock On defining integers and proving arithmetic circuit lower bounds.
\newblock {\em Computational Complexity}, 18(1):81--103, 2009.

\bibitem[Cai07]{C07}
Jin-Yi Cai.
\newblock {$\mathrm{S}_2^p \subseteq \mathrm{ZPP}^{\mathrm{NP}}$}.
\newblock {\em Journal of Computer and System Sciences}, 2007.

\bibitem[CGI{\etalchar{+}}16]{CGIMPS16}
Marco~L. Carmosino, Jiawei Gao, Russell Impagliazzo, Ivan Mihajlin, Ramamohan
  Paturi, and Stefan Schneider.
\newblock {Nondeterministic Extensions of the Strong Exponential Time
  Hypothesis and Consequences for Non-reducibility}.
\newblock In {\em {ITCS}}, 2016.

\bibitem[Che25]{Che25}
Lijie Chen.
\newblock {Nondeterministic Quasi-Polynomial Time is Average-Case Hard for
  {ACC} Circuits}.
\newblock {\em {SIAM} Journal on Computing}, 2025.

\bibitem[CHLR23]{CHLR23}
Yeyuan Chen, Yizhi Huang, Jiatu Li, and Hanlin Ren.
\newblock {Range Avoidance, Remote Point, and Hard Partial Truth Tables via
  Satisfying-Pairs Algorithms}.
\newblock In {\em STOC}, 2023.

\bibitem[CHLR26]{CHLR26}
Lijie Chen, Shuichi Hirahara, Zeyong Li, and Hanlin Ren.
\newblock {Symmetric Exponential Time Requires Near-Maximum Circuit Size}.
\newblock {\em Journal of the ACM}, 2026.

\bibitem[CIKK15]{CIKK15}
Marco~L. Carmosino, Russell Impagliazzo, Valentine Kabanets, and Antonina
  Kolokolova.
\newblock {Tighter Connections between Derandomization and Circuit Lower
  Bounds}.
\newblock In {\em RANDOM}, 2015.

\bibitem[CKS18]{CKS18}
Chi-Ning Chou, Mrinal Kumar, and Noam Solomon.
\newblock {Hardness vs Randomness for Bounded Depth Arithmetic Circuits}.
\newblock In {\em CCC}, 2018.

\bibitem[Coo73]{C73}
Stephen~A. Cook.
\newblock {A Hierarchy for Nondeterministic Time Complexity}.
\newblock {\em J. Comput. Syst. Sci.}, 7(4):343--353, 1973.

\bibitem[CR20]{CR20}
Lijie Chen and Hanlin Ren.
\newblock Strong average-case circuit lower bounds from non-trivial
  derandomization.
\newblock {\em Electron. Colloquium Comput. Complex.}, {TR20-010}, 2020.

\bibitem[CRTY23]{CRTY23}
Lijie Chen, Ron~D. Rothblum, Roei Tell, and Eylon Yogev.
\newblock {On Exponential-time Hypotheses, Derandomization, and Circuit Lower
  Bounds}.
\newblock {\em Journal of the {ACM}}, 2023.

\bibitem[DL78]{DL78}
Richard~A. DeMillo and Richard~J. Lipton.
\newblock {A Probabilistic Remark on Algebraic Program Testing}.
\newblock {\em Information Processing Letters}, 1978.

\bibitem[DST24]{DST24}
Pranjal Dutta, Nitin Saxena, and Thomas Thierauf.
\newblock {Weighted Sum-of-Squares Lower Bounds for Univariate Polynomials
  Imply {VP} {\(\not =\)} {VNP}}.
\newblock {\em Computational Complexity}, 2024.

\bibitem[FGHK16]{FGHK16}
Magnus~Gausdal Find, Alexander Golovnev, Edward~A. Hirsch, and Alexander~S.
  Kulikov.
\newblock A better-than-$3n$ lower bound for the circuit complexity of an
  explicit function.
\newblock In {\em {FOCS}}, pages 89--98, 2016.

\bibitem[For24]{F24}
Michael~A. Forbes.
\newblock {Low-Depth Algebraic Circuit Lower Bounds over Any Field}.
\newblock In {\em CCC}, 2024.

\bibitem[FPdJdV15]{FPV15}
Herv{\'{e}} Fournier, Sylvain Perifel, and R{\'{e}}mi de~Joannis~de Verclos.
\newblock {On fixed-polynomial size circuit lower bounds for uniform
  polynomials in the sense of Valiant}.
\newblock {\em Information and Computation}, 2015.

\bibitem[FS13]{FS13}
Michael~A. Forbes and Amir Shpilka.
\newblock {Quasipolynomial-Time Identity Testing of Non-Commutative and
  Read-Once Oblivious Algebraic Branching Programs}.
\newblock In {\em FOCS}, 2013.

\bibitem[FSW09]{FSW09}
Lance Fortnow, Rahul Santhanam, and Ryan Williams.
\newblock {Fixed-Polynomial Size Circuit Bounds}.
\newblock In {\em {CCC}}, 2009.

\bibitem[GGNS23]{GGNS23}
Karthik Gajulapalli, Alexander Golovnev, Satyajeet Nagargoje, and Sidhant
  Saraogi.
\newblock Range avoidance for constant-depth circuits: Hardness and algorithms.
\newblock In {\em {RANDOM}}, 2023.

\bibitem[GKSS22]{GKSS22}
Zeyu Guo, Mrinal Kumar, Ramprasad Saptharishi, and Noam Solomon.
\newblock {Derandomization from Algebraic Hardness}.
\newblock {\em {SIAM} Journal on Computing}, 2022.

\bibitem[GKST18]{GKST18}
Alexander Golovnev, Alexander~S. Kulikov, Alexander~V. Smal, and Suguru Tamaki.
\newblock {Gate elimination: circuit size lower bounds and \#SAT upper bounds}.
\newblock {\em Theoretical Computer Science}, 2018.

\bibitem[GLV24]{GLV24}
Karthik Gajulapalli, Zeyong Li, and Ilya Volkovich.
\newblock {Oblivious Complexity Classes Revisited: Lower Bounds and
  Hierarchies}.
\newblock In {\em FSTTCS}, 2024.

\bibitem[GLW22]{GLW22}
Venkatesan Guruswami, Xin Lyu, and Xiuhan Wang.
\newblock Range avoidance for low-depth circuits and connections to
  pseudorandomness.
\newblock In {\em {RANDOM}}, 2022.

\bibitem[GM15]{GM15}
Oded Goldreich and Or~Meir.
\newblock {Input-Oblivious Proof Systems and a Uniform Complexity Perspective
  on P/poly}.
\newblock {\em {ACM} Transactions on Computation Theory}, 2015.

\bibitem[GP18]{GP18}
Joshua~A. Grochow and Toniann Pitassi.
\newblock {Circuit Complexity, Proof Complexity, and Polynomial Identity
  Testing: The Ideal Proof System}.
\newblock {\em Journal of the {ACM}}, 2018.

\bibitem[HNOS96]{HNOS96}
Lane~A. Hemaspaandra, Ashish~V. Naik, Mitsunori Ogihara, and Alan~L. Selman.
\newblock {Computing Solutions Uniquely Collapses the Polynomial Hierarchy}.
\newblock {\em {SIAM} Journal on Computing}, 1996.

\bibitem[HS80]{HS80}
Joos Heintz and Claus-Peter Schnorr.
\newblock {Testing Polynomials Which Are Easy to Compute}.
\newblock In {\em STOC}, 1980.

\bibitem[HWY10]{HWY10}
Pavel Hrubes, Avi Wigderson, and Amir Yehudayoff.
\newblock {Relationless Completeness and Separations}.
\newblock In {\em {CCC}}, 2010.

\bibitem[Hya79]{H79}
Laurent Hyafil.
\newblock On the parallel evaluation of multivariate polynomials.
\newblock {\em SIAM Journal on Computing}, 8(2):120--123, 1979.

\bibitem[IKW02]{IKW02}
Russell Impagliazzo, Valentine Kabanets, and Avi Wigderson.
\newblock In search of an easy witness: exponential time vs. probabilistic
  polynomial time.
\newblock {\em Journal of Computer and System Sciences}, 2002.

\bibitem[ILW23]{ILW23}
Rahul Ilango, Jiatu Li, and R.~Ryan Williams.
\newblock {Indistinguishability Obfuscation, Range Avoidance, and Bounded
  Arithmetic}.
\newblock In {\em STOC}, 2023.

\bibitem[IP01]{IP01}
Russell Impagliazzo and Ramamohan Paturi.
\newblock {On the Complexity of $k$-SAT}.
\newblock {\em Journal of Computer and System Sciences}, 2001.

\bibitem[JMV18]{JMV18}
Hamidreza Jahanjou, Eric Miles, and Emanuele Viola.
\newblock Local reduction.
\newblock {\em Information and Computation}, 2018.

\bibitem[JS12]{JS12}
Maurice~J. Jansen and Rahul Santhanam.
\newblock {Stronger Lower Bounds and Randomness-Hardness Trade-Offs Using
  Associated Algebraic Complexity Classes}.
\newblock In {\em {STACS}}, 2012.

\bibitem[Kan82]{K82}
Ravi Kannan.
\newblock {Circuit-Size Lower Bounds and Non-Reducibility to Sparse Sets}.
\newblock {\em Information and Control}, 1982.

\bibitem[KI04]{KI04}
Valentine Kabanets and Russell Impagliazzo.
\newblock {Derandomizing Polynomial Identity Tests Means Proving Circuit Lower
  Bounds}.
\newblock {\em Computational Complexity}, 2004.

\bibitem[KKMP21]{KKMP21}
Robert Kleinberg, Oliver Korten, Manolis Mitropolsky, and Christos~H.
  Papadimitriou.
\newblock Total functions in the polynomial hierarchy.
\newblock In {\em {ITCS}}, 2021.

\bibitem[Koi05]{K05}
Pascal Koiran.
\newblock Valiant's model and the cost of computing integers.
\newblock {\em Computational Complexity}, 13(3-4):131--146, 2005.

\bibitem[Kor21]{K21}
Oliver Korten.
\newblock {The Hardest Explicit Construction}.
\newblock In {\em {FOCS}}, 2021.

\bibitem[KP11]{KP11}
Pascal Koiran and Sylvain Perifel.
\newblock Interpolation in {Valiant's} theory.
\newblock {\em Computational Complexity}, 20(1):1--20, 2011.

\bibitem[KP24]{KP24}
Oliver Korten and Toniann Pitassi.
\newblock Strong vs. weak range avoidance and the linear ordering principle.
\newblock In {\em {FOCS}}, pages 1388--1407, 2024.

\bibitem[KS19]{KS19}
Mrinal Kumar and Ramprasad Saptharishi.
\newblock {Hardness-Randomness Tradeoffs for Algebraic Computation}.
\newblock {\em Bulletin of {EATCS}}, 2019.

\bibitem[KST19]{KST19}
Mrinal Kumar, Ramprasad Saptharishi, and Anamay Tengse.
\newblock {Near-optimal Bootstrapping of Hitting Sets for Algebraic Circuits}.
\newblock In {\em {SODA}}, 2019.

\bibitem[KvMS12]{KMS12}
Jeff Kinne, Dieter van Melkebeek, and Ronen Shaltiel.
\newblock {Pseudorandom Generators, Typically-Correct Derandomization, and
  Circuit Lower Bounds}.
\newblock {\em Computational Complexity}, 2012.

\bibitem[LFKN92]{LFKN92}
Carsten Lund, Lance Fortnow, Howard~J. Karloff, and Noam Nisan.
\newblock {Algebraic Methods for Interactive Proof Systems}.
\newblock {\em Journal of the {ACM}}, 1992.

\bibitem[LMP19]{LMP19}
Guillaume Lagarde, Guillaume Malod, and Sylvain Perifel.
\newblock Non-commutative computations: lower bounds and polynomial identity
  testing.
\newblock {\em Chicago Journal of Theoretical Computer Science}, 2019.

\bibitem[LST25]{LST25}
Nutan Limaye, Srikanth Srinivasan, and S{\'{e}}bastien Tavenas.
\newblock {Superpolynomial Lower Bounds Against Low-Depth Algebraic Circuits}.
\newblock {\em Journal of the {ACM}}, 2025.

\bibitem[LY22]{LY22}
Jiatu Li and Tianqi Yang.
\newblock {$3.1n-o(n)$} circuit lower bounds for explicit functions.
\newblock In {\em {STOC}}, pages 1180--1193, 2022.

\bibitem[Mal03]{M03}
Guillaume Malod.
\newblock {\em Polyn{\^o}mes et coefficients}.
\newblock PhD thesis, {Universit{\'e} Claude Bernard Lyon~1}, France, 2003.

\bibitem[MP08]{MP08}
Guillaume Malod and Natacha Portier.
\newblock Characterizing {V}aliant's algebraic complexity classes.
\newblock {\em Journal of Complexity}, 24(1):16--38, 2008.

\bibitem[MS26]{MS26}
Om~Swostik Mishra and Rahul Santhanam.
\newblock An explicit {$3.5n-o(n)$} size lower bound against general {Boolean}
  circuits.
\newblock {\em Electron. Colloquium Comput. Complex.}, TR26-260, 2026.

\bibitem[MW18]{MW18}
Cody Murray and R.~Ryan Williams.
\newblock Circuit lower bounds for nondeterministic quasi-polytime: an easy
  witness lemma for {NP} and {NQP}.
\newblock In {\em {STOC}}, 2018.

\bibitem[Nis91]{N91}
Noam Nisan.
\newblock {Lower Bounds for Non-Commutative Computation}.
\newblock In {\em {STOC}}, 1991.

\bibitem[{Ope}26]{OpenAI26}
{OpenAI}.
\newblock The exponential time hypothesis is false.
\newblock Preprint, \url{https://github.com/openai/math}, 2026.

\bibitem[RS05]{RS05}
Ran Raz and Amir Shpilka.
\newblock Deterministic polynomial identity testing in non-commutative models.
\newblock {\em Computational Complexity}, 2005.

\bibitem[RSW22]{RSW22}
Hanlin Ren, Rahul Santhanam, and Zhikun Wang.
\newblock On the range avoidance problem for circuits.
\newblock In {\em {FOCS}}, pages 640--650, 2022.

\bibitem[San09]{Santhanam09}
Rahul Santhanam.
\newblock {Circuit Lower Bounds for Merlin--Arthur Classes}.
\newblock {\em {SIAM} Journal on Computing}, 2009.

\bibitem[Sax14]{S14}
Nitin Saxena.
\newblock {Progress on Polynomial Identity Testing - {II}}.
\newblock {\em CoRR}, abs/1401.0976, 2014.

\bibitem[Sch80]{S80}
Jacob~T. Schwartz.
\newblock {Fast Probabilistic Algorithms for Verification of Polynomial
  Identities}.
\newblock {\em Journal of the {ACM}}, 1980.

\bibitem[Sel94]{Sel94}
Alan~L. Selman.
\newblock {A Taxonomy of Complexity Classes of Functions}.
\newblock {\em Journal of Computer and System Sciences}, 1994.

\bibitem[Sha49]{Sha49}
Claude~E. Shannon.
\newblock The synthesis of two-terminal switching circuits.
\newblock {\em Bell System Technical Journal}, 28(1):59--98, 1949.

\bibitem[Sho05]{S05}
Victor Shoup.
\newblock {\em {A Computational Introduction to Number Theory and Algebra}}.
\newblock Cambridge University Press, 2005.

\bibitem[SS95]{SS95}
Michael Shub and Steve Smale.
\newblock On the intractability of {Hilbert's} {Nullstellensatz} and an
  algebraic version of ``{NP} $\ne$ {P}?''.
\newblock {\em Duke Mathematical Journal}, 1995.

\bibitem[Str73]{S73}
Volker Strassen.
\newblock Vermeidung von divisionen.
\newblock {\em Journal f{\"u}r die reine und angewandte Mathematik}, 1973.

\bibitem[Str75]{S75}
Volker Strassen.
\newblock {Die Berechnungskomplexit{\"{a}}t der symbolischen Differentiation
  von Interpolationspolynomen}.
\newblock {\em Theoretical Computer Science}, 1975.

\bibitem[SY10]{SY10}
Amir Shpilka and Amir Yehudayoff.
\newblock {Arithmetic Circuits: {A} survey of recent results and open
  questions}.
\newblock {\em Foundations and Trends in Theoretical Computer Science}, 2010.

\bibitem[Tod91]{T91}
Seinosuke Toda.
\newblock {{PP} is as Hard as the Polynomial-Time Hierarchy}.
\newblock {\em {SIAM} Journal on Computing}, 1991.

\bibitem[Tra84]{Tra84}
Boris~A. Trakhtenbrot.
\newblock {A Survey of {R}ussian Approaches to \emph{Perebor} (Brute-Force
  Search) Algorithms}.
\newblock {\em Annals of the History of Computing}, 1984.

\bibitem[Tse83]{T83}
Grigori~S. Tseitin.
\newblock On the complexity of derivation in propositional calculus.
\newblock In {\em {Automation of reasoning: 2: Classical papers on
  computational logic 1967--1970}}. Springer, 1983.

\bibitem[Uma03]{Uma03}
Christopher Umans.
\newblock {Pseudo-random generators for all hardnesses}.
\newblock {\em Journal of Computer and System Sciences}, 2003.

\bibitem[Val79a]{V79a}
Leslie~G. Valiant.
\newblock {Completeness Classes in Algebra}.
\newblock In {\em {STOC}}, 1979.

\bibitem[Val79b]{V79}
Leslie~G. Valiant.
\newblock {The Complexity of Computing the Permanent}.
\newblock {\em Theoretical Computer Science}, 1979.

\bibitem[Val82]{V82}
Leslie~G. Valiant.
\newblock {Reducibility by algebraic projections}.
\newblock {\em L’Enseignement Mathematique}, 1982.

\bibitem[Vin05]{V05}
N.~V. Vinodchandran.
\newblock {A Note on the Circuit Complexity of {PP}}.
\newblock {\em Theoretical Computer Science}, 2005.

\bibitem[VSBR83]{VSBR83}
Leslie~G. Valiant, Sven Skyum, S.~Berkowitz, and Charles Rackoff.
\newblock F{ast Parallel Computation of Polynomials Using Few Processors}.
\newblock {\em {SIAM} Journal on Computing}, 1983.

\bibitem[Weg87]{W87}
Ingo Wegener.
\newblock {\em The complexity of Boolean functions}.
\newblock Wiley-Teubner, 1987.

\bibitem[Wil13]{W13}
Ryan Williams.
\newblock {Improving Exhaustive Search Implies Superpolynomial Lower Bounds}.
\newblock {\em {SIAM} Journal on Computing}, 2013.

\bibitem[Wil14]{W14}
Ryan Williams.
\newblock {Nonuniform ACC Circuit Lower Bounds}.
\newblock {\em Journal of the {ACM}}, 2014.

\bibitem[Z{\'{a}}k83]{Z83}
Stanislav Z{\'{a}}k.
\newblock {A Turing Machine Time Hierarchy}.
\newblock {\em Theor. Comput. Sci.}, 26:327--333, 1983.

\bibitem[Zip79]{Z79}
Richard Zippel.
\newblock Probabilistic algorithms for sparse polynomials.
\newblock In {\em {EUROSAM}}, 1979.

\end{thebibliography}

\end{document}